\newcommand\ignore[1]{}

\documentclass[11pt, oneside]{article}   	
\usepackage{geometry}                		
\usepackage[parfill]{parskip}    		
\usepackage{graphicx}				
\usepackage{color, colortbl}								
\usepackage[left]{lineno}
\usepackage{amssymb}
\usepackage{amsmath}
\usepackage{amssymb}
\usepackage{graphicx}
\usepackage{hyperref}
\usepackage{float}
\usepackage{fullpage}
\usepackage{braket}
\usepackage{url}
\usepackage{cite}
\usepackage{subcaption}
\usepackage{upgreek}
\usepackage{outlines}
\usepackage{braket}
\usepackage{appendix}
\usepackage{xspace}
\usepackage{tikz}
\usepackage{placeins}
\usepackage{multirow}
\usepackage{sidecap}
\usepackage{soul}
\usepackage{enumitem}

\usetikzlibrary{matrix,arrows,decorations.pathmorphing}
\usetikzlibrary{patterns}
\usetikzlibrary{trees}
\usetikzlibrary{decorations.pathmorphing}
\usetikzlibrary{decorations.markings}
\usetikzlibrary{positioning,arrows}
\usetikzlibrary{decorations.pathreplacing}
\usetikzlibrary{decorations.pathmorphing}
\usetikzlibrary{decorations.markings}

\usepackage{etoolbox} 

\newcommand*\linenomathpatch[1]{%
	\cspreto{#1}{\linenomath}%
	\cspreto{#1*}{\linenomath}%
	\csappto{end#1}{\endlinenomath}%
	\csappto{end#1*}{\endlinenomath}%
}

\linenomathpatch{equation}
\linenomathpatch{gather}
\linenomathpatch{multline}
\linenomathpatch{align}
\linenomathpatch{alignat}
\linenomathpatch{flalign}

\newcommand\be{\begin{equation}}
\newcommand\ee{\end{equation}}
\newcommand\bea{\begin{eqnarray}}
\newcommand\eea{\end{eqnarray}}

\newcommand{\bdm}{\begin{displaymath}}
\newcommand{\edm}{\end{displaymath}}
\newcommand\nn{ \nonumber\\}

\numberwithin{equation}{section}
\numberwithin{figure}{section}

\def\C{\mathbb{C}}

\renewcommand{\epsilon}{\varepsilon}

\usepackage{authblk}

\renewcommand{\phi}{\varphi}

\graphicspath{{/}{fig/}{rootFigs/}}

\title{Embedding Space Approach to Lorentzian CFT Amplitudes\\ and \\Causal Spherical Functions}

\author[*]{Pulkit Agarwal}
\author[**]{Richard Brower}
\author[+]{Timothy Raben}
\author[++]{Chung-I Tan}
\affil[*]{Centro de Física do Porto, Faculdade de Ci\^{e}ncias da Universidade do Porto, Portugal}
\affil[**]{Boston University, Boston, MA 02215}
\affil[+]{Michigan State University} 
\affil[++]{Brown University, Providence, RI 02912}

\begin{document}

\maketitle

\begin{abstract}
	Conformal Field Theory  in a Minkowski setting is discussed  in an embedding space approach, paying special attention to  causality constraints for four-point amplitudes. The physics of dilatation and Lorentz boost is emphasized in specifying the non-compact Maximal Abelian subgroup (MASG) of $SO(d,2)$. Reduction of a Conformal Field Theory (CFT) four-point amplitudes as  functions of cross ratios is shown to be equivalent to enforcing $H$ bi-invariance, i.e., $F(hgh')=F(g)$, with $g\in SO(d,2)$ and $H$ an appropriate subgroup. Causality is imposed by introducing appropriate semigroups. Causal zonal spherical functions are constructed, making contact with Minkowski conformal blocks introduced previously.
\end{abstract}
\pagenumbering{gobble}

\newpage
\tableofcontents
\pagenumbering{gobble}

\newpage
\setlength{\parskip}{.1in}

\setlength{\parindent}{0.2 in}
\newpage

\section{Introduction}\label{sec:intro}

\pagenumbering{arabic}
\flushbottom

Most current CFT studies have  been carried out in a Euclidean framework, i.e., working with the group $SO(d+1,1)$. There has been increasing recent interest in the relation between CFT in an Euclidean setting to its counter part in a Minkowski setting. Most notably is the work of Caron-Huot  \cite{caron2017analyticity}, where  a connection was made between OPE expansion coefficients in terms of Euclidean conformal blocks with an integral over matrix elements of double-commutators (to be precise, relating to a ``double discontinuity"). Many other interesting  recent works have also appeared. In particular, Simmons-Duffin, Stanford and Witten \cite{simmons2018spacetime} offer a Minkowski spacetime derivation of Caron-Huot's result and discusses the causal structures involved. By introducing a confinement scale or considering current flows, ``CFT scattering'', an intrinsically Minkowski phenomenon,  has  also been an  area of active research~\cite{Brower:2006ea,Cornalba:2006xm,Cornalba:2006xk,Brower:2007qh,Brower:2007xg,Cornalba:2007zb,Cornalba:2007fs,Hofman:2008ar,Strassler:2008bv,Cornalba:2008qf,Cornalba:2009ax,Costa:2012cb,Brower:2014wha,Belitsky:2013xxa,Belitsky:2013ofa,Banks:2009bj,Nally:2017nsp,Raben:2018rbn,carmi2020conformal}. Two important related areas are the Out of Time Order Correlators (OTOCs) \cite{Maldacena:2015waa,Gu:2018jsv,Gu:2021xaj,Shenker:2014cwa} in models like the SYK \cite{kitaev1,kitaev2,Maldacena:2016hyu,Polchinski:2016xgd}, and celestial holography \cite{raclariu2021lectures,strominger2018lectures,Ball:2018prg}. In many of these and related works the issue of CFT Regge behavior has begun to play an increasingly important role.

In Minkowski space, a double-commutator, e.g.,  $\langle 0|[\phi(x_4),\phi(x_1)] [\phi(x_2),\phi(x_3)]|0\rangle$, can be identified with an appropriate  discontinuity of a four-point time-ordered correlators. Such time-ordered amplitude, $\langle 0|T(\phi(x_4)\phi(x_1) \phi(x_2)\phi(x_3))|0\rangle$, defines a Green's function, and  it can be used to study the ``$s$-channel" scattering process, 
\be
\Phi(1) + \Phi(3) \rightarrow \Phi(2) + \Phi(4)\, . \label{eq:4phi}
\ee
In a theory with explicit particle interpretation, such time-ordered amplitudes are equivalently referred to as off-shell scattering amplitudes. We shall use these notions (Lorenztian CFT correlators and scattering amplitudes) interchangeably for our study. (A concrete example to consider is  a scattering amplitude involving currents.)  Unlike its Euclidean counter part, {\it these amplitudes are complex in the scattering regions}~\footnote{An Euclidean fourpoint amplitude is real and can be analytically continued into Lorentzian regions. It remains real so long as all points are spacelike separated, i.e., $[\phi(x_i),\phi(x_j)]=0$, (see Fig. \ref{fig:crsym}).  Regions where $[\phi(x_i),\phi(x_j)]\neq 0$ can be reached by crossing kinematic (lightcone) singularities, e.g., boundaries of regions $M_s$, $M_t$ and $M_u$ in Fig. \ref{fig:crsym}. Scattering regions correspond to non-vanishing double-commutators.}.

The main goal of this paper is to make precise  the claim made in Refs.  \cite{Raben:2018rbn,kravchuk2018light,mack2007simple} that {\it  Lorentzian CFT 4-point amplitudes can be represented  directly in terms of the principal series representation of $SO(d,2)$ and to construct causal representation functions appropriate for Lorentzian physics directly from an embedding space approach}, (see Eq. (\ref{eq:double-mellin}) below).  

In much of the previous Lorentzian CFT work, results are obtained by working in a Euclidean framework and then performing an analytic continuation to Minkowski space at the end, e.g., \cite{Costa:2012cb,Mack:2019akh}. This in principle engages in group complexification.  A serious attempt in connecting Euclidean CFTs to their Minkowski counter parts from an axiomatic approach has begun recently by Kravchuk, Qiao and Rychkov \cite{Kravchuk:2020scc,Kravchuk:2021kwe}, (see additional references therein).

In a  CFT, correlation functions can be described via a ``partial wave analysis'' where the dynamics of the process are captured by partial wave amplitudes and the kinematics is encoded in the \emph{conformal blocks} (see \cite{qualls2015lectures,rychkov2017epfl} for a more complete introduction). A most intriguing result in \cite{caron2017analyticity} is the finding that  it is possible to analytically continue Euclidean partial wave amplitude, a projection initially defined  over the Euclidean region, Fig. \ref{fig:crsym}(a), as a function of complex angular momentum, $\ell$. The procedure models after the technique of ``Froissart-Gribov" continuation known in standard Regge theories. That is, instead of ordinary Legendre polynomials, $P_\ell(z)$, with $\ell$ integral, and $-1<z<1$, the projection is analytically continued to $Q_\ell(z)$, Legendre function of the second kind, with $\ell$ complex over the range $1<z<\infty$. The new projection is over non-vanishing ``double discontinuities" in Lorentzian regions and can be identified with ``double-commutators" \cite{caron2017analyticity,simmons2018spacetime}.

An initial direct examination of conformal blocks in a Minkowski setting was carried out in  \cite{Raben:2018rbn}, which plays the analogous role of $Q_\ell(z)$ in Regge  theories.  A related work using the analytic continuation approach can also be  found in Kravchuk and Simmons-Duffin ~\cite{kravchuk2018light}, where the importance of a direct Minkowski approach was also emphasized. This perspective  has  previously  been discussed   by Mack~\cite{mack2007simple}.

\begin{figure}
\centering
\includegraphics[width=0.5\textwidth]{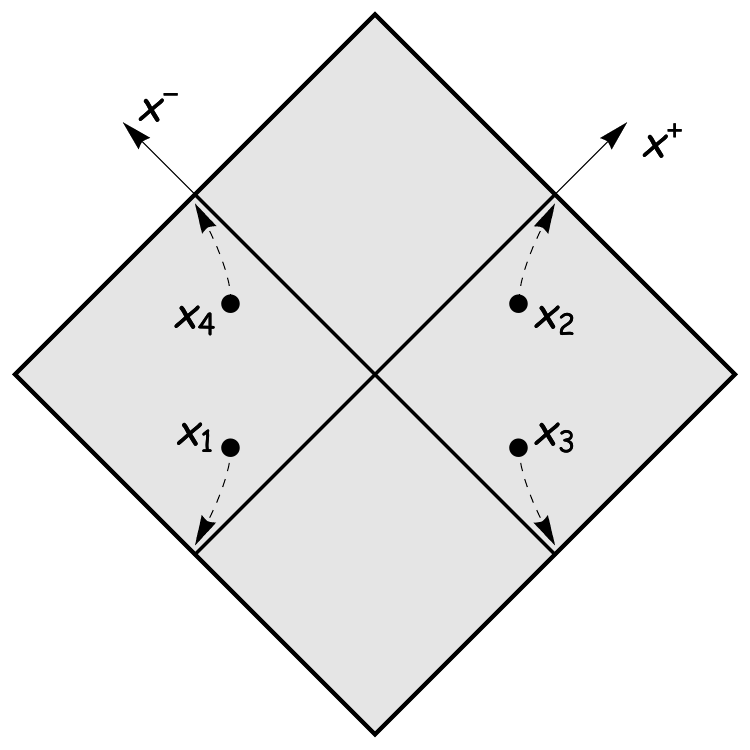}
\caption{\label{fig:DLC-limit}The $ s $-channel scattering region. Switching $ x_3\leftrightarrow x_4 $ leads to the $ u $-channel scattering. Figure taken from Ref. \cite{Brower:2014wha}. }
\end{figure}

Note that a double-commutator  is non-vanishing only in  certain configurations in Minkowski space. Consider $\langle 0|[\phi(x_4),\phi(x_1)] [\phi(x_2),\phi(x_3)]|0\rangle$; it is non-vanishing, 
\be
\langle 0|[\phi(x_4),\phi(x_1)] [\phi(x_2),\phi(x_3)]|0\rangle\neq 0 , \label{eq:scausality}
\ee
only when $x_{14}$ and $x_{23}$ are both time-like separated. Since a four-point correlator can also be defined outside these regions, they are therefore defined piecewise over different kinematic regions by use of step functions, and Minkowski amplitudes can be thought of as \textit{distributions}.  (The issue of analytic continuation outside of physical scattering regions will be touched upon briefly  in the discussion section relating to dispersion relations.) The configuration of space-time points, (\ref{eq:scausality}), signifies  ``{\bf causality}" at work, a concept which is meaningful only in a Minkowski setting. Fig. \ref{fig:DLC-limit} provides a kinematic interpretation in terms of a lightcone diagram, which can also be associated with the notion of what we will refer to as ``Minkowski $t$-channel OPE"  limit, by identifying with lightcone singularities at $x_{12}^2=x_{34}^2=0$. (For the kinematic region involved, see discussion in Sec. 2.2.)
This also ties in with Regge limits for CFT~\cite{Brower:2006ea,Cornalba:2006xm,Cornalba:2006xk,Brower:2007qh,Brower:2007xg,Cornalba:2007zb,Cornalba:2007fs,Hofman:2008ar,Strassler:2008bv,Cornalba:2008qf,Cornalba:2009ax,Costa:2012cb,Brower:2014wha,Belitsky:2013xxa,Belitsky:2013ofa,Banks:2009bj,Nally:2017nsp,Raben:2018rbn,carmi2020conformal,kravchuk2018light}. We shall refer to Eq. (\ref{eq:scausality})  as the associated causal condition for the ``$s$-channel" scattering process, ($1+3\rightarrow 2+4$)~\footnote{In an Euclidean OPE/boostrap treatment, with conformal invariants $(u,v)$ conventionally defined as in Eq. (\ref{eq:invcr}), one normally associates the  limit  $u\rightarrow 0, v\rightarrow 1$, (labelled as point $T$ in Figs. \ref{fig:crsym}(a) and \ref{fig:crsym2}),  with  the s-channel OPE and denotes the partition $(12)(34)$ as the s-channel. The limit of $u\rightarrow 1, v\rightarrow 0$, for the partition $(14)(23)$, is associated with t-channel OPE, and the limit $u\rightarrow \infty, u/v\rightarrow 1$ with the partition $(23)(14)$. However, for scattering, we adopt the convention where partition $(13)(24)$ designates the s-channel process, i.e., (\ref{eq:4phi}), and reserve partition $(12)(34)$ for the t-channel scattering process and   $(14)(23)$ for the u-channel.}.

In a Minkowski treatment, for a fourpoint function, there are $6$ independent Lorentz and translational invariant separations, $x_{ij}^2$,  and each one can be spacelike or timelike. There are therefore a total of $2^6=64$ distinct configurations. We refer to these as causal orientations. In this paper, for (\ref{eq:scausality}), we further restrict our problem by holding all other $x_{ij}^2>0$,  in order to give us a minimal setup required for studying the ``double discontinuities" introduced in \cite{caron2017analyticity}.   This also allows  one to associate  a  double-commutator with the imaginary part of a normalized dimensionless scalar amplitude, (\ref{eq:reduced}).
  
  Physically, this kinematics can also be understood by analogy with that for off-shell Compton scattering, $\gamma^*(1) + \gamma^* (3) \rightarrow \gamma^* (2) + \gamma^* (4)$ (ignoring spin degrees of freedom),  
i.e., $x_{14}$ and $x_{32}$ time-like, while keeping all other Lorentz invariants space-like, $x_{ij}^2>0$ and $x_i^2>0$,   $i,j=1,2,3,4$, as typically done in \cite{Brower:2006ea,Cornalba:2006xm,Cornalba:2006xk,Brower:2007qh,Brower:2007xg,Cornalba:2007zb,Cornalba:2007fs,Hofman:2008ar,Strassler:2008bv,Cornalba:2008qf,Cornalba:2009ax,Costa:2012cb,Brower:2014wha,Belitsky:2013xxa,Belitsky:2013ofa,Banks:2009bj,Nally:2017nsp,Raben:2018rbn,carmi2020conformal}. This corresponds to our causal condition (\ref{eq:scausality}). With $x_i^2$  fixed, translational invariance becomes less apparent. It can be restored formally by introducing a fiducial point $x_5$ so that $x_i^2\rightarrow x_{i5}^2$. (See analogous treatment in Ref. \cite{simmons2018spacetime}.)
The corresponding ``$u$-channel" process,   ($1+ 4\rightarrow 2 +  3$ by $3\leftrightarrow 4$, i.e., $s$-$u$ crossing), takes place in the region where 
\be
\langle 0|[\phi(x_3),\phi(x_1)] [\phi(x_2),\phi(x_4)]|0\rangle\neq 0\, . \label{eq:ucausality}
\ee
Both  regions can be identified with Minkowski $t$-channel OPE point, $T'$~\footnote{ In  extending to Lorentzian limits, the Minkowski limit of cross-ratios $u\rightarrow 0, v\rightarrow 1$,  with $\sqrt v=-1$, is labelled as point $T'$ in Fig. \ref{fig:crsym2}. One can reach $T'$ from $T$ by various holonomically equivalent paths in terms of the standard pair of complex  variables $(z,\bar z)$~\cite{Cornalba:2006xm,Cornalba:2006xk,Cornalba:2007zb,Cornalba:2007fs,Cornalba:2008qf,Cornalba:2009ax,Costa:2012cb,caron2021dispersive,qiao2022classification}. These limits can be characterized more precisely by Eqs. (\ref{eq:MCB1}) and (\ref{eq:ECB1}). For further clarification, see Appendix A.1 of Ref. \cite{Raben:2018rbn}.}.

For a scattering amplitude defined in terms of a time-ordered Green's function, 
its imaginary part in a \emph{causal} domain, e.g., $x_{14}^2<0, x_{23}^2<0$, can be identified with  the double commutator,
\be
{\rm Im} T(x_i)= \frac{\langle 0|[\phi(x_4),\phi(x_1)] [\phi(x_2),\phi(x_3)]|0\rangle} {\langle \phi(x_1) \phi(x_2)\rangle \langle\phi(x_3)\phi(x_4)\rangle }\, .  \label{eq:reduced}
\ee
where we have normalized   the amplitude  by $\langle \phi(x_1) \phi(x_2)\rangle \langle\phi(x_3)\phi(x_4)\rangle$.  The amplitude  is thus dimensionless. With $x_{12}$ and $x_{34}$ spacelike separated, the normalization factor is real;  this choice is appropriate for $s$- and $u$-channel scattering as specified by causal conditions (\ref{eq:scausality}) and (\ref{eq:ucausality}). With $x_4$ in the future of $x_1$ and $x_2$ in the future of $x_3$, this defines a amplitude with ${\rm Im} T(x_i)>0$  in the $s$ channel region~\footnote{To be explicit, for region described by Eq. \ref{eq:scausality}, there are  two equivalent time-ordering configurations, i.e., $x_1<x_3<x_4<x_2$ vs. $x_3<x_1<x_2<x_4$. However, in terms of $(\sqrt u,\sqrt v)$, these regions are equivalent. They can be identified separately in terms of our group theoretic motivated variables, $(y,\eta)$.}. This choice also maintains $s$-$u$ symmetry. By focusing on $s$- and $u$-channel scatterings, 
our treatment  is not totally $s$-$t$-$u$ symmetric~\footnote{See studies on CFT scattering based on gauge/string duality, e.g., \cite{Brower:2006ea,Cornalba:2006xm,Cornalba:2006xk,Brower:2007qh,Brower:2007xg,Cornalba:2007zb,Cornalba:2007fs,Cornalba:2008qf,Cornalba:2009ax,Costa:2012cb,Brower:2014wha}. In terms of a new set of variables, $(w,\sigma)$, Eq. (\ref{eq:newvariables}), $s$-$u$ symmetry, e.g., $x_3$ and $x_4$ exchange in Fig. \ref{fig:DLC-limit}, simply corresponds to $w\leftrightarrow -w$ interchange, with $\sigma$ fixed. See Fig.  \ref{fig:crsymb}.}.  Causal region for $t$-channel scattering can be treated by permutation. This asymmetric setup is important  to accomplish the main purpose of this paper which is to directly work with CFT in a Minkowski framework using a group theoretic approach.

Lorentzian CFTs are invariant under the symmetry group $SO(d,2)$, which has a set of commuting generators forming a Cartan subalgebra. In particular, $SO(d,2)$ contains a rank-2 non-compact Maximal Abelian Sub-Group ({\bf MASG})~\footnote{The MASG should  be understood as the non-compact subgroup of the Cartan subgroup of $SO(d,2)$  and this will be clarified further in Appendices \ref{app:Iwasawa} and \ref{app:embed}. This usage is conventional.},
\begin{equation}\label{eq:MASG}
A=SO(1,1)\times SO(1,1) \, .
\end{equation}
By relating the MASG to the physics of dilatation and Lorentz boosts~\footnote{It should also be pointed out that there are other possible choices for the MASG. In this study, we have made a choice appropriate by the physics of Lorentzian CFTs.  This choice will subsequently be used to define the restricted root space of $SO(d,2)$.}, it was suggested in  \cite{Raben:2018rbn} that ${\rm Im} T(x_i)$, identified as a discontinuity,  can be  represented in terms of the {\bf principal series representation} of $SO(d,2)$~\cite{Toller3,Vilenkin:1991,Vilenkin1993,Vilenkin:1992,knapp1986,knapp2002lie}
\begin{equation}
{\rm Im} T(x_i) =  \int_{-i \infty}^{i \infty} \frac {d \widetilde \ell }{2\pi i} \int_{-i \infty}^{i \infty} \frac {d \widetilde  \Delta }{2\pi i} \, \, a^{(\pm)}(\widetilde  \ell ,\widetilde  \Delta) \,{\mathcal G}_{(\widetilde\ell,\widetilde \Delta; 0)} (g) \,.\label{eq:double-mellin}
\end{equation}
This equation is of the double-Mellin type~\footnote{By invoking analyticity and dispersion integral, a similar representation can equally be obtained  for the full amplitude, i.e., both the real and imaginary parts. In Regge theories, this amounts to supplying an additional kernel, i.e., ``signature factor", $(e^{-i\pi \ell}\pm 1)/\sin \pi \ell$, (and  also introducing subtractions), see \cite{Raben:2018rbn}.}.  It was advocated for a direct treatment of Minkowski CFT in \cite{Raben:2018rbn}, with Regge asymptotic limit  re-interpreted as an appropriate Minkowski OPE limit.  The discussion was often guided by intuitions gained from studies based on gauge/string duality, e.g., \cite{Brower:2006ea}, and also by \cite{Cornalba:2006xm,Cornalba:2006xk,Brower:2007qh,Brower:2007xg,Cornalba:2007zb,Cornalba:2007fs,Hofman:2008ar,Strassler:2008bv,Cornalba:2008qf,Cornalba:2009ax,Costa:2012cb,Brower:2014wha}.

In this paper, we present a proper interpretation for this representation, Eq. (\ref{eq:double-mellin}), by adopting an embedding space approach, focusing on the scattering region specified by the causal condition, (\ref{eq:scausality}). This is the first of two papers in our current effort on a more direct approach to Minkowski CFT. In this first part, we identify the representation functions (or group harmonics) $ \mathcal{G}_{\widetilde\Delta,\widetilde\ell; 0}(g) $ as {\bf ``causal" zonal spherical functions}~\cite{Hilgert:1996,Vilenkin:1991,Vilenkin1993,Vilenkin:1992,helgason2022groups} for $SO(d,2)$, and associate them with Minkowski conformal blocks discussed in \cite{Raben:2018rbn}~\footnote{The term spherical functions is used in several different contexts in the mathematical literature. We reserve the phrase spherical functions for right-invariant functions. This is closer to the familiar physics literature, since $ Y_{\ell,m} $ are called spherical harmonics for $ SO(3) $ and are associated with right-invariant functions.}. A second paper currently being prepared  \cite{Agarwal:2024} will focus on the inversion aspect of Eq. (\ref{eq:double-mellin})~\footnote{Replacing ${\mathcal G}_{(\widetilde\ell,\widetilde \Delta; 0)} (g)$ in Eq. (\ref{eq:double-mellin}) simply by the leading exponential factors, e.g., Eq. (\ref{eq:double-mellin-2}), it would become a conventional double-Mellin representation. In the context of induced representations for non-compact groups, this transition is referred to by some as  ``Harish transform", which can also be considered as a generalized Radon transform.}.

This paper consists of  two main parts; a short outline is provided below:

\begin{enumerate}[font=\bfseries]
\item \textbf{
	Minkowski Fourpoint Functions and Causality:}
\begin{itemize}
	\item 
	In Sec. \ref{sec:kinematics}, {\it 
		we focus on specifying causal scattering regions for  (\ref{eq:scausality}) and (\ref{eq:ucausality}) in terms of  a set of group theoretic motivated variables}, Eq. (\ref{eq:newvariables}), 
$w\equiv \frac{1-\sqrt v}{\sqrt u}$, and $  \sigma \equiv  \frac{1+\sqrt v}{\sqrt u}$. 
By directly relating to $(\sqrt u,\sqrt v)$, they  resolve the multi-sheeted structure of the $ u$-$v $ plane, and they delineate clearly transitions from the Euclidean region  to Minkowski regions, (see Figs. \ref{fig:crsym}, \ref{fig:crsym2} and \ref{fig:crsymb}).
	
\item   In Sec. \ref{sec:4point},  {\it  we examine consequences of conformal invariance  and causality for Minkowski CFT in an embedding space treatment}. After reducing a 4-point correlator to a function over a single-copy of $SO(d,2)$, $F(g)$,  it remains a function  of $(d+2)(d+1)/2$ variables. For CFTs in a Minkowski setting, the residual symmetry  can be identified as ``$ H $ bi-invariance", with $H$ appropriate subgroup, leading to 
\begin{equation} 
	F(g) = F(hah') = F(a),\nonumber
\end{equation}
where  $a\in A$, the maximal abelian subgroup, Eq. (\ref{eq:biH}). Enforcing  $ H $ bi-invariance leads $F(g)$ acting on two-sided coset~\footnote{A two-sided coset is often denoted by $ H\backslash  G/H$ in mathematical literature, with $G/H$ as ``left-coset". We will adopt the alternative notation $G//H$ for two-sided coset. This bi-invariance property has been discussed by Schomerus et al. \cite{Schomerus2017,Schomerus2018,Isachenkov2018,Buric2020,Buric:2022ucg} for Euclidean CFTs. See \textbf{Note Added} at the end of the paper for a discussion on the relationship between their work and ours.\label{foot:bi-cost}}, thus reducing  $F$ to   a function of two conformal cross ratios.   When causality is imposed, additional structure emerges, e.g., various regions in the $ w$-$\sigma $ plane are covered via semigroups.  This leads to $H$ bi-invariant zonal spherical functions associated with each representation of $SO(d,2)$~\footnote{Similar type of analysis has been carried out  in early years with emphasis on Lorentz group, e.g., \cite{Hilgert:1996,Vilenkin:1992,Toller:1973,Viano:1980,Faraut:1986}.}.  These functions are necessarily constructed piecewise in different regions of the $ w$-$\sigma $ plane. (See also \cite{Schomerus2017,Schomerus2018,Isachenkov2018,Buric2020,Buric:2022ucg}).

\end{itemize}

\item \textbf{
Representation Theory of Semigroups:}
\begin{itemize}
\item In Sec. \ref{sec:induced},  {\it we discuss induced representations  for 
	$SO(d,2)$ via Iwasawa decomposition and the principal series.} 
The principal series representations for $SO(d,2)$ can be labelled by the unitary irreducible representation of its MASG via induced representation, leading to  Eq. (\ref{eq:double-mellin}), with purely imaginary representation labels $\vec{\widetilde \lambda}= (\widetilde \ell, \widetilde \Delta)$. 

The induced representation  involves  introducing  a multiplier factor, Eq. (\ref{eq:invF}),  which can be interpreted as adding real parts to $\vec{\widetilde \lambda}$. 
These real parts, can be associated with a Weyl vector, $\vec \rho$, leading to a pair of complex representation labels, $\vec \lambda=-\vec \rho+\vec{\widetilde \lambda}$, Eq. (\ref{eq:lambda}). 
This Weyl vector is in turn fixed by examining the positive root-space structure~\footnote{A comparable analysis from an Euclidean perspective has also be carried out in \cite{kravchuk2018light}.}, dictated by causal {\it semigroup} consideration, leading  to a consistent identification with the conventionally defined variables $(\ell, \Delta)$, Eq. (\ref{eq:connection}), ($
	\ell  = -(d-2)/2 + \widetilde \ell $ and $ \Delta= d/2 + \widetilde \Delta $).
   This in turn leads to the desired boundary condition for Minkowski conformal blocks Eq.  (\ref{eq:MCB}), (see Ref. \cite{Raben:2018rbn}), defined in the causal region $1<\sigma<w<\infty$,
\begin{equation}
	G^{(M)}_{(\Delta, \ell)}(u,v) \sim \left(\sqrt{u}\right)^{1-\ell} 	\left(\frac{1-v}{\sqrt u}\right)^{1-\Delta}\, ,\label{eq:MCB1}
\end{equation}
and Euclidean conformal blocks, Eq.  (\ref{eq:ECB}), defined in the  region $1<w<\sigma<\infty$~\footnote{The original derivation in \cite{Dolan:2011dv}, in terms of $(u,v)$ variables,  extends to all four regions in Fig. \ref{fig:crsym}.},  
\be
	G^{(E)}_{(\Delta, \ell)}(u,v) \sim \left(\sqrt{u}\right)^{\Delta} 	\left(\frac{1-v}{\sqrt u}\right)^{\ell}\,  \label{eq:ECB1}.
\ee
	Note that these are formally related by  $(\ell,\Delta)$ and $(1-\Delta, 1-\ell)$ swap, so necessary in various related treatments, e.g., \cite{caron2017analyticity,simmons2018spacetime}.
		
	\item   In  Sec. \ref{sec:sphr},{\it  we construct causal zonal spherical functions which serve as group harmonics for the principal series, Eq. (\ref{eq:double-mellin}).} An explicit construction of casual zonal spherical functions is introduced for $SO(d,2)$, expressed as  integrals over the subgroup $H$, Eq. (\ref{eq:genindch}), 
\begin{equation}
	\varphi_{{\lambda}}(a )  =\varphi_{{\lambda}}(h_1 a h_2) = \int_{H} e^{\vec \lambda \cdot \vec {t}(h_1a)} \text{d}h_1, \nonumber
\end{equation}
where as before,  $\vec \lambda = -\vec \rho+\vec{\widetilde{\lambda}}$. 
The integrand is referred to as the inductive character, with $\vec t(a)=(y,\eta)$~\footnote{The factor ${\vec \lambda} \cdot {\vec {t}(h_1a)}$ in the equation above, typically expressed by an abbreviated shorthand notation in mathematical literature, has to be understood in terms of Iwasawa decomposition, i.e., $\vec {t}(h_1a)=(y_I,\eta_I)$. The subgroup $a$ is parametrized in a Cartan decomposition, Eq. (\ref{eq:biH}).}. In particular,  explicit evaluation of  Eq. (\ref{eq:genindch}) for $d=2$ leads to causal Minkowski conformal  blocks satisfying boundary condition, Eq.  (\ref{eq:MCB}), in agreement with that derived in \cite{Raben:2018rbn}.
\end{itemize}

\end{enumerate}

We end here by a short summary on what we consider to be the key significant results of this paper.  (1) Our discussion on the Lorenztian kinematics in terms of a real slice of the two-dimensional $(\sqrt u, \sqrt v)$ plane, perhaps known to experts in other alternative representation, is a useful advance. We provide a direct framework in addressing novel issues, e.g., Regge behavior for CFTs,  by avoiding the considerable complications of analytic continuation from Euclidean region.   (2) The use of embedding space approach allows a more direct linear realization for conformal symmetry via principal series representation. However, there exists a significant complication when the group involved is of split-rank 2. 
We have taken a crucial first step in addressing this new issue, e.g., root-space consideration. Further work to characterize the entire cross-ratio space in these terms will be needed. (3) It is possible to explore conformal block structure via differential equation approach by continuation and by considering alternative boundary conditions, i.e., Eq. (\ref{eq:MCB1}). We have provided an alternative integral approach which not only satisfy the desired differential equation but also the boundary condition dictated by causality requirement. (4) We have raised in the discussion section new issues which have not been addressed adequately in the past Lorentzian approach. Some of these will be touched on in our follow-up paper. We shall return to provide further comments in Sec.  \ref{sec:conclusion}.

Some of the materials in this study do not typically appear in standard CFT physics literature. In the appendices, a deliberate effort is made on being comprehensive rather than brief. Appendix-\ref{app:notations} summarizes our notational conventions. Appendix-\ref{app:MinkowskiCFT} includes  clarification  on the kinematics of a 4-point amplitude  as a function of various  equivalent sets of conformal invariant variables. Appendix-\ref{app:Iwasawa} is a brief discussion on the structure of semi-simple Lie groups and, for $SO(d,2)$,  the notion of MASG, Eq. (\ref{eq:MASG}).  
Other appendices provide supplemental discussions to the text proper. Appendix \ref{app:embed} summarizes some of the key issues involved in the construction of \textit{causal symmetric spaces} and how lightcone causality is understood using \textit{semigroups}. In Appendix \ref{app:sphr}, we  extend the discussion to non-compact groups, paying special attention to how ``causality" enters.

\section{CFT Amplitudes in Minkowski Limits}\label{sec:kinematics}
Minkowski scattering kinematics has often been illustrated schematically by the use of a light-cone diagram, e.g. Fig. \ref{fig:DLC-limit}. This necessarily requires adopting a Lorentzian signature. However, this illustration necessitates specifying a frame and it also lacks precision since the diagram involves several causal configurations of the four points which are physically distinct \cite{Cornalba:2006xm,Cornalba:2008qf,Brower:2014wha,Raben:2018rbn,simmons2018spacetime,kravchuk2018light}. It is therefore important to make precise the frame of reference that we shall use by setting up an explicit coordinate representation.

The main goal of this section is to introduce this frame via a new set of group theoretically motivated conformal invariant variables, $(w,\sigma)$. These variables defined in terms of $\sqrt u$-$\sqrt v$ by Eq. (\ref{eq:newvariables}), cover the whole $\sqrt u$-$\sqrt v$ plane, Fig. \ref{fig:crsym}. They help us generalize the radial quantization picture used commonly in Euclidean CFTs to the Lorentzian limit. This allows a more precise specification for the regions of interest such as those specified by Eqs. (\ref{eq:scausality}) and (\ref{eq:ucausality}). This shall also prepare us for an embedding space treatment in the subsequent sections when we make the group theoretic meaning of these variables explicit.

\subsection{Kinematics of Causal Scattering Regions} \label{sec:beyondEuclidean}

\begin{figure}[ht]
	\centering
	\begin{subfigure}{0.4\textwidth}
		\includegraphics[height=\textwidth]{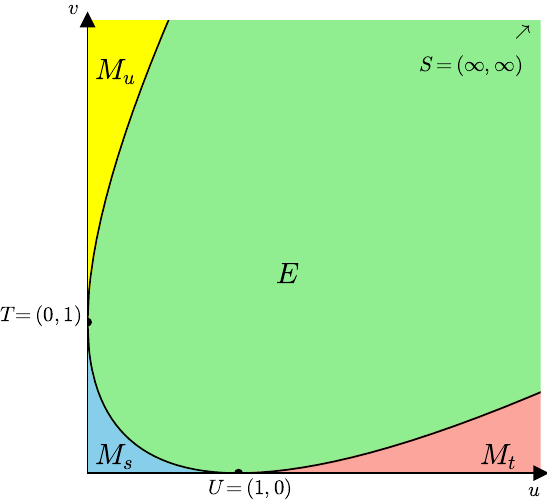}
		\caption{}
	\end{subfigure}
	\hfil \hspace{20pt}
	\begin{subfigure}{0.4\textwidth}
		\includegraphics[height=\textwidth]{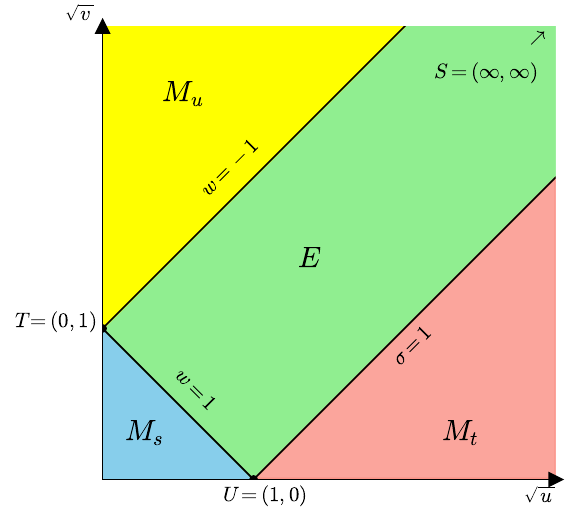}
		\caption{}
	\end{subfigure}
	\caption{Euclidean region $E$ (green) and  three Lorentzian extensions,  $M_s$, $M_u$, $M_t$,  
	Fig. \ref{fig:crsym}(a).  $T$ and $U$ and $S$ are three Euclidean OPE points. corresponding to limits ($u\rightarrow 0, v\rightarrow 1$), ($u\rightarrow 1, v\rightarrow 0$) and $(u\rightarrow \infty, u/v\rightarrow 1$)  respectively. 	 In  Fig. \ref{fig:crsym}(b), straight line segments separating these regions, labelled by variables $(w,\sigma)$.}\label{fig:crsym}
\end{figure}

In a standard Euclidean treatment, after removing a  kinematic factor,  as done in (\ref{eq:reduced}), conformal invariance leads to a reduced dimensionless scalar function of two independent cross ratios, ${\cal G}(u,v)$,  
$
u={x_{12}^2} {x_{34}^2} / {x_{13}^2}{x_{24}^2} $ and $  v={x_{23}^2}{x_{14}^2}/{x_{13}^2}{x_{24}^2},
$ 
with $u,v>0$. With $x_{ij}^2$ defined with Euclidean signature, this scalar function is well-known to  be further restricted kinematically,  due to Schwarz inequality~\cite{Dolan:2011dv}.  The curve bounding this Euclidean region-$E$, (green shaded region in Fig. \ref{fig:crsym}(a)),  is quadratic in $u$ and $v$, and it can be factorized in terms of $\sqrt u$ and $\sqrt v$, i.e.,
$
(\sqrt{u}-\sqrt{v}-1)(\sqrt{u}+\sqrt{v}-1)(\sqrt{u}-\sqrt{v}+1)(\sqrt{u}+\sqrt{v}+1)=0  
$.
The first quadrant  of the $u$-$v$ plane  is now divided into four separate regions, with straight-line segments separating them in the $\sqrt u$-$\sqrt v$ plane, Fig. \ref{fig:crsym}(b). 

It is also important to note that Euclidean CFT scalar amplitudes, e.g., ${\cal G}(u,v)$, can be expressed in terms of OPE expansions in region $E$. There are three options, associated  with $s$-, $t$- and $u$-channels, leading to three scalar amplitudes, ${\cal G}^{(s)}$, ${\cal G}^{(t)}$ and ${\cal G}^{(u)}$. They are simply related by kinematic factors, which are singular at $u, v=0, \infty$. These expansions can be  associated with Euclidean OPE points~\footnote{For  identical scalar primaries  $\phi$ with conformal dimension $\Delta_0$,  with our choice of variables and labelling order, they are related to  standard convention by  $ {\cal G}^{(s)}=  {\cal G}^{(t)}_{standard}=\langle \phi(1)\phi(2)\phi(3)\phi(4)\rangle/\langle \phi(1)\phi(3)\rangle \langle \phi(2)\phi(4)\rangle $ with $\langle \phi(i)\phi(j)\rangle \equiv x_{ij}^{-2\Delta_0}$.   Correspondingly, 
${\cal G}^{(t)}= {\cal G}^{(s)}_{standard}=u^{2\Delta_0}{\cal G}^{(s)}$ 
and $G^{(u)}= {\cal G}^{(u)}_{standard}=v^{2\Delta_0}{\cal G}^{(s)}$ respectively. 
It is useful to remind ourselves that these are analytic within the region $E$, and they are  related kinematically. They are singular  at respective OPE points, $S$, $T$ and $U$. They can also be extended analytically to regions $M_s$, $M_t$, and $M_u$, without crossing singularities on the boundaries in  Fig. \ref{fig:crsym}(a) or Fig. \ref{fig:crsym}(b).},
labelled as $S$, $T$ and $U$ in Fig. \ref{fig:crsym}(a).

\paragraph{Natural Lorentzian Extension of Euclidean CFT:} 

It is natural to first extend a scalar CFT amplitude, e.g., ${\cal G}^{(t)}(u,v)$, from region $E$ to cover the whole first quadrant of the $\sqrt u$-$\sqrt v$ plane,  i.e., extending into  regions $M_s,M_u,M_t$   by analytic continuation {\it without}  crossing the expected singularities at $ u,v=0 $, (and also at $+\infty$).  However, to keep $x_{ij}^2>0$, i.e., all points  spacelike separated,  it necessitates in  a Wick rotation so that   $x^2=x^\mu x_\mu=\eta^{\mu\nu}x_\mu x_\nu$, defined with {\it Lorentzian signature}, ${\rm dia} \,\eta=(-,+,\cdots,+)$.
In this {\it first stage of extension}, we have
\be
\sqrt u=\frac{\sqrt {x_{12}^2} \sqrt{x_{34}^2} }{\sqrt {x_{13}^2}\sqrt{x_{24}^2} }\quad \quad {\rm and} \quad \quad\sqrt v=\frac{\sqrt{x_{23}^2}\sqrt{x_{14}^2}}{\sqrt{x_{13}^2}\sqrt{x_{24}^2}}, \label{eq:invcr}
\ee
 defined with positive square roots for all $\sqrt {x_{ij}^2}$. With this convention, the first quadrant of $u$-$v$ plane is in 1-to-1 correspondence with that for the $\sqrt u$-$\sqrt v$ plane.  

From a group theoretic perspective, moving away from region $E$  corresponds to  a shift from $SO(d+1,1)$ to $SO(d,2)$.   
However,  regions $ M_s,M_u,M_t $ do not correspond to {\it causal scattering regions} since all points remain space-like separated. We  will loosely refer to them  as  {\bf Minkowski non-causal scattering regions}. In particular, with $0<\sqrt u,\sqrt v<\infty$,  region $M_s$ is bounded, with  $0<\sqrt u+\sqrt v <1$, with  $M_u$ and $M_t$ unbounded, as shown in Fig. \ref{fig:crsym}(b).  

 In place of $(u,v)$, it is often convenient to treat Euclidean four-point functions, e.g.,  ${\cal G}^{(t)}(u,v)$, in terms of   other equivalent variables. The most often used is a pair $(z,\bar z)$  where  $u= z \bar z$ and $v=(1-z)(1-\bar z)$, with $z^*=\bar z$ in  the Euclidean region $E$. 
 To continue outside of the Euclidean region, it is necessary to  treat $z$ and $\bar z$ as independent complex variables, with  ${\cal G}^{(t)}(u,v)$ having singularities at $z\, {\rm or} \, \bar z\, =0,1$. To reach $ M_s,M_u,M_t $, one needs to approach the limit $z$ and $\bar z$ on the real axis, $-\infty< z, \bar z<\infty$. There are many {\it holonomically equivalent}  paths by staying on the ``physical sheet" without  crossing the singular lines at $u, v=0$ or at $+\infty$~\footnote{See  Appendix \ref{app:MinkowskiCFT} for the notion of ``first sheet" or ``physical sheet", and also for the equivalence among various sets of variables, e.g., $(\sqrt u,\sqrt v)$, $(z,\bar z)$, $(\rho,\bar \rho)$ and  $(q,\bar q)$. (The set $(q,\bar q)$, introduced in Ref. \cite{Raben:2018rbn}, is simply related to $(z,\bar z)$ by $(q=2/z-1,\bar q=2/\bar z-1)$.  They  are not to be confused with that introduced by Zamolodchikov, which are widely used in the 2d CFT literature.)}.  This initial transition from $E$ to regions $ M_s,M_u,M_t $ is smooth and can be understood in terms of Wick rotations.   To move out of these regions, i.e.,  $\sqrt u$ and/or $\sqrt v$ take on negative signs, however, one must cross ``lightcone singularities", i.e., holonomies around $u,v=0,\infty$,  which we discuss next.

\paragraph{Causal Scattering Regions:}

\begin{figure}
\centering
\includegraphics[width=0.6\textwidth]{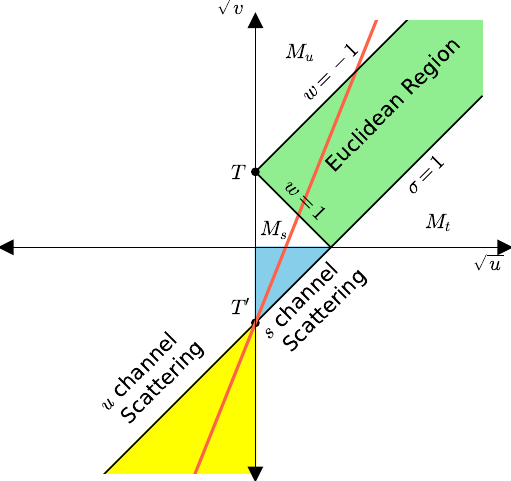}
\caption{\label{fig:crsym2}Extension into $ \sqrt{u}$-$\sqrt{v} $ plane, with $s$- and $u$-channel  causal scattering regions  shown, shaded blue and yellow respectively. }
\end{figure}

To reach {\it causal scattering regions}, it is necessary to cross singular lines at $\sqrt u=0$ and/or $\sqrt v=0$, (or at $+\infty$), thus extending from the first quadrant to the whole $\sqrt u$-$\sqrt v$ plane, Fig. \ref{fig:crsym2}, i.e.,  with $\sqrt u$ and $\sqrt v$ real, but allowed to take on both signs.   There can again be many holonomically equivalent paths. Consider $s$-channel causal scattering, (\ref{eq:4phi}), subject to causal constraint  (\ref{eq:scausality}). 
With both ${x_{14} }$ and ${x_{23}}$  timelike, (i.e., both $\sqrt {x_{14}^2}$ and  $\sqrt {x_{23}^2}$ purely imaginary), by keeping  both having the same sign, it  leads to  $\sqrt u>0$ and $\sqrt v<0$,  i.e., the fourth quadrant in the $\sqrt u$-$\sqrt v$ plane. 
To fix our convention for  the two-dimensional real $\sqrt u$-$\sqrt v$ plane, we move to $\sqrt {x_{ij}^2}$  on the positive imaginary axis, when $x_{ij}^2$ is timelike, i.e.,  $\sqrt {x_{ij}^2}= +i   \big|\sqrt {x_{ij}^2}\big |$. (With this convention, it also fixes the sign for the discontinuity in (\ref{eq:reduced}) to be positive. This issue is closely tied to crossing symmetry and dispersion relations, a problem that should be addressed elsewhere.)  

Staying with  $s$-channel causal scattering, let us focus on the bounded region  in the fourth quadrant,
$0<\sqrt u+|\sqrt v|<1$, (shaded blue in Fig. \ref{fig:crsym2} and will  be designated as $M_s^{(t)}$ in what follows).  This  region is bounded by three line segments. Of particular interest is the segment $L_s$: $0<\sqrt u<1$, $\sqrt v=0$, which separates it from the non-causal region, $M_s$.
It can be shown that  this region corresponds to where the causal condition (\ref{eq:scausality}) holds. This can best be carried out in a generalized antipodal frame. (See Secs. \ref{sec:antipodal} and \ref{sec:Wick}. The superscript for $M_s^{(t)}$ will also be explained shortly.) 
Region $M_s^{(t)}$ will be referred to as the $s$-channel causal region, with $M_s$ as the $s$-channel non-causal region.

Similarly, the $u$-channel causal region, specified by (\ref{eq:ucausality}), corresponds to having  $\sqrt u<0$ and $\sqrt v<0$, i.e., in the third quadrant, with  $1< |\sqrt v|-|\sqrt u|<\infty$.  This region will be designated as $M_u^{(t)}$, (shaded yellow region in Fig. \ref{fig:crsym2}). The corresponding line segment $L_u$ separating it from non-causal region $M_u$ is less explicit in Fig. \ref{fig:crsym2}. It  corresponds to the limit $\sqrt u, \sqrt v \rightarrow \pm \infty$, with $\sqrt v/\sqrt u >1$. (Line segments $L_s$ and $L_u$  can be  better described  in terms of group theoretic variables $(w,\sigma)$ to be introduced below.  They  are shown  symmetrically  in Fig. \ref{fig:crsymb}.  In particular, our group theoretic approach maintains $s$-$u$ symmetry.) The straight-line (colored red) in Fig. \ref{fig:crsym2} corresponds to $\sqrt v= -1+\sigma \sqrt u$, with a fixed slope, $1<\sigma<\infty$, (in Fig.  \ref{fig:crsymb}, the corresponding line is identified).

The corresponding  region  $M^{(t)}_t$ for causal $t$-channel scattering lies in the second quadrant, (not shown in Fig. \ref{fig:crsym2} but can be read off   Fig. \ref{fig:crsym22}   in Sec. \ref{sec:conclusion}). 
It can be better understood intuitively in a ``direct-channel picture". It can also be treated in terms of group theoretic variables $(w,\sigma)$, to be introduced next. 
However, since it does not directly lead to a principal series representation, we will not deal directly with this  region in what follows.

 \paragraph{Minkowski OPE:}
 In analogy with Euclidean OPE, it is equally meaningful to introduce OPE for Minkowski scattering~\cite{Brower:2006ea,Cornalba:2006xm,Cornalba:2006xk,Brower:2007qh,Brower:2007xg,Cornalba:2007zb,Cornalba:2007fs,Hofman:2008ar,Strassler:2008bv,Cornalba:2008qf,Cornalba:2009ax,Costa:2012cb,Brower:2014wha,Raben:2018rbn}.   We will return to this in Sec. \ref{sec:conclusion}. Here we simply mention the fact that the antipodal frame adopted here can be  associated  with Minkowski $t$-channel OPE~\cite{Raben:2018rbn},  thus  the superscript for $M_s^{(t)}$, $M_u^{(t)}$ and $M_t^{(t)}$.    Regions $M_s^{(t)}$, $M_u^{(t)}$ can also be identified with Minkowski $t$-channel OPE point~\footnote{Traditionally, OPEs are defined in a Euclidean setting when positions of operator insertions approach each other, for example $x_{12}^2,x_{34}^2\rightarrow0$. In a Minkowski setting, these null separations correspond physically to the forward scattering limit $x_{12}^2,x_{34}^2\rightarrow0^+$ of a Regge expansion.  \label{fn:MinOPE}}, $T'=(0,-1)$, as indicated in Fig. \ref{fig:crsym2}. More importantly, for causal scatterings, Minkowski OPE's can in turn be associated with Regge limits for CFT~\cite{Brower:2006ea,Cornalba:2006xm,Cornalba:2006xk,Brower:2007qh,Brower:2007xg,Cornalba:2007zb,Cornalba:2007fs,Hofman:2008ar,Strassler:2008bv,Cornalba:2008qf,Cornalba:2009ax,Costa:2012cb,Brower:2014wha,Raben:2018rbn}.

\subsection{Group Theoretic Motivated Variables}  \label{sec:antipodal}

\begin{figure}[ht]
\centering
\includegraphics[width=.6\textwidth]{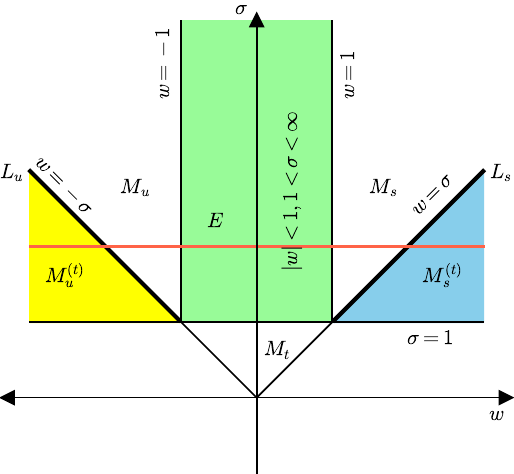}
\caption{A portion of the  $ w$-$\sigma $ plane covering regions for the first  quadrant of the $\sqrt u$-$\sqrt v$ plane, ($E, M_s,M_u,M_t$) and the two causal regions, ($M_s^{(t)},M_u^{(t)}$). In the graph, $s$-$u$ crossing symmetry, ($w\leftrightarrow -w$),  is manifest. Transition lines $L_s$ and $L_u$ are clearly shown.} 
\label{fig:crsymb}
\end{figure}

Instead of $\sqrt u$-$\sqrt v$, let us define a new set of variables
\be
w\equiv \frac{1-\sqrt v}{\sqrt u}, \quad \sigma \equiv  \frac{1+\sqrt v}{\sqrt u} \quad\quad \Leftrightarrow \quad\quad  \sqrt u = \frac {2 }{\sigma+w}, \quad \sqrt v = \frac {\sigma-w}{\sigma+w}.
\label{eq:newvariables}
\ee
Since $(w,\sigma)$ is given in terms of $(\sqrt u,\sqrt v)$ by construction, they provide a natural extension into the whole $\sqrt u$-$\sqrt v$ plane, with focus on the Minkowski OPE point $T'$.  The map  is 1-to-1, as can be seen as follows. From $\sqrt v=-1+\sigma \sqrt u$, straight lines through $T'$ with slope $-\infty<\sigma<\infty$  cover the entire $\sqrt u$-$\sqrt v$ plane.  It is also easy to verify that the range for each line covers $-\infty<w<\infty$.   In Fig. \ref{fig:crsym2}, such a line with $1<\sigma<\infty$ is shown.  Correspondingly, this line with $\sigma>1$ is shown in Fig. \ref{fig:crsymb}.  We emphasize the important fact  that   $s$-$u$ crossing, $(u,v) \leftrightarrow  (u/v,1/v)$,  corresponds  to
$
(w,\sigma)  \leftrightarrow  (-w,\sigma)
$~\footnote{In Ref. \cite{Raben:2018rbn}, this was discussed  via another set of variables, also denoted as $(w,\sigma)$, which we will designated as $(w',\sigma')$ here. They are  defined by  $w'=\sqrt {q\bar q}$ and $\sigma'= (\sqrt{q/\bar q} + \sqrt {\bar q/q})/2$, with $(q,\bar q)$ simply related to $(z, \bar z)$. They can also be expressed easily in terms of $(\rho,\bar \rho)$. For practical purpose, $(w',\sigma')\simeq (w,\sigma)$, in the relevant limit of $(q, \bar q)$ large or  $(z, \bar z)$ small. See Table \ref{tab:vars} in Appendix \ref{app:MinkowskiCFT} for relations to group parameter $y$ and $\eta$.}. Of course, manifest crossing symmetry proves useful for discussing dispersion relations and inversion. A related discussion for $ d=1 $ is presented in \cite{Mazac:2018qmi}.
 
 These variables can be understood intuitively in a generalized antipodal frame and they help to pave the way for group theoretic analysis in sections to come.

\paragraph{Regions $E, M_s,M_u,M_t$:}

These regions  now correspond to
$
E: (-1<w<1,  1<\sigma<\infty),$
$M_s: (1<w<\infty, 1<\sigma<w),$
$M_u: (-\infty<w<-1, 1<\sigma <|w|)$ and 
$M_t: ( -1<w<1,  0<\sigma<|w|)$,  shown in  Fig. \ref{fig:crsymb}), with  line segments $L_s$ and $L_u$ simply labelled. 
Note also that,  the Euclidean $t$-channel OPE point, $ T=(\sqrt{u},\sqrt{v})=(0,1) $, has moved to $\sigma=+\infty$, with $w/\sigma\rightarrow 0$. It  can be reached from regions $E$, $M_s$ and $M_u$.

\paragraph{Causal Scattering Regions:}
Causal scattering regions  we will concentrate on are
\bea
 M_s^{(t)}\quad (s{\rm-channel}): & 1<w<\infty, & 1<\sigma<w<\infty,\nn
  M_u^{(t)}\quad (u{\rm-channel}):& -\infty<w<-1, & 1<\sigma<|w|<\infty.
\eea
The $s$-channel causal region is bordered by the non-scattering Minkowski region, $M_s$, by the line segment $L_s$, $1<\sigma=w<\infty$. This also corresponds to $ \sqrt{v}=0 $,  Fig. \ref{fig:crsym2}. Similar remarks also apply for the $u$-channel scattering, with $w\rightarrow -w$.

Minkowski $t$-channel OPE can be identified with  $T'=(\sqrt u, \sqrt v )=(0,-1)$, conjugate to Euclidean $t$-channel OPE point $T$~\footnote{That is, $T'$ and $T$ coincide in the traditional $(u,v)$ plane, Fig. \ref{fig:crsym}(a), but holonomically inequivalent, as shown in  Fig. \ref{fig:crsym2}. See also footnote \# \ref{fn:MinOPE}.}.  It can also be approached in two limits,   $w\rightarrow \pm\infty$,  with $1<\sigma<|w|$  fixed, from regions $M^{(t)}_s$ and $M^{(t)}_u$ respectively. (Limits taken along the horizontal  line (colored red) in Fig.  \ref{fig:crsymb}.)

\paragraph{Generalized Antipodal Frames:}\label{sec:GAF}

Our new variables $(w,\sigma)$ are conformal invariants.  One can provide them with different interpretations as they take on values in different ranges.
Observe that the upper half plane above $\sigma=1$ is divided into five regions.  For region $E$, it can best be interpreted  by adopting the {\bf Euclidean  Antipodal Frame} 
where~\cite{Hogervorst:2013sma,brower2021radial}
\be
-1\leq  w= \cos \theta \leq 1 \quad {\rm and} \quad 1<\sigma=\cosh \tau\, <\infty\, .  \label{eq:regionE}
\ee
(In fact, this is nothing but the equivalent of ``CM frame", i.e., $x_1=-x_2$ and $x_3=-x_4$.)

For non-causal regions $M_s$, $M_u$ as well as for causal regions $M^{(t)}_s$, $M^{(t)}_u$, we adopt {\bf Minkowski Antipodal frame}, (also known as the {\bf double-lightcone} frame in Ref. \cite{Raben:2018rbn}), where, for the light-cone components, $x^\pm_1=-x^\pm_2$ and $x^\pm_3=-x^\pm_4$, and, for transverse components, $x_{\perp,1}=x_{\perp,2}$ and $x_{\perp,3}=x_{\perp,4}$. It follows that $(w,\sigma)$ can be parametrized by
\be
1<|w|= \cosh y <\infty \quad{\rm and} \quad 1<\sigma =  \cosh \eta <\infty.
\label{eq:sigma}
\ee
(See Appendix \ref{app:B2} and Ref. \cite{Raben:2018rbn}.)  We shall show in Sec. \ref{sec:4point} how to attach group theoretic interpretations for all these  regions.

\subsection{Wick Rotation and Causal Scattering vs Non-Causal Regions}\label{sec:Wick}

Comparing (\ref{eq:regionE}) with (\ref{eq:sigma}), the transition from $|w|<1$ to $1<|w|$ can be interpreted  as a {\bf Wick rotation}.  That is, 
by substituting $ \theta\rightarrow iy $ or $ \theta\rightarrow \pi+ iy $, 
one moves from  $ 0\leq | w|\leq 1$  to  $1<|w|$. 
Group theoretically, this corresponds to a  formal continuation from $SO(d+1,1)$ to $SO(d,2)$.   Kinematically, one moves from $E$ to $M_s$ or $M_u$ initially.

However, various Wick rotations  take on different interpretations in different frames.
Consider first transitions from $E$ to $M_s$ or $M_u$. Immediately after a Wick rotation, by crossing $|w|=1$, region $E$ is smoothly connected to the region $1<|w|<\sigma<\infty$, i.e., non-causal   regions $M_s$ and $M_u$. Next, by subsequently crossing $|w|=\sigma$, it is possible to move to $1<\sigma< |w|<\infty$, the causal scattering regions $M^{(t)}_s$ and $M^{(t)}_u$, i.e., crossing line segments $L_s$ or $L_u$. These are singular lines.

Consider $ \theta\rightarrow iy $ first.  In  the Minkowski antipodal frame, a quick check, using Eqs. (\ref{eq:Santipodal}) and (\ref{eq:Santipodalperp}), establishes that 
\be
x_{14}^2= x_{23}^2 =  \sigma -w=\cosh \eta-\cosh y,       \label{eq:condition}
\ee
with all other $x_{ij}^2$ spacelike. The Minkowski non-scattering region $ M_s $ corresponds to  $ x_{14}^2= x_{23}^2>0$, leading to
$
0<y< \eta<\infty  \label{eq:scausalcrossing}
$,
consistent with $0<\sqrt u$, $0<\sqrt v$ and $0<\sqrt u+\sqrt v<1$, i.e., remaining in the 1st quadrant. 

Since the line segment $L_s$ corresponds to $w=\sigma$, 
 crossing $L_s$ enters the causal region $M_s^{(t)}$.  With $\sigma<w$, i.e.,  $\cosh \eta< \cosh y$, the causal condition is achieved with  $ x_{14}^2= x_{23}^2<0$. The {\bf causality  constraint} can also be expressed as
\be
0<\eta<y<\infty. \label{eq:semi}
\ee

Consider next   $ \theta\rightarrow \pi+iy $.  A similar analysis also follows for the $u$-channel causal region, e.g., with $w=\cosh( y -{i\pi})=-\cosh y<0$, 
\be
x^2_{13}=x_{24}^2=\sigma+w=  \cosh \eta-\cosh y, \label{eq:ucondition}
\ee
all other $x_{ij}^2$ spacelike, and consistent with the $u$-channel causal constraint when (\ref{eq:scausalcrossing}) is imposed. Moving into $u$-channel causal region $M_u^{(t)}$ corresponds to crossing the line segment $L_u$, leading again to causal constraint, (\ref{eq:scausalcrossing}).

The most significant fact  here is the presence of two unbounded parameters, $y$ and $\eta$ as well as the causal restriction, (\ref{eq:semi}). As emphasized in \cite{Raben:2018rbn}, variables $y$ and $\eta$ are nothing but group parameters 
characterizing boosts and dilations in the context of $SO(d,2)$. (See Secs.  \ref{sec:4point} and \ref{sec:induced}.) We will also show in Sec. \ref{sec:4point} that the causal restriction, (\ref{eq:semi}),  interpreted group theoretically, leads to the structure of a ``semi-group".

Before leaving this section, we point out that transition from region $E$ to $M_t$ can be  interpreted via a double Wick rotations,  or a ``twisted" $SO(d,2)$,  i.e., moving from $0<|w|<1$  and $1<\sigma<\infty$ to $0<|w|<1$ and $|w|<\sigma<1$, with both $w$ and $\sigma$ ``compact". (This can be better understood in an embedding space treatment.) 
The transition from $M_t$ to $M_t^{(t)}$ can be accomplished by changing the sign of $\sigma$. (Equivalently, $\sqrt u\rightarrow \sqrt{u'}=-\sqrt u/\sqrt v$, $\sqrt v\rightarrow \sqrt {v'}=-1/\sqrt v$.) This can again be carried out via more involved Wick rotations. We  will not follow up on these possibilities here since it does not lead to the principal series representation for $SO(d,2)$.  

We also note that the relation between the invariant Euclidean four-point functions, ${\cal G}(u,v)$  and our Minkowski discontinuity functions, ${\rm Im}\, T$, has traditionally been discussed   via analytic continuation, e.g., via $(z,\bar z)$. In \cite{caron2017analyticity}, an important identification between ${\rm Im}\, T$ and a ``double discontinuity", ${\rm dDisc} G$, was made.  This relation is briefly reviewed in Appendix \ref{app:dDisc} in connection with crossing line segments $L_s$ and $L_u$ and illustrated by the Ising and generalized mean-field models. In the following sections, we focus mostly on the causal regions  where  Minkowski discontinuity functions, ${\rm Im}\, T$, are defined, i.e., they vanish outside their respective defining regions.  We will return to  this issue  in Sec. \ref{sec:conclusion} and also in \cite{Agarwal:2024}. In this paper, we will  focus  mostly on the  kinematic aspect of this transition.

\section{Four-point Functions, \texorpdfstring{$H$}{H}-bi-Invariance, and Semigroups}\label{sec:4point}

In this section, we begin the effort  to understand  CFT scattering in the context of group functions. In Sec. \ref{sec:kinematics}, we, in essence,  described the left hand side of Eq. (\ref{eq:double-mellin}) as functions over conformal invariants, e.g., by adopting the antipodal frame, the Minkowski analog to the setup for radial quantization in the Euclidean setting. Describing the right hand side of Eq. (\ref{eq:double-mellin}), requires an understanding on the class of group functions that can be associated with causal fourpoint functions. In other words, does understanding the group theory associated with the right hand side of Eq. (\ref{eq:double-mellin}) naturally lead to the condition in Eq. (\ref{eq:semi})?

This is answered in several steps. Our key results for this section can be stated as follows:

\begin{enumerate}
\item  An invariant fourpoint scalar amplitude, after lifting to embedding space, can be treated as a  group function,  $F(g)$,  over $G=SO(d,2)$.  A reduction to cross ratios can be identified group theoretically as ``$ H $-bi-Invariance", $F(h g h')=F(g)$, with an appropriate subgroup $H\subset SO(d,2)$. This subgroup  can be found as the isometry of certain basepoints in the embedding space. (Not surprisingly, it is directly related to that identified by Mack \cite{mack2007simple} in his discussion on Minkowski space-time symmetry.)  We prove this through a formal reduction using a particular coordinate realization appropriate for Minkowski $ t $-channel OPE, and also just by counting that showcase that the number of degrees of freedom after imposing $ H $-bi-invariance is 2, for general dimensions, $d\geq 2$.

\item  Causality in this context can be intuitively understood in terms of lightcones that can group theoretically be identified with ``semigroups", $S\subset SO(d,2)$, a proper subset. We show that this semigroup admits a Cartan-like decomposition $ HA_+H $, with the restriction to $A_+$ given precisely by Eq. (\ref{eq:semi}). That is, our causal condition can be understood as semigroup restrictions due to causality.  Ultimately, in studying Eq. (\ref{eq:double-mellin}), we want to construct representation functions over this semigroup.
\end{enumerate}
We begin by a short review of embedding space. The first part listed above will be dealt with in Secs. \ref{sec:emantipodal} and \ref{sec:HiInv}. The second will be discussed in Sec. \ref{sec:causalsemi}.

\subsection{Embedding Space}\label{sec:embedding}

Following the same procedure as  done for $SO(d+1,1)$, e.g., Ref. \cite{Dobrev:1977}, we again introduce  $ (d+2) $-dimensional embedding space, 
$
\xi=(X_{-2},X_{-1}; X_0,\cdots,X_{d-1})\label{eq:embedding}
$, which provides  a linear realization for
$SO(d,2)$. The group 
leaves quadratic form 
$
B(\xi,\xi)=\xi^T \eta \xi =-X_{-2}^2+X_{-1}^2-X_0^2+X_{d-1}^2 +X_{b_\perp}^2
$
invariant, where $\eta$ is a diagonal metric with two time-like directions, 
$
{\rm diag}\, \eta = (-1, 1;-1,1,\cdots,1)  
$.
Note that we have adopted a particular notation by separating the ``conformal coordinates'' $(X_{-2}, X_{-1})$ from the Poincar\'e coordinates $(X_0,X_{d-1},X_{b_\perp})$, with  $X_{b_\perp} = (X_1,X_2,\cdots,X_{d-2})$ are the $(d-2)$-dimensional transverse coordinates.

\paragraph{Null-Cone  and Physical Slice:}

Now we consider a restriction to  the conic section of a null sub-manifold, 
\be
B(\xi,\xi)=0,  \label{eq:CFTcone}
\ee
which allows a projective identification with the  Minkowski space, i.e., 
\be
x_\mu = \frac{X_{\mu}}{X_{-2}+ X_{-1}},  \label{eq:projective}
\ee
for $\mu=0,1,2,\cdots,d-1$.  In particular, a ``boost" (dilatation)~\footnote{We use the term ``boost" as a catch all phrase for hyperbolic rotations, and ``rotations" will refer to all circular rotations. For example, a boost in $ (X_{-2},X_{-1}) $ direction is physically associated with a dilatation whereas a boost in $ (X_0,X_3) $ direction is a Lorentz boost.} where $(X_{-2}+X_{-1})$ changes by a scale $\lambda^{-1}$ while leaving $X_\mu$ unchanged, leads to scaling in physical coordinates, $x_\mu \rightarrow \lambda x_\mu $. This reduces the $(d+2)$-dimensional embedding manifold down to $d$-dimensional spacetime.
Further, we also have that:
$x^2 = (X_{-2}- X_{-1})/(X_{-2}+X_{-1}).$
Note also that scaling all embedding coordinates uniformly by a  factor, i.e., 
$ \xi \rightarrow  a\xi, \, \forall\ a>0  $ leads  to the same point on the physical spacetime manifold. (This is a classic property of projective spaces.) Therefore, we can choose to set $ X_{-2}+X_{-1} = 1 $ leading to a parametrization of the null cone given by:
\begin{equation}
\xi = \bigg(\frac{1+x^2}{2},\ \frac{1-x^2}{2};\ x_\mu\bigg).  \label{eq:CFTslice}
\end{equation}
This can be identified as a parabolic ``slice" (section) of the null cone containing the ``basepoint" $ \xi_0=(1/2,1/2;0, 0, \cdots) $~\footnote{This parabola is seen when we go to the ``lightcone'' in $(X_{-2},X_{-1})$, that is, $ (X_{-2}+ X_{-1},X_{-2}-X_{-1};X_\mu) = (1,x^2;x_\mu) $. This is a \textit{maximal} parabola in the sense that $ x^2 $ involves all remaining free parameters.}. In the embedding space, we have that:
\begin{align}\label{eq:embed-sep}
\xi_a\cdot\xi_b &=-\frac{1}{2}|x_a-x_b|^2.
\end{align}

The basepoint $\xi_0$, up to a scale, corresponds to the coordinate origin in the physical space, $x_\mu=0$. It can play a special role in providing intuition for introducing \emph{causality}, e.g., via the notion of a forward lightcone. For our purposes, it is also useful to consider different basepoints, e.g., Eq. (\ref{eq:basepointab}), which correspond to adopting different slices in CFT applications. We will return to this in the next section.  

\paragraph{Coset Spaces and Group Decompositions:}
In standard field theory, the identification of the Lorentz invariant spacetime manifold $ R^{1,3} $ is done through  the Poincar\'e group via a semidirect product. In the conformal group $ SO(d,2) $, this identification can best be carried out  in the so-called Gauss group decomposition $ G = Q_0^+H_0Q_0^- $. That is,  a separation into translations, $ Q_0^+ $, Lorentz group plus dilatation, $ H_0 $, and  special conformal transformations, $ Q_0^- $. The spacetime manifold is then identified as a coset $ G/H_0Q_0^- $~\footnote{This identification is also emphasized by Mack in \cite{mack2007simple}. Our $ H_0Q_0^- $ is what he calls $ H $.}.

The subgroup $ H_0Q_0^- $ is the isometry group of a basepoint $x_\mu=0$ in physical space-time, which corresponds to $ \xi_0 = \{1,1;0,0,\cdots\} $ in the embedding space~\footnote{Note that in the embedding space, the isometry is upto an overall scaling of the point. This is because our identification of the spacetime manifold is a cross-section of the null cone in the embedding space with a projective identification (see Eq. \ref{eq:projective}). This subtlety becomes important as it requires us to introduce our ``doubling procedure'' in Sec. \ref{sec:sphr}.\label{fn:isometry}}. Treating  $ H_0$ and $Q_0^- $  separately can be cast in the language of symmetric spaces where we divide the algebra into three parts $ ({\cal Q}_0^+,{\cal H}_0,{\cal Q}_0^-) $ by their spectrum $ (+1,0,-1) $, relative to the generator of dilatations $ D $ by commutators.  (See Appendix \ref{app:embed} for details.)  For $d=2$, ${\cal H}_0$ is particularly simple, consisting of a pair of commuting generators $\{D,L\}$, while for $d = 4$, it is given by Eq. (\ref{eq:HforSapcetime}).

In Section \ref{sec:induced}, we will introduce another decomposition, the Iwasawa decomposition, which is important  in defining unitary irreducible representations for semi-simple groups. 
Here, we will consider instead  a generalization of a group decomposition by Euler angles,   familiar in the case of  compact groups, e.g., $ SO(3) $.  

In most standard mathematical literature on semi-simple Lie groups~\cite{hermann1966}, one starts with a decomposition for its Lie algebra, $ {\cal G} = {\cal K}\oplus {\cal P} $,  with  $ {\cal K} $ for  the generators of the maximal compact subgroup  and $ {\cal P} $  for the non-compact generators~\footnote{See Appendix \ref{app:Iwasawa}.  This is also known as Cartan's Theorem and this decomposition is also referred to as ``Cartan decomposition" for the Lie algebra. This is not to be confused with the usage of ``Cartan-like" decomposition for the Lie group to be  introduced below.}.   These follow the commutation relations
\begin{equation}\label{eq:symm}
[{\cal K},{\cal K}]\subset {\cal K};\quad [{\cal K},{\cal P}]\subset {\cal P};\quad [{\cal P},{\cal P}]\subset {\cal K}\, .
\end{equation}
Using these commutators, one can show that all group elements carry the decomposition $ G = KAK $, where $ K $ is the group associated with $ {\cal K} $ and $ A $ is a maximal abelian subgroup in $ P $~\footnote{As pointed out earlier, $A$ is the non-compact subgroup in the Cartan subgroup of $SO(d,2)$. See App. \ref{app:symmetricspace} for a more detailed handling of how $A$ is fixed.}, which is associated with $ {\cal P} $. This group decomposition can be considered as a generalization of decomposition by Euler angles for compact groups, and will also be referred to as ``Cartan decomposition". The key idea is that all elements of $ P $ can be generated by conjugating elements of $ A $ with a rotation. That is, $ p = kak^{-1} $. 

This can be further generalized for all symmetric spaces. In particular, we can again consider the decomposition $ {\cal G} = {\cal H}_0\oplus {\cal Q}_0 $, where  $ {\cal Q}_0 ={\cal Q}_0^++{\cal Q}_0^-$, and follow similar commutation relations as Eq. (\ref{eq:symm}), with $ {\cal K}\rightarrow{\cal H}_0 $ and $ {\cal P}\rightarrow{\cal Q}_0 $. This can be formalized by introducing an involution, (see Appendix \ref{app:embed}). Under such an involution, all group elements carry a decomposition of the form $ G = H_0A_0H_0 $, i.e., Cartan-like decomposition. These types of  generalized decompositions will play an important role in our harmonic analysis leading to zonal spherical functions.

\subsection{Embedding Space Realization of Antipodal Frame}\label{sec:emantipodal}

In an embedding space approach, an antipodal frame of Sec. \ref{sec:kinematics} can be introduced  by identifying  two special basepoints $ \xi_a$ and $\xi_b $.  These basepoints can be reached directly from the standard basepoint $\xi_0$ mentioned in Sec. \ref{sec:embedding}, by conformal transformations. To be precise, the basepoints in Eq. (\ref{eq:basepointab}) can be reached from $\xi_0$ by rotation of $\pm \pi/2$ in the $(X_{-1},X_{d-1})$ plane. (This also turns out to be important for our causal consideration in Sec. \ref{sec:causalsemi}.)  We demonstrate below that this leads to the amplitude as a function over a single copy of the \emph{maximal abelian subgroup} (MASG), $ A $, i.e., a function of group parameters $y$ and $\eta$. This choice of maximal abelian subgroup generated by Lorentz boost $ L $ and dilatation $ D $ is a physical choice we shall explain shortly.

To simplify the discussion, we shall stay with $ d = 2$.  
Consider  two special basepoints:
\begin{equation}
\xi_a = \{1,0;0,1\}, \quad {
\rm and} \quad \xi_b = \{1,0;0,-1\}.\label{eq:basepointab}
\end{equation}
In physical coordinates, $(t,z)$, they correspond to $x_a=(0,1)$ and $x_b=(0,-1)$ respectively. The action of group elements $ a_{\text{left}},a_{\text{right}}\in A $, generated by the maximal abelian subalgebra identified in Sec. \ref{sec:induced}, $ {\cal A}= \{L,D\} $, leads the antipodal configuration where we identify: 
\begin{equation}
\xi_1 = a_{\text{left}}\cdot \xi_b;\quad \xi_2 = a_{\text{left}}\cdot \xi_a;\quad
\xi_3 = a_{\text{right}}\cdot \xi_a;\quad \xi_4 = a_{\text{right}}\cdot \xi_b.
\end{equation}
To be precise, $a_{\text{left}} =a_L(y_l) a_D(\eta_l)$ consists of a Lorentz boost and a scale transformation, and similarly for $a_{\text{right}}(y_r,\eta_r) $. This parametrization connects directly with that discussed in Sec. \ref{sec:kinematics} and given by Eq. (\ref{eq:Santipodal}). Given $\xi_a$, the second point $\xi_b$ is also fixed by our choice of antipodal frame. In what follows, when speak of $\xi_a$, the point $\xi_b$ is implied.

The fourpoint function is a scalar physical quantity and is invariant under a global conformal transformation. That is:
$
F(\xi_1,\xi_2,\xi_3,\xi_4)=F(g\cdot\xi_1,g\cdot\xi_2,g\cdot\xi_3,g\cdot\xi_4)$. If we choose $ g = a^{-1}_{\text{right}} $, this function reduces to a function of a single group element such that
$ F(\xi_i)=  F(a(y,\eta)\xi_a,a(y,\eta)\xi_b,\xi_a,\xi_b) $,
where $ a(y,\eta) = a^{-1}_{\text{right}}a_{\text{left}}\in A $ with $ y = y_l-y_r $ and $ \eta = \eta_l-\eta_r $.  
That is, it is a function over a single copy of $A$ only, $F(\xi_i) \rightarrow F(a(y,\eta))$, and therefore reduces to a function of just 2 variables. An equivalent identification of this reduction to two variables is through cross ratios. By evaluating  Eq. (\ref{eq:invcr}) in embedding space using Eq. (\ref{eq:embed-sep}), one arrives at 
\begin{equation}\label{eq:uv}
\sqrt u = \frac{2}{\cosh \eta + \cosh y};\quad  \sqrt v = \frac{\cosh \eta - \cosh y}{\cosh \eta + \cosh y}, 
\end{equation}
justifying the earlier claim with $(w,\sigma)$, (\ref{eq:newvariables}), as group theoretic motivated variables. In this view, cross ratios are a simple change of variables. This however covers only a small subset of possible fourpoint configurations. The challenge then is to show that any generic configuration can always be reduced to the one discussed here. We showcase how this can be done next.

\subsection{$ H $-bi-Invariance}\label{sec:HiInv}

The analysis of Sec. \ref{sec:emantipodal} amounts to the demonstration that reducing a fourpoint function to that of  conformal cross ratios is equivalent to  reducing   a group function to that over the MASG, $A$. 
More general fourpoint configurations can be considered by extending the  lifting process via  $ a_{\text{left,right}} $ to elements over the full group $ g_{\text{left,right}} $. These configurations can be reached by starting with basepoints (\ref{eq:basepointab}), i.e., \begin{equation}\label{eq:points}
\xi_1 = g_{\text{left}}\cdot \xi_b ;\quad
\xi_2 = g_{\text{left}}\cdot \xi_a ;\quad
\xi_3 = g_{\text{right}}\cdot \xi_a ;\quad
\xi_4 = g_{\text{right}}\cdot \xi_b ;
\end{equation}
where $ g_{\text{left}},g_{\text{right}} $ can be any element $ G $. This again leads to a reduction of a fourpoint amplitude to a function of a single copy of $SO(d,2)$, $F(g)$, with $g = g^{-1}_{\text{right}}g_{\text{left}}$. Incidentally, we note that this also removes the restriction for $d=2$. We shall make further comments on $ d>2 $ in Sec. \ref{sec:CFTd}. 

As was discussed in Sec. \ref{sec:embedding}, $g$ can always be expanded in a Cartan-like decomposition, $g=hah'$, $h,h'\in H$.  By choosing  the subgroup $H$ as a subgroup of isometries of the basepoints, 
we can  show that $F(g)$ is invariant when left- and right-conjugated, i.e., 
\be
F(hgh')=F(g).
\ee
(See discussion leading to Eq. (\ref{eq:H2}) and also the following subsection.) Functions with this property are called $ H $-bi-invariant. It follows, for this class of functions,
\be
F(g) = F(hah') = F(a).   \label{eq:biH}
\ee

Let us  provide an explicit demonstration, we note that  we can decompose the group elements as $ g_{\text{left}} = h_l a_{\text{left}}h'_l $ and $ g_{\text{right}} = h_r a_{\text{right}}h'_r $.  It follows from the reduction to $g = g^{-1}_{\text{right}}g_{\text{left}}$ described above that 
\begin{equation}
F(g) = F(hah'\cdot\xi_b,hah'\cdot\xi_a,\xi_a,\xi_b) = F(ha\cdot\xi_b,ha\cdot\xi_a,\xi_a,\xi_b)
\end{equation}
where we have used the $ H $-invariance property of $ \xi_{a,b} $. That is, under a shift by $ h\in H $, the basepoints $ \xi_a,\xi_b $ remain fixed upto an overall scaling. However, in the embedding space, since the cross ratios are given by
$ u = (\xi_1\cdot \xi_2\ \xi_3\cdot \xi_4)/(\xi_1\cdot \xi_3\ \xi_2\cdot \xi_4)$ and $v = (\xi_1\cdot \xi_4\ \xi_2\cdot \xi_3)/(\xi_1\cdot \xi_3\ \xi_2\cdot \xi_4)$, 
the scaling drops out, making them invariant under shifts by $ h $.  A final global shift by $ h^{-1} $ of this function, i.e., $ F(g)\rightarrow F(h^{-1}g) $, proves the $ H $-bi-invariance property of the fourpoint function, Eq. (\ref{eq:biH}).

Stated differently, a scalar fourpoint amplitude reducing to a function of   conformal invariant cross ratios, e.g., $(u,v)$, can be  identified with a  function of a single copy of $G$ which is $ H $-bi-invariant, which also reduces to be a function of the MASG, $A$. 
 The subgroup $ A $ is parametrized by 2 variables which are related directly to cross ratios by Eq. (\ref{eq:uv}).

One might wonder if the $ H $-bi-invariance discussion makes sense from a purely counting perspective. That is, is requiring $ H $-bi-invariance sufficient to restrict us to a function of two variables for general $ d $? It is instructive to consider this purely in terms of reduction of variables without requiring additional constraints. 

After reducing to a single copy of the group, the fourpoint function, as a group function,  is a function of $ (d+2)(d+1)/ 2 $ independent variables. The subgroup $ H $ has $d(d-1)/2 +1 $ independent generators. Note that $ H $ is conjugate to $ H_0 $ by a discrete rotation, with $ H_0 $  made up of the Lorentz group and dilatations (see App. \ref{app:embed}). However, there is a $ (d-2)(d-3)/2 $ dimensional subgroup of rotations $ R \in H $ that commutes with our MASG $ A $. Therefore, in the group decomposition
\begin{equation}
G = H_1 A H_2, \label{eq:Cartanlike}
\end{equation}
we can commute those rotations through $ A $. Imposing $ H $-bi-invariance therefore has $ (d-2)(d-3)/2 $ fewer constraints. That is, total number of independent variables after imposing $ H $-bi-invariance is:
\begin{equation}\label{eq:count}
\frac{(d+2)(d+1)}{2}-\left(\frac{d(d-1)+2}{2}+\frac{d(d-1)+2}{2}-\frac{(d-2)(d-3)}{2}\right) = 2.
\end{equation}
This is the expected reduction of variables to 2 cross ratios. This counting applies to both the Euclidean and Minkowski settings.

Our discussion of $ H $-bi-invariance is in contrast with possible utility of the traditional Cartan decomposition $ G = KAK $, that might be more familiar from standard mathematical literatures. With  $g=hah'=ka'k'$, one finds
\begin{equation}
f(g) =f(hah')=f(a)= f(ka'k') \neq f(a').
\end{equation}
This showcases an important feature of our discussion.  Although both representations are allowed, i.e., $ f(g) = f(ka'k') = f(hah') = f(a) $, only the element $ a\in A $ parametrizes the cross ratios. Since  $ a'\neq a $ in general, the cross ratios are still  functions of $ k,a',k' $.

For Euclidean signature $SO(d+1,1)$, $K $-bi-invariant functions reduce to a function of a single variable.  Although four-point functions are not $ K $-bi-invariant in the CFT setting, a class of functions that are $ K $-bi-invariant do feature prominently in AdS/CFT studies under other contexts. An example of particular relevance here is that of Euclidean $ \text{AdS}^E_{d+1} = SO(d+1,1)/SO(d+1) $. Geometrically, this is identified with the two sheeted hyperboloid $ H_{d+1} $ and can be studied using various familiar models of hyperbolic geometry such as the upper half plane or the Poincar\'e disk. True AdS, however, still requires a non-compact $ H $ by definition.

\subsection{Causal Semigroups}\label{sec:causalsemi}

As stressed in the Introduction, a key difference between Minkowski and Euclidean CFT involves causal constraints. In Sec.  \ref{sec:kinematics}, in terms of variables $\{w,\sigma\}$, we have identified kinematically causal scattering regions where Eqs. (\ref{eq:scausality}) or (\ref{eq:ucausality}) hold, leading to 
(\ref{eq:scausalcrossing}). From a group theoretic perspective, in an embedding-space treatment,  understanding these constraints leads to new elements: {\it causal symmetric spaces}~\cite{Hilgert:1996} and {\it semigroups}~\cite{Toller:1973,Viano:1980,Faraut:1986}.  
A thorough discussion on causal symmetric spaces is beyond the scope of this study. A summary is provided in Appendix \ref{app:embed}.  Instead, we present   below  various new ideas intuitively, leading to the emergence of a causal semigroup.

Starting with an appropriately chosen basepoint,  we have demonstrated that a CFT fourpoint functions can be lifted to be a group function, $F(g)$, with $g\in G=SO(d,2)$.  By adopting an antipodal frame, one can move to cover different regions in the cross ratio space, expressed in terms of $\{w,\sigma\}$, as illustrated in Sec. \ref{sec:kinematics}. Our choice of  basepoints, $\xi_a$ and $\xi_b$, Eq. (\ref{eq:basepointab}), corresponds to starting from $(0,1)$ and $(0,-1)$ in physical space for $(t,z)$.    A group action can move these points in physical space, for example the moving coordinates in the lightcone diagram, Figs. \ref{fig:DLC-limit} and \ref{fig:antipodal}. We are interested in using the group action to move in different regions in Fig. \ref{fig:crsymb}, starting with $w=\sigma=1$.  We are interested in finding restriction on group parameters which map within the causal region. 

As demonstrated earlier, conformal invariance is now realized as $H$ bi-invariance, i.e., it is a function of a MASG, $A$. Clearly, a minimal requirement for mapping within the causal region is the restriction to a subset of MASG, $A^+(y,\eta)$, where
\begin{equation}
y-\eta > 0; \quad y>0 \, . \label{eq:causalconstraint} 
\end{equation}
(This restriction can be made more mathematically precise by introducing additional concepts of causal symmetric spaces and semigroups~\cite{Hilgert:1996,Toller:1973,Viano:1980,Faraut:1986}.)

Given a basepoint, the key step involved for imposing causal constraint is by  invoking forward lightcone.   
Recall that, in an embedding space approach,  the spacetime manifold can be identified with the coset space $ {\cal M} = G/H_0Q_0^- $ defined against the basepoint $ \xi_0 = \{1,1;0,0\} $,  which corresponds to the space-time origin. Relative to  $\xi_0$,   a \textit{causal} map can be made moving to each point in its future lightcone. Restricting to the forward light cone of $\xi_0$ requires the constraints  $ z^2-t^2 < 0 $ for $ 0 < t $. A parametrization of the group elements that preserves this map is $ S_0 = H_0\ e^{tT_t}, t>0 $, where $ H_0 $ contains the Lorentz group and dilatation, and $ T_t $ is the generator of time translations. (This map doesn't carry any information about the trajectory and therefore does not contain the inverse of its elements.)  Elements of $S_0$  that obey this restriction therefore form a \textit{causal semigroup}.  
The group $ G $ carries the decomposition $ H_0A_0H_0 $ where $ {\cal A}_0 = \{B_{0,-1},B_{-2,3}\} $ with parameters $ (y_0,\eta_0) $ respectively. Restricting to the forward lightcone leads to $ S_0 = H_0A_0^+H_0 $ where $ A_0^+ $ is a restriction to $ y_0-\eta_0>0 $ with $ y_0>0 $.  (See Appendix  \ref{app:semigroup} for more details.)

The relevant basepoint $\xi_a$,  for our  fourpoint function, can be mapped from $\xi_0$  by a discrete rotation. A corresponding causal structure about $\xi_a$ can now be constructed by applying this discrete rotation. In particular, this    takes $ A_0 $ to 
\be
A = \{L,D\}.
\ee
Furthermore, it maps the subgroup $H_0$ to $H$, given by  (\ref{eq:H2}) for $d=2$ and (\ref{eq:H4}) for $d=4$. Note that $H$ also serves as a subgroup of the isotropy for our basepoint, as claimed. 
More explicitly, the resulting semigroup is  
\be
S = HA^+H. 
\ee
Here $ A^+ $ is a restriction of MASG by
$y-\eta > 0; \quad y>0 $, (\ref{eq:causalconstraint}).
This semigroup preserves the causal ordering of our four points.  (See Appendix  \ref{app:SG4point} for more details.)

A partial wave expansion of the imaginary parts of a four-point function are therefore identified with the $ H $-bi-invariant representation functions with support only over a causal semigroup. 
This is one of our key insights. This restriction was discussed in Sec. \ref{sec:kinematics} via kinematics. In our current approach, we arrive at the same restriction by working directly with four copies of the group and using global conformal invariance to restrict to a single copy along with the basepoint $ \xi_a = \{1,0;0,1\} $.  We will return to the case $d=4$ in Sec. \ref{sec:CFTd}.

\section{Principal Series  for \texorpdfstring{$SO(d,2)$}{SO(d,2)}}\label{sec:induced}

Having discussed the space of functions relevant for expansion in the causal scattering region, we have recast a problem of studying fourpoint functions into group theoretic terms. Let us turn to the problem of constructing a basis using group theoretic technologies. We begin in this section the task of clarifying the claim  \cite{Raben:2018rbn} that the four-point discontinuity function, ${\rm Im}\, T(X_i)$, can be expressed  in terms the principal series of unitary irreducible representation of $SO(d,2)$, leading to a double-Mellin-like representation, Eq. (\ref{eq:double-mellin}). This is done by working in the $ (d+2) $-dimensional  embedding space, thus providing a linear realization for the group $SO(d,2)$. 

We will see that causal considerations emphasized so far fix how positive roots are organized, which fixes the real parts of the principal series representation labels. This in turn provides the necessary connection with Minkowski conformal blocks discussed in \cite{Raben:2018rbn}. 

\subsection{Induced Representation} 

Building group representations for compact groups by starting with a subgroup is known as an {\it induced representation} \cite{Tung:1985}. This procedure can also be applied to non-compact Lie group $G$ \cite{hermann1966,knapp1986,knapp2002lie,helgason2022groups}. The starting point is the {\it Iwasawa decomposition}, 
\be
G=P K=N A K, \label{eq:Iwasawa}
\ee
where a noncompact maximal abelian subgroup (MASG), $A$, plays a central role, (see Appendix \ref{app:Iwasawa} for convention adopted here). The subgroup $ K $ is the maximal compact subgroup and $ N $ is a set of nilpotents characterized by the set of positive roots~\footnote{The root space of a group is characterized by the eigenvalue of generators over the Cartan subalgebra of the group (see Sec. \ref{sec:ReturnRoots} and App. \ref{app:Iwasawa}). The group $SO(d,2)$ has what is called the $B_2$ root system. See Fig. \ref{fig:roots} for an illustration of the root space and positive roots.} of the algebra \footnote{A subtle point here is that $P\neq NA$, with its generators satisfying commutation relations Eq. (\ref{eq:symm}). We are using group notations and all elements of $P$ have an Iwasawa decomposition $NAK$. In terms of group elements, $g = pk = nak'$ where $k,k'\in K$. In what follows, we will also consider a subset $P_0=\{p=na\} \subset P$.}. One begins by constructing  an unitary representation $\pi^{(s)}$ for $A$. An induced representation, $U^{(s)}$, can next be constructed, by introducing ``multiplier factors". The best known example for such approach was first  carried out  for 3-d Lorentz group, $SO(2,1)$, by Bargmann~\cite{Bargmann:1947} (see also \cite{lang,sugiura}), with the  Lorentz boosts in the z-direction serving as the MASG, i.e.,  $A=SO(1,1)$. Here, we generalize to the conformal group.

For $SO(d,2)$,  $A=SO(1,1)\times SO(1,1)$, which is of {\it split-rank} 2.  We shall adopt a ``gauge choice" by identifying $A$ with a Lorentz boost (with its generator denoted by $L=L_{tz}={\cal L}_{0,d-1}$) and dilatation (with generator denoted $D={\cal L}_{-2,-1}$), which are parameterized by $y$ and $\zeta$ respectively.  In general, there are many equivalent choices for a MASG, all with different ``orderings" of root vectors~\footnote{Different ordering of root vectors for the same choice of Cartan subalgebra naturally leads to a different set of nilpotents that appear in Eq. (\ref{eq:Iwasawa}). These different choices are related by a discrete set of rotations that form the group of Weyl transformations (see for example Table 1 and also Sec. 2.3 of \cite{kravchuk2018light}). This will be a key point in what follows. (Also see App. \ref{app:Iwasawa}.)}  for the Iwasawa decomposition. This freedom leads in general to a Weyl group and has been discussed in \cite{kravchuk2018light}.  In Sec. \ref{sec:4point}, we have seen that both $A$ and its ordering are in fact completely determined by the identification of the spacetime manifold by cosets and by a causal requirement.

Let us focus first on the unitary representation for  the MASG.  For a function over $A$ only,  group actions lead to two independent  Fourier integrals, $-\infty< k_y, k_\zeta<\infty$,
\be \label{eq:fourier}
F(a(y,\zeta))=\int_{- \infty}^{ \infty} \frac {d k_y}{2\pi } \int_{-\infty}^{ \infty} \frac {d k_\zeta}{2\pi } \,  e^{i k_y\, y } \, e^{-i k_{\zeta}\,\zeta} \, \widetilde F(k_y, k_\zeta)
\ee
where $a(y,\zeta)=e^{y L_{zt}}\, e^{ \zeta D }\in A=SO(1,1)\times SO(1,1)$ and $\widetilde F(k_y, k_\zeta)$ its Fourier transform.  It is conventional to express  the integrals along the imaginary axes, with 1-d unitary representations: $L_{zt}\rightarrow \widetilde  \ell = i k_y$ and $D\rightarrow -\widetilde \Delta  =- i k_\zeta$, leading to a double mellin-like representation, 
\be
\int_{- \infty}^{ \infty} \frac {d k_y}{2\pi } \int_{-\infty}^{ \infty} \frac {d k_\zeta}{2\pi } \,  e^{i k_y\, y } \, e^{-i k_{\zeta}\,\zeta}  \rightarrow \int_{-i \infty}^{i \infty} \frac {d \widetilde \ell }{2\pi i} \int_{-i \infty}^{i \infty} \frac {d \widetilde  \Delta }{2\pi i} \,  e^{ \widetilde \ell y } \, e^{-\widetilde \Delta \,\zeta} \, .  \label{eq:double-mellin-2}
\ee
The sign convention is adopted for later convenience. Although notationally unfortunate, there is good reason for this. The equation $ \Delta = d/2 + \widetilde{\Delta} $ lends a natural interpretation to $ \Delta $ as an analytic continuation of $ \widetilde{\Delta} $.

In what follows, we shall  designate this unitarity representation for $A$ by  $\pi^{(\widetilde \ell, \widetilde \Delta)}(a(y,\zeta))=e^{\widetilde \ell\,\, y - \widetilde \Delta \, \, \zeta} $, with $\widetilde \ell$ and $ \widetilde \Delta$  purely imaginary. 
In contrast, for an Euclidean CFT, i.e., working with $SO(d+1,1)$ where the MASG $A$ is 1-dim, it leads to a single Mellin-like representation.

An induced representation, $U^{(\widetilde \ell, \widetilde \Delta)}(g)$, can be constructed  by  extending $\pi^{(\widetilde \ell, \widetilde \Delta)}$  to the whole group via Iwasawa decomposition.
A trivial first step is to extend over a subset of $P$,  $P_0=\{p=na\}$.   With $\pi^{(\widetilde \ell, \widetilde \Delta)}(n)=1$, for $n\in N$, this leads to   
$\pi^{(\widetilde \ell, \widetilde \Delta)}(na)=\pi^{(\widetilde \ell, \widetilde \Delta)}(a)$. To proceed,  one needs to  define  a functional space, ${\cal F}$, with an inner product, on which $G$ acts.  An induced representation, $U^{(\widetilde \ell, \widetilde \Delta)}(g)$,  can be constructed via a ``shift" procedure, (for a general discussion, see, for instance, Sec. 2.1.6 of \cite{Vilenkin:1991}),  leading to linear maps, $U^{(\widetilde \ell, \widetilde \Delta)}(g): F\rightarrow F_g$,  preserving group  structure $U(g_2)U(g_1) =U(g_2g_1)$.

Let us first provide a quick description on the shift procedure. Given a function $F$ over $G$,  we identify each as a vector, labelled by  its vector component $F(g)$. Since $G$ acts transitively on the embedding space,  we can associate each with a function over the embedding space,  $F(g)\leftrightarrow f(\xi)$, with $\xi=g \xi_0$ and $\xi_0$ a fixed basepoint.  We next introduce a shifting operation.  There is an option of ``left-shift" vs ``right-shift".   
For ``left-shift", by a group element $g_1$,  $F\rightarrow F_{g_1}$, with  $f_{g_1}(\xi)$ specified  by its action on the embedding space, $f_{g_1}(\xi)=f(g_1^{-1}\xi)$.  
More compactly, the action of this left-shift operator can be expressed as
\be
U_L(g_1) F(g) \equiv F_{g_1}(g) \equiv  F( g_1^{-1} g ). \label{eq:Rtranslation}
\ee
A right-shift operator can also be defined accordingly with $U_R(g_1) F_R(g) \equiv F_{g_1,R}(g) \equiv  F_R(  g g_1)$.  Either approach can be adopted in  dealing with the situation of appropriate bi-invariance relative to a subgroup, (see footnote-\ref{foot:bi-cost}), which we will turn to in the next section.

Completing the specification of this map and the proof of unitary irreducibility   involves certain mathematical gymnastics, (see, for instance, ~\cite{knapp1986,knapp2002lie}); we provide only a brief description here. Consider a space  ${\cal F} $ of complex-valued function over $G$ with the following invariant property 
\be
{\cal F}^{(\widetilde \ell, \widetilde \Delta)} = \{ F : G\rightarrow  \mathbb{C}\, | F(p g) =\phi( \rho (a)) \pi^{(\widetilde \ell, \widetilde \Delta)}(a) F(g), \quad \forall \,\, p=n a \in P_0\}  \label{eq:invF}
\ee
where $\phi( \rho (a)) $ serves as a multiplier factor, as done by Bargmann in ~Ref. \cite{Bargmann:1947}~\footnote{In defining ${\cal F}^{(\widetilde \ell, \widetilde \Delta)}$, it is sufficient to restrict here to $P_0$. This is because, as required below, shifting $F(k)$ by an element of $P$ can be re-expressed as a shift of $F(k')$ by an element of $P_0$. Once  $F(k)$ is defined for all $k\in K$, it is sufficient to define multipliers over $P_0$ only.}. This multiplier factor  is a scalar function to be specified below and elaborated further in Sec. \ref{sec:WeylVector} in relation to root-space structure for $SO(d,2)$.

 Because of  the invariance property specified in Eq. (\ref{eq:invF}),  it renders  ${\cal F}^{(\widetilde \ell, \widetilde \Delta)}$ effectively as functions over the subgroup $K$, (given Iwasawa decomposition, $g=nak$ and the multiplier factor, $F$  is fixed once  $F(k)$ is specified~\footnote{There is an option of expressing Iwasawa decomposition either as $g=nak$ of $kan$. The choice  (\ref{eq:invF} ) prepares for a representation via right-shift. As pointed earlier, one arrives at the same result when considering bi-invariant functions at the end.
}.  
It is then meaningful to introduce a norm with  Haar measure $d\mu(k)$, over $K$,
thus elevating $\mathcal{F}$ to a Hilbert space. To establish $U^{(\widetilde \ell, \widetilde \Delta)}(g)$ as a unitarity representation, one demands, for  $U^{(\widetilde \ell, \widetilde \Delta)}(g) F(k)=F_g(k)$,  
\be
\int_{K} | F_g(k)|^2 d \mu(k)= \int_{K} | F(k')|^2 d \mu(k')\, .
\ee
where $F_g$ is given by a right-shift.  With $g$ given in an Iwasawa expansion, $F_g(k)$ can be re-written in terms of $F(k'(g))$ and a multiplier factor, with $k'(g)$ shifted from $k$. This  variable change  introduces a jacobian, $d\mu(k)/d\mu(k')$. The scalar factor $\phi( \rho (a))$ is introduced precisely to compensate  this extra contribution.   When this is done, it can be shown that this leads to a unitary irreducible representation for $SO(d,2)$.  

Since our group is of split rank 2, the jacobian, $d\mu(k)/d\mu(k')$, thus  the multiplier factor, $\phi^{(\widetilde \ell, \widetilde \Delta)}( \rho (a))$, depends on how the   root-space is ordered~\cite{knapp1986,knapp2002lie}.  Before constructing our representation functions of interest, let us clarify how these representation labels are understood in the context of roots ordering, and their impact on the leading behavior of conformal blocks. We show below that the proper ordering  in turn leads to a particular relation between representation labels  $(\widetilde \ell, \widetilde \Delta)$ and the conventionally defined $(\ell, \Delta)$ in an Euclidean treatment, Eq. (\ref{eq:connection}), 
\be
\ell  = -(d-2)/2 + \widetilde \ell , \quad \quad \Delta= d/2 + \widetilde \Delta. \nonumber
\ee

\subsection{Multiplier Factor, Roots and Weyl-Vector}\label{sec:WeylVector}

Let us provide a more precise determination for the multiplier $\phi(\rho(a))$. For $SO(d+1,1)$, $A$ is 1-dim, generated by $D$, i.e., $a(\zeta)=e^{-\zeta D}$, and the multiplier $\phi(\rho(a))$  also serves as the standard scaling factor, $ \phi(\rho(a(\zeta)))=e^{-\rho(a(\zeta))}=  e^{ -c \,\zeta/2}$, with $c$ a real constant. When unitarity is enforced, the constant $c$ is fixed by the above mentioned jacobian, with $c$ given by the dimension $d$, leading to  
$
\phi(\rho(a(\zeta)))=e^{-\rho(a(\zeta))}=  e^{ -d \,\zeta/2} \equiv e^{ -(d/2) \,\log a},  
$
where we have also adopted another (abused) short-hand  notation, $\zeta=\log a $.  
The factor $d/2$ can be combined with $\widetilde \Delta$ to form a single complex variable $\Delta$, considered as the analytically continued conformal dimension from Euclidean analysis.  To be precise, we identify
$
\Delta = d/2 \pm \widetilde \Delta
$
where at this point we have allowed the freedom of defining $\widetilde \Delta$ up to a sign choice~\footnote{For $d=1$, there is an option for  $A$ depending on the choice of Minkowski or Euclidean $CFT_1$, leadging to $\log a =  \,\zeta$ or $\log a =  \, y$, e.g., for SYK-like models, Ref. \cite{Raben:2018rbn}.}.

We next generalize the notation of multiplier to higher dimensional MASG that is appropriate for our Minkowski case. For $SO(d,2)$, in imposing unitarity, a combined treatment of the jacobian associated with the boost, $y$, and scaling, $\zeta$,  is required. Since these two operations commute, one has
\be
\phi(\rho(a))=e^{-\rho(a(y,\zeta))}= e^{-(c_y y + c_\zeta \zeta)/2}\equiv e^{-\vec \rho \cdot \vec t(a)} \equiv e^{-\langle \rho , t(a) \rangle }  
\label{eq:multiplier-2}
\ee
where, for later convenience, we have introduced additional  short-hand notations, e.g.,   
\be
\vec \rho=(c_y/2, c_\zeta/2), \quad {\rm and} \quad  \vec t(a(y,\zeta)) =(y,\zeta),  \label{eq:rho}
\ee
with  $\vec \rho \cdot \vec t(a)\equiv \langle \rho , t(a) \rangle \equiv (c_y y + c_\zeta \zeta)/2$.   This $\rho$-vector, (or {\bf Weyl-vector}),  will play an important role in what follows~\footnote{We mention the notation $ \langle \rho , t(a) \rangle $ as a way of clarifying the notation typically used in mathematical literature.}.

\begin{table}
	\centering
	\begin{tabular}{ |c|c|c|c|c| } 
		\hline
		\textbf{Ordering of Roots} & $ \boldsymbol{c_y} $ & $ \boldsymbol{c_\zeta} $ & \multicolumn{2}{c|}{\textbf{Leading Behavior}}\\
		\hline \hline
		\rowcolor{yellow}
		$(L,D)$ & $d $ & $d-2$ & $ \ell-1$ & $1-\Delta $ \\ \hline
		$(L,-D)$ & $d$ & $d-2$ & $ \ell-1$ & $\Delta-d+1 $\\ \hline
		$(-L,D)$ & $d$ & $d-2$ & $ 1-d-\ell$ & $1-\Delta $\\ \hline
		$(-L,-D)$ & $d $ & $d-2$ & $ 1-d-\ell$ & $\Delta-d+1 $ \\ \hline
		\rowcolor{yellow}
		$(D,L)$ & $d-2$ & $d$ & $ -\Delta$ & $\ell $\\ \hline
		$(D,-L)$ & $d-2$ & $d$ & $ -\Delta$ & $2-d-\ell $\\ \hline
		$(-D,L)$ & $d-2$ & $d$ & $ \Delta-d$ & $\ell $\\ \hline
		$(-D,-L)$ & $d-2$ & $d$ & $ \Delta-d$ & $2-d-\ell $\\ \hline
	\end{tabular}
	\caption{Coefficients for the group parameters in Weyl vector $ \vec \rho $, (\ref{eq:rho}), change with a choice of ordering of the root vectors. The highlighted rows correspond to Minkowski scattering and non-scattering settings respectively. Similar structure was also highlighted in \cite{kravchuk2018light,Isachenkov2018} Table 1.}
	\label{tab:order}
\end{table}

\paragraph{Positive Roots:} It is important to stress that $\vec \rho$ depends on the choice of the MASG. 
It also  depends on how the ``positive root vectors" are handled in organizing the group algebra. This option on how to specify positive roots can be understood intuitively  from a Gram-Schmidt  perspective. Geometrically, this requires specifying the order of leading root vectors. 
As will be explained in the following sections, when applied to CFT, physics consideration will determine the ordering of positive roots. This in turn will be correlated with the leading asymptotic behaviors of the associated conformal blocks~\cite{Raben:2018rbn}. 
Depending on the ordering for the leading positive root, the allowed values for $\vec \rho$ are given in Table \ref{tab:order}.

To be more precise, the representation functions are labelled by two complex numbers that we write in the shorthand notation
\begin{equation}\label{eq:lambda}
	\vec{\lambda} = \vec{\widetilde{\lambda}} - \vec{\rho}
\end{equation}
where $ \vec{\widetilde{\lambda}} = (\widetilde{\ell},-\widetilde{\Delta}) $. As will be explained in the next section, in the Minkowski scattering setting of interest to us, the roots are ordered $ (L,D) $. The Weyl vector, as is indicated in Tab. \ref{tab:order}, is given by $ \vec{\rho} = \{d/2, (d-2)/2\} $. Therefore, the representation functions are labelled by
\begin{equation}
	\vec{\lambda}^M = 
	\left(\widetilde{\ell}; -\widetilde{\Delta}\right)-
	\left(\frac{d}{2}; \frac{d-2}{2}\right) 
	= \left(\widetilde{\ell}-\frac{d}{2}; -\widetilde{\Delta}-\frac{d-2}{2}\right)
\end{equation}
where the superscript $ M $ denotes that these are labels appropriate for the Minkowski causal scattering case. Similarly, for the non-scattering case, the roots are ordered $ (D,L) $ and the representation labels are given by:
\begin{equation}
	\vec{\lambda}^E = 
	\left(\widetilde{\ell}; -\widetilde{\Delta}\right)-
	\left(\frac{d-2}{2}; \frac{d}{2}\right) 
	= \left(\widetilde{\ell}-\frac{d-2}{2}; -\widetilde{\Delta}-\frac{d}{2}\right).
\end{equation}
The resulting zonal spherical function in this case  can be analytically continued to the standard Euclidean conformal blocks where $\Delta$ is real and positive, with $\ell$ non-negative integers. This will be discussed further in \cite{Agarwal:2024}.
Identification with the traditional representation labels can now be made by recognizing that $ \vec{\lambda}^E = (\ell, -\Delta) $ which leads to
\be
\ell  = -(d-2)/2 + \widetilde \ell , \quad \quad \Delta= d/2 + \widetilde \Delta. \label{eq:connection}
\ee
(Note that Eq. (\ref{eq:connection}) is defined upto a sign of $ (\widetilde{\ell},\widetilde{\Delta}) $. As already mentioned earlier, here we have implicitly made a choice.)

Note also that this is precisely the identification required in matching the quadratic Casimir obtained from an Euclidean analysis, $C(\ell, \Delta)= -[\Delta (\Delta-d) + \ell (\ell +d-2) ]$, with $\Delta$ real and positive and $\ell$ integral~\footnote{For $SO(d+1,1)$, the physics of CFT dictates the choice of identifying $\Delta$ as the eigenvalue for the dilatation and $\ell$ as the integral representation label  for $SO(d)$. It can shown that there is one more independent quartic Casimir, which can also be expressed in terms of $\Delta$ and $\ell$. Therefore, $\Delta$ and $\ell$ are independent.}. By analytic continuation via (\ref{eq:connection}), one obtains
\be \label{eq:casimir}
C(\ell, \Delta)
=-\widetilde \Delta^2 -\widetilde \ell^2+d^2/4+ (d-2)^2/4.
\ee
 More interestingly, we can also translate the labels for representation functions in the Minkowski scattering region into these traditional labels to get $ \vec{\lambda}^M = (\ell-1,1-\Delta) $, which recovers the expected leading behavior for Minkowski conformal blocks  in a Lorentzian OPE expansion by closing appropriate contours~\cite{Raben:2018rbn}.

\subsection{Induced Representation for Semigroups and Root Space}\label{sec:ReturnRoots}

Induced representations, as introduced so far, rely on the identification of Iwasawa decomposition of the group $ G = N^+AK $, where we write $ N^+ $ to emphasize that these are nilpotents associated with the set of positive roots.  For this case, we noted that the identification of $ N^+ $ relies on the ordering of root vectors. For finite dimensional Lie algebras such as $ SO(d,2) $, the vectors in this space are discrete objects in the sense that they are finitely numbered. We can therefore divide them into two equal subspaces by identifying a hyperplane perpendicular to some root (See Appendix \ref{app:Iwasawa} and Fig. \ref{fig:roots}). We can call the subspace containing the specified root to be the positive set, whereas the complement subspace is negative set. The choice of hyperplane can be described by an appropriate Weyl transformation as illustrated in Table \ref{tab:order}. Ultimately, the representation functions constructed via the Iwasawa decomposition cover the full group and an equivalence between representations can be established.

\begin{figure}[ht]
	\centering
	\begin{subfigure}{0.4\textwidth}
		\includegraphics[width=\textwidth]{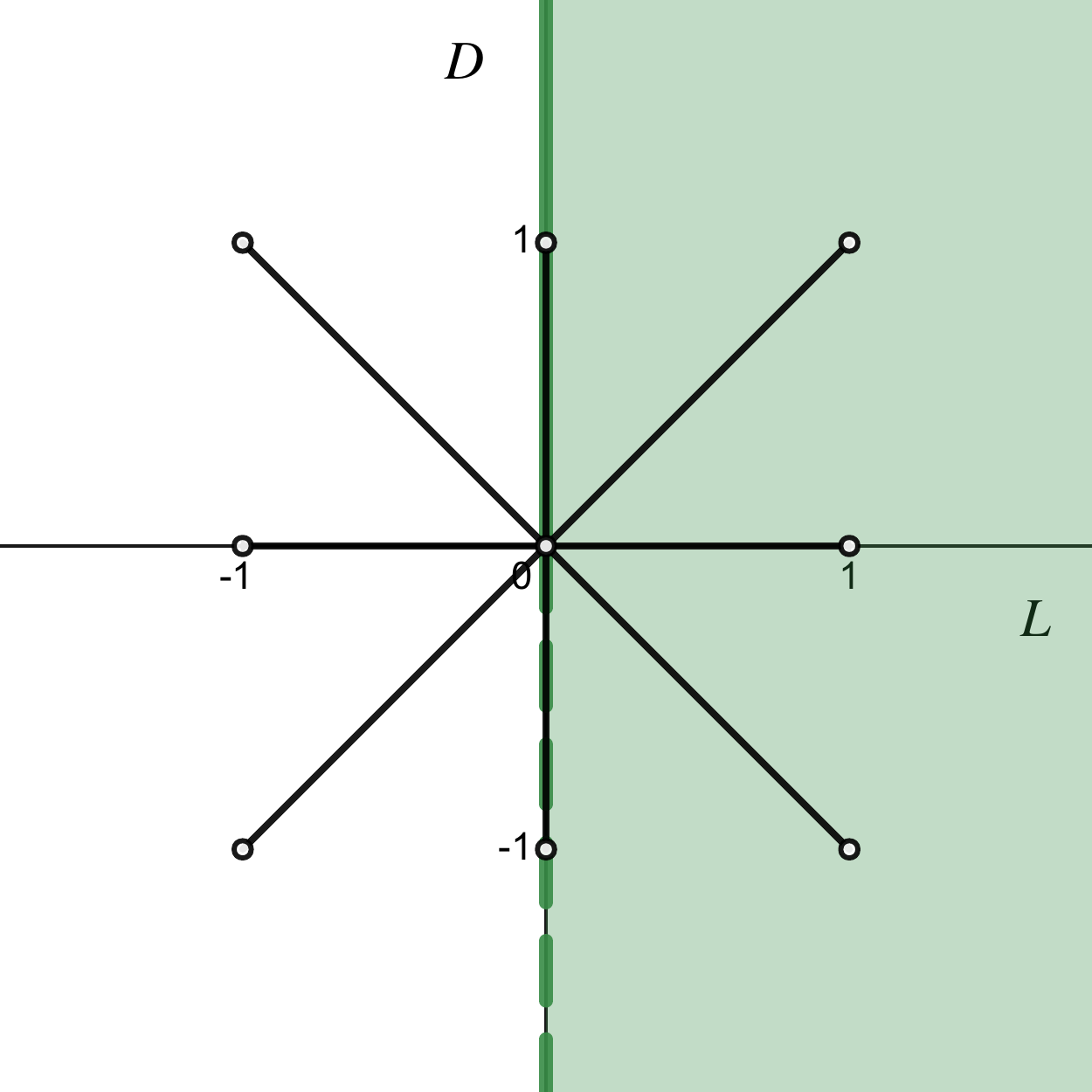}
		\caption{}
	\end{subfigure}
	\quad\quad\quad\quad\quad\quad
	\begin{subfigure}{0.4\textwidth}
		\includegraphics[width=\textwidth]{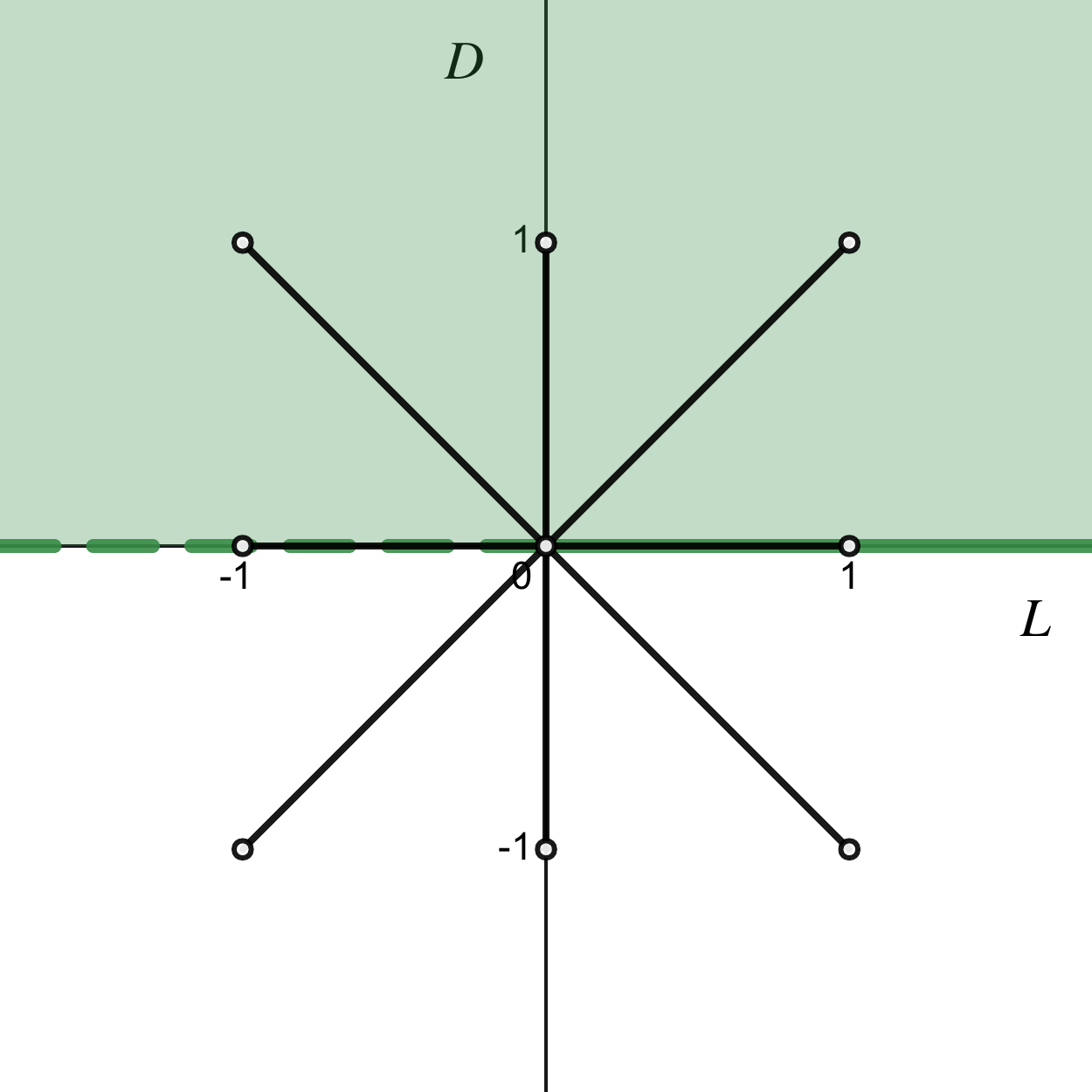}
		\caption{}
	\end{subfigure}
	\caption{The restricted root space diagram for $ SO(d,2) $ is of type $B_2$. The green shaded region contains the set of positive roots. Figure-a corresponds to ordering $ (L,D) $ whereas figure-b corresponds to $ (D,L) $. The roots on the solid boundary of this region are included whereas those on the dashed boundary are excluded. }
	\label{fig:roots}
\end{figure}

However, the equivalence in representations is not true for representation functions we wish to define over the \textit{semigroup}. This freedom in the choice of positive roots $ N^+ $ no longer exists for the semigroup and the ordering of roots is completely determined by $ S $ \cite{Hilgert:1996}. That is, for $ S\subset N^+AH $ to hold in the scattering region, $ N^+ $ is the set of positive roots defined with $ L $ as the leading component of root vectors. Intuitively, the idea is that combinations of elements in $ N^+ $ can \textit{at most} correspond to null translations on the lightcone to stay within the semigroup (see Table \ref{tab:roots} and App. \ref{app:root-2} for an illustration for $d=2$). The element against which the positive roots are defined is therefore called the cone generating element \cite{Hilgert:1996}. (A discussion on this element and how it is completely fixed is presented in App. \ref{app:embed}.)

\begin{table}[ht]
	\centering
	\begin{tabular}{ |c|c| } 
		\hline
		\textbf{Roots} & \textbf{Generators} \\
		\hline \hline
		$(1,-1);(-1,1)$ & $T_t+T_z;C_t+C_z$ \\ \hline
		$(1,1);(-1,-1)$ & $C_t-C_z;T_t-T_z$ \\ \hline
	\end{tabular}
	\caption{Roots for $SO(2,2)$ ordered as $(L,D)$. Positive roots that belong in $N^+$ are therefore $\{T_t+T_z,C_t-C_z\}$ (see Appendix \ref{app:Iwasawa}).}
	\label{tab:roots}
\end{table}

For our choice of basepoint $ \xi_a = \{1,0;0,1\}$ in $d=2$, the Weyl vector is given by
\begin{equation}
	\vec{\rho} = \frac{1}{2}\sum_{\alpha\in\alpha_+}m_\alpha \vec{\alpha} = \frac{1}{2}(2,0)
\end{equation}
where $m_\alpha$ is called the multiplicity of roots, with the element $ L $ dictating that the roots be ordered as $ (L,D) $. The Weyl vector is defined as the half sum of positive roots. As seen in Table \ref{tab:roots}, for $d=2$, $m_\alpha = 1$. Therefore, the multiplier  in the Minkowski scattering region is  $ \varphi(\rho(a)) = e^{-y} $.  That is, as established in Sec. \ref{sec:kinematics}, this  applies when $1<\sigma< w<\infty $.

\subsection{Additional Issues for Minkowski  CFT Applications:}  \label{sec:additions}

To  apply the principal series representation, Eq. (\ref{eq:double-mellin}) for Minkowski CFT scattering, as emphasized in Secs.  \ref{sec:kinematics} and  \ref{sec:4point}, it is necessary to move kinematically into appropriate causal regions, e.g., as indicated in Fig. \ref{fig:crsymb}. It is also, in an embedding space approach, equivalent to implement $H$-bi-invariance in order to reduce ${\rm Im}\, T_s$ to a function of cross ratios. Some of these issues  were already emphasized in Ref. \cite{Raben:2018rbn}, where the double sheeted structure of the $ u$-$v $ plane was resolved by taking square roots.  We shall address briefly here some additional issues, which are not central to our current focus in this study but nevertheless important. We will keep the discussion here brief.

We have thus far focussed primarily  on the 
$s$-channel discontinuity, ${\rm Im}\, T_s$, now with an added subscript. Because of $w\leftrightarrow -w$ symmetry, it is simple to combine $s$-channel and $u$-channel discontinuities by forming signatured discontinuities ${\rm Im}\, T^{(\pm)}(w,\sigma)= {\rm Im} \,T_s(w,\sigma)\pm {\rm Im} \,T_u(-w,\sigma)$.  This corresponds to the introduction of signatured amplitudes $T^{(\pm)}(w,\sigma)$, as done in classical Regge theory, which, at fixed $\sigma$, can be represented by one-side dispersion integrals, (with branch cut along $\sigma<w<\infty$). This approach allows one to deal with more general CFT applications with mixed $s$-$u$ symmetry, e.g.,  see  \cite{Brower:2008cy,Brower:2014wha}. 

The use of a dispersion integral is only formal unless one can control the asymptotic behavior of its discontinuity, i.e., ${\rm Im}\, T^{(\pm)}(w,\sigma)$, as $w\rightarrow \infty$. The required convergence condition is generically not met  for CFTs. Here, we accept, minimally, the standard assumption of polynomial-boundedness, as in a more axiomatic approach, e.g., \cite{Kravchuk:2020scc,Kravchuk:2021kwe}. More  fruitfully, we adopt that followed from model studies, e.g., that based on string-gauge duality~\footnote{Discussions for the issue of asymptotic bounds for CFT amplitudes can be found in \cite{Brower:2006ea,Maldacena2016,Penedones:2019tng} (see additional references therein and also \cite{Cornalba:2006xm,Cornalba:2006xk,Brower:2007qh,Brower:2007xg,Cornalba:2007zb,Cornalba:2007fs,Hofman:2008ar,Strassler:2008bv,Cornalba:2008qf,Cornalba:2009ax,Costa:2012cb,Brower:2014wha,Belitsky:2013xxa,Belitsky:2013ofa,Banks:2009bj,Nally:2017nsp,Raben:2018rbn,carmi2020conformal}). From a Regge perspective, this is due to the intercept of the so-called ``Pomeron" being greater than 1 but bounded by 2 \cite{Brower:2006ea,Kotikov:2004er}.  For a historical perspective on the Pomeron and Bootstrap, see \cite{Tan:2022} and other contributions  in \cite{brink2022}.}, leading to ${\rm Im}\, T^{(\pm)}(w,\sigma)=0(w^{1-\delta})$, with $0\leq\delta<1$ . Under such assumption, a dispersion integral can be constructed with one or two subtractions.

Conversely, it can also be shown, with $\vec \rho$ fixed,  the space of functions defined by double-Mellin representation in general are of the class $O(w^{-d/2})$, for $w\rightarrow \infty$.  (See discussion below.) The resulting  partial-wave amplitude $A(\widetilde \ell,\widetilde \Delta)$ would be analytic when $ {\rm Re} \,\widetilde \ell>0$~\footnote{We defer on how to define the partial-wave amplitude with an illustration with $d=1$ to Sec. \ref{sec:more}. The problem of square integrability to define functional spaces with an inner product is dealt with appropriate generalizations of the Plancherel theorem. A discussion for Riemannian spaces can be found in \cite{helgason2022groups,Anker2005}.}.  
Given ${\rm Im}\, T^{(\pm)}(w,\sigma)=0(w^{1-\delta})$, and $0\leq \delta<1$, this condition is again not met for CFTs. 
This can also be handled by a subtraction. An alternative approach is to  allow $A(\widetilde \ell,\widetilde \Delta)$ having isolated singularities  in the right-half $\widetilde \ell$-plane, while distorting the contour for  the $\widetilde \ell$ integral in the principle series representation,  Eq. (\ref{eq:double-mellin}), by staying to the right of these singularities,  as discussed in Ref. \cite{Raben:2018rbn}. (One can interpret  this procedure adiabatically by first keeping $\delta$ sufficiently negative so that all singularities in $\widetilde \ell$ lie in the left-half plane and next, by increasing $\delta$ with a finite number  of the leading singularities cross the line ${\rm Im} \widetilde \ell=0$. This leads to a contour distortion, which is effectively equivalent to a  subtraction procedure. This is typically the accepted procedure, e.g., \cite{Brower:2006ea,Cornalba:2006xm,Cornalba:2006xk,Brower:2007qh,Brower:2007xg,Cornalba:2007zb,Cornalba:2007fs,Hofman:2008ar,Strassler:2008bv,Cornalba:2008qf,Cornalba:2009ax,Costa:2012cb,Brower:2014wha,Belitsky:2013xxa,Belitsky:2013ofa,Banks:2009bj,Nally:2017nsp,Raben:2018rbn,carmi2020conformal},     in accounting for the leading Regge poles.)  An explicit illustration, in the case of Minkowski $CFT_1$, i.e., $d=1$, will be provided at the end of next section.

The proper identification of the asymptotic limit also ties in with our treatment of root-space ordering, which we turn to next. The defining property of the representation functions on the null cone is the homogeneity condition of Eq. \ref{eq:invF}. In the next section, we will discuss further on how $H$-bi-invariant functions are constructed using these scaling conditions. Their asymptotic behavior can however be studied immediately.

There are 8 distinct root orderings for the $B_2$ root-system (listed in Table \ref{tab:order}) that are related by Weyl transformations. The multiplier factors of Eq. (\ref{eq:invF}) are given by
\begin{equation}
	e^{\vec\lambda\cdot \vec t} = e^{(\vec{\widetilde{\lambda}}-\vec\rho)\cdot \vec{t}} = e^{(-c_y/2+\widetilde{\ell})y} e^{(-c_\zeta/2-\widetilde{\Delta})\zeta}\ .
\end{equation}
These multipliers dictate the dependence of the scattering amplitudes on $\Delta,\ell$ once we close the contour to pick up the poles along spectral curves. This gives us the leading exponents tabulated in Table \ref{tab:order} under different root orderings. (For our causal basis, this leads to $c_y=d$ and $c_\zeta=d-2$.)

An important question to ask about our basis functions is whether they are invariant under some subgroup of the Weyl group. It can be shown that, for the semigroup, the property of $H$-bi-invariance allows for invariance under a subgroup of Weyl transformations~\footnote{In the language of \cite{kravchuk2018light}, the transform $\mathrm{S}_\Delta\in H$, whereas $\mathrm{L},\mathrm{S}_J\notin H$, where $H$ is as defined in the scattering region. Therefore, $H$-bi-invariant functions are invariant only under the scale shadow transform $\mathrm{S}_\Delta$ which takes $D\rightarrow-D$. The transform $\mathrm{L}$ takes us to the non-scattering region, where the ordering of roots is $(D,L)$. Since $\mathrm{S}_J = \mathrm{L}\mathrm{S}_\Delta\mathrm{L}$, the appropriate zonal spherical functions in the non-scattering region are symmetric under $L\rightarrow-L$.}. In particular, we expect for our causal semigroup at $d=2$ that the zonal spherical functions would be even under the Weyl transformation that takes $D\rightarrow-D$. This can be seen in our basis functions of Eq. (\ref{eq:d2conformalblock}) which are invariant under $\widetilde{\lambda}_2\rightarrow-\widetilde{\lambda}_2$. Spectral curves for Minkowski CFTs discussed in \cite{Raben:2018rbn} are symmetric in $\widetilde{\Delta}$ for this reason. Similarly, since the positive root space is not invariant under $L\rightarrow-L$, the spectral curves do not have the corresponding symmetry in $\widetilde\ell$.

Therefore, this symmetry of the basis functions is central to the issue of constructing a complete and orthonormal basis along with dictating some natural assumptions we make about the singularity structure of the partial wave amplitudes. Much in analogy with the Legendre functions of the second kind  $Q_\mu(z)$, representation functions over semigroups in general do not have simple orthogonality relations which makes their usage in a traditional partial wave analysis challenging. As was discussed in \cite{Raben:2018rbn}, this can dealt with by taking a linear combination of blocks and their shadows, leading to Legendre functions of the first kind $P_\mu(z)$ via Eq. (\ref{eq:PQ}). These issues will be discussed in Sec. \ref{sec:more} for $d = 1$ and dealt with in more detail in \cite{Agarwal:2024}.

In general, one can construct functions that are invariant under the entire Weyl group by taking appropriate linear combinations of the different $H$-bi-invariant functions with distinct root orderings
\begin{equation}
	{\cal G}_{\text{symmetric}} = {\cal G}_{L,D}\oplus{\cal G}_{-L,D}\oplus{\cal G}_{D,L}\oplus{\cal G}_{-D,L}\ .
\end{equation} 
Each term in this combination has distinct asymptotic behavior and, in a Mellin representation, a natural plane in which the contour can be closed at infinity (with required subtractions discussed previously). Since the four-point function itself should be $H$-bi-invariant, only the appropriate terms should contribute to the final result. Therefore, the partial wave amplitudes would be meromorphic in the region associated with the representation functions dictated by the correct causal semigroups while being analytic everywhere else.

\section{Spherical Functions and Semigroups}\label{sec:sphr}

The unitary irreducible representation (UIR) also provides the opportunity to represent functions over a group in terms of a set of properly chosen basis functions by specifying ${\cal F}^{(\widetilde \ell, \widetilde \Delta)}$ when additional symmetry is involved; this leads to harmonic analysis over the group. The conventional procedure for specifying group harmonics appropriate for CFT fourpoint functions is to consider the second order partial differential equation which follows from the quadratic Casimir operator.  Zonal spherical functions can be considered as the solutions to the radial component of Casimir equations by imposing appropriate boundary conditions.  

Instead, we show in this section, due to the structure of semi-group and $H$-bi-invariance, Eq. (\ref{eq:biH}),  the desired   causal zonal spherical functions for $SO(d,2)$ can be expressed in an integral representation
\be
\varphi_{{\lambda}}(a )  =\varphi_{{\lambda}}(h_1 a h_2) = \int_{H} e^{(-\vec{\rho}+\vec{\widetilde{\lambda}})\cdot\vec{t}(h_1a)} \text{d}h_1, \label{eq:genindchH}
\ee
with $\vec{\widetilde{\lambda}}=(\widetilde \ell, -\widetilde \Delta)$.  These can be evaluated and identified with Minkowski conformal blocks~\cite{Raben:2018rbn}. We begin by a short discussion on group harmonics, both from the perspective of coset construct and from a geometrical perspective of  scaling over null-cone. We next provide a more concrete illustration for the case of $SO(1,2)$, before turning to the case of $SO(d,2)$ by introducing a ``doubling" procedure.

\subsection{A First Look at Group Harmonics} 
\label{sec:example}

\paragraph{Coset Construct for $SO(3)$:}
UIRs for $SO(3)$ correspond to having  the Casimir taking on a value $\lambda=\ell(\ell+1)$, $\ell$ a non-negative integer. As is well understood, spherical functions for $SO(3)$ are the standard spherical harmonics, $Y_{\ell,m}(\theta,\phi)=  D^{\ell}_{mn}(\phi,\theta,\psi)\Big |_{n=0}$, i.e., a restricted subset of  Wigner $D$-functions. They  can be considered as  group functions defined over the coset $SO(3)/SO(2)$, thus the  absence of dependence on the Euler angle $\psi$. As   such,  they are right $K$-invariant, with $K=SO(2)$. 

Zonal spherical functions are  the Legendre polynomials, $P_{\ell}(\cos \theta)=D^{\ell}_{00}(\phi,\theta,\psi)$. They are functions defined over two-sided cosets, thus independent of both Euler angles $\phi$ and $\psi$. As group functions,  they  are $K$-bi-invariant.  They are given by matrix elements $\langle \ell, 0| g(\phi,\theta,\psi)|\ell,0\rangle$ and admit   an integral representation,
$
P_\ell(\cos\theta)=   2\pi \int_0^{2\pi} W(\theta, \phi; \ell)\text{d}\phi, \label{eq:Legendre}
$
where
\be
W(\theta, \phi; \ell)=(\cos\theta+i\sin\theta\cos\phi)^{\ell}\equiv e^{i \ell \theta'(\theta, \phi)}.\label{eq:Wcompact}
\ee
For reason to be explained shortly, it  will be referred to as the {\it  inductive character} for $SO(3)$.

\paragraph{Coset Construct for $SO(2,1)$ or $SO(1,2)$:}

 Consider next zonal spherical functions for slightly more involved examples. As a  direct extension from $SO(3)$,  we can consider coset construct for either  $SO(2,1)$ or $SO(1,2)$. Each contains a compact subgroup $K=SO(2)$ and also  two non-compact subgroups $H=SO(1,1)$. 
 As already mentioned in the Introduction, $ SO(1,2)$ and $SO(2,1)$ are  identical in their mathematical structure. However, they can take on different interpretation depending on physics applications.  We shall focus  on $SO(1,2)$ in what follows.
 
 \paragraph{Geometrical Perspective and Scaling:}
 
 Let us next provide a geometrically intuitive way of understanding induced representations from the perspective of homogeneous functions. To be  definite, consider $SO(1,2)$ where homogeneous functions can be described over a null submanifold in the embedding space with metric $(-1,1;-1)$, 
\begin{equation}
-\xi_1^2+\xi_2^2-\xi_3^2 = 0,   \label{eq:d1nullcone}
\end{equation}
 invariant under simultaneous scaling $\xi_i\rightarrow  a \, \xi_i$. (For this section, we shall denote embedding coordinates by $\xi_i$, $i=1,2,3$, with $\xi_1$ always timelike.)  
Projective geometry is the natural language of these submanifolds. 

We start by identifying an appropriate projective plane that  ``slices'' the null cone, e.g., by fixing one of the coordinates.
Representation functions over this conic section, (or slice), can be written down first and then ``lifted up'' to the full cone by scaling. In particular, we specify this scaling by a non-compact boost parameter $\eta$ in the $(\xi_1,\xi_2)$ plane. This corresponds to fixing the MASG, $A$, as  $B_{12}(\eta)$, i.e., a boost in the 1-2 plane. There is one remaining homogeneous coordinate, which can be chosen to be  $x=\xi_3/(\xi_1+\xi_2)$ and it can be identified with Minkowski time, $t$, appropriate for  Minkowski $CFT_1$~\footnote{By switching to $SO(2,1)$, while staying with $B_{12}$ as the MASG, the physical coordinate, $x=\xi_3/(\xi_1+\xi_2)$, is now Euclidean, appropriate for Euclidean $CFT_1$. 
}.  

Given a chosen  conic section, a subgroup can be identified which acts transitively. There are two remaining subgroups, one compact, $K=SO(2)$, and one non-compact, $H=SO(1,1)$.  Each choice leads to a different coset construct, i.e.,  $SO(1,2)/SO(2)$ and $SO(1,2)/SO(1,1)$ respectively. Functions over these cosets  are right-$K$ and right-$H$ invariant respectively. One can construct next two-sided cosets, $SO(1,2)//SO(2)$ and $SO(1,2)//SO(1,1)$. Functions on these double cosets are $K$-bi-invariant and $H$-bi-invariant respectively.

\subsection{Space of Homogenous Functions  for  $ SO(1,2)$}\label{sec:so12}

To illustrate how to construct zonal spherical functions, we consider first  the  case of compact subgroup $K=SO(2)$ by identifying the inductive character via  the procedure of  homogeneous functions, (see Appendix \ref{app:F1}).  The result is analogous to that for $SO(3)$, (Eq. (\ref{eq:Wcompact})). We next consider the case of non-compact subgroup, $H=SO(1,1)$, appropriate for Minkowski $CFT_1$.  The feature of ``causal effect" begins to emerge~\footnote{The concepts of causal spherical function was first considered in \cite{Viano:1980,Faraut:1986} by extending beyond the null cone, e.g., cosets defined over either $AdS$ or $dS$. A brief discussion on this perspective is provided in Appendix \ref{app:sphr}.}. This case   also directly serves as a stepping stone for our desired generalization for $d\geq 2$. 

\paragraph{Compact Subgroup:}    Let us begin by specifying a  conic section, (a slice  cutting through the null cone),   $\Gamma_1=\left(\xi_1/\xi_2,\; 1,\; \xi_3/\xi_2\right)= \left(\cos (\varphi),\; 1,\; \sin (\varphi)\right)$,  with $\varphi\in [0,2\pi]$, and $\xi_2>0$ fixed. (See Appendix \ref{app:F1}) The subgroup $ K =R_{13}=SO(2)$, rotations in the 1-3 plane, acts transitively on $\Gamma_1$. 
Consider next a basis for the space of homogeneous functions defined over the null cone, labelled by $\lambda\in \C$ and $m\in \mathbb{Z}$, 
\be
f_{\lambda,m}(\xi)= \xi_2^\lambda  \left(\frac{\xi_1+i\xi_3}{ \xi_2}\right)^{m} =\xi_2^{\lambda} e^{im  \varphi} \ . \label{eq:homogeneousK}
\ee
The restriction to the  slice $\Gamma_1$ corresponds to $\xi_2=1$. Conversely, one can lift functions on $\Gamma_1$ to that over the null-cone by the scaling factor $\xi_2^\lambda$.

Let us next provide a connection between  the space of homogeneous functions for $SO(1,2)$, defined by (\ref{eq:homogeneousK}), with  the  induced unitary representation, Eq. (\ref{eq:invF}).   With Iwasawa decomposition, $g=nak$, functions over $K$, i.e., $ F(k) = e^{im\phi} $,  is lifted up to the $SO(1,2)$, $F(n a k)$,  by the multiplier factor $\varphi(\rho(a))\pi(a) = e^{({-\rho +\widetilde\lambda})\eta_I} $, where $\eta_I$ is the boost parameter for $a$.  The representation is labelled by a complex $\lambda= -\rho+\widetilde{\lambda}=-1/2+\widetilde{\lambda}$. (The fact that $\rho=1/2$ will be shown below to be appropriate for $d=1$. Here $\lambda$ can be identified with either $\Delta$ or $\ell$.) Note that we have added a subscript for the boost parameter, $\eta_I$, reminding ourselves that this is done through an Iwasawa decomposition, $ G = NAK $. Therefore, the representation on the group  is given by
\begin{equation}
{F}(g)= e^{(-\rho +\widetilde{\lambda})\eta_I}e^{im\phi}= e^{\lambda\eta_I}e^{im\phi} .  \label{eq:InduceK}
\end{equation}
where $g=n(x) a(\eta_I) k(\phi)$.  The multiplier factor, $e^{\lambda\eta_I}$, can be identified with the scaling factor $\xi_2^\lambda$, from the perspective of homogenous functions.

On the coset space $ SO(1,2)/SO(2) $, we have that $ g k = g $. That is, all group elements on the coset are characterized by $ g\in NA $. Much in analogy with $ SO(3)/SO(2) $, this can be achieved by taking $ m = 0 $ in  Eqs. (\ref{eq:homogeneousK}) and (\ref{eq:InduceK}). This yields the subspace of functions that are right-invariant under $ K $.  

For the subspace of functions that are $ K $-bi-invariant, the representation functions live on a two sided coset such that $ k_1gk_2 = g $,  again in the sense of  cosets. It is therefore useful to discuss this space of functions in the Cartan decomposition $ G = KAK $, i.e., in terms of an Eulerian-like parametrization for each group element, $g=k_1(\phi) a(\eta) k_2(\psi)$.  Since $\eta_I$ is independent of $\psi$ due to right $K$-invariance,  the key idea is to write the multiplier $ e^{(-\rho+\widetilde{\lambda})\eta_I} $ in terms of a function over $ k_1(\phi) a(\eta) $. 
We can then project out the left invariant function by integrating over $ k_1$. This is carried out  in Appendix \ref{app:F1}, leading to $K$-bi-invariant zonal spherical functions, Eq.  (\ref{eq:Pcompact}), 
\be
\varphi_{{\lambda}}(a )  =\varphi_{{\lambda}}(k_1 a k_2)= P_{\lambda}(\cosh \eta)=
 \int_0^{2\pi} d\phi  (\cosh \eta + \cos \phi \sinh \eta)^{\lambda}.
 \ee
 The integrand  is  the inductive character for $SO(1,2)$ over $SO(2)$, (it is also occasionally referred to as the Poisson kernel \cite{helgason2022groups,Faraut:1986,Vilenkin1993,Vilenkin:1992}.)
 Mathematically, the inductive character can be written as
\be
W_K(\cosh \eta, \phi; \lambda) = e^{\lambda  \eta_I(k_1a)}=(\cosh \eta + \cos \phi \sinh \eta)^{\lambda}, \label{eq:genindchK}
\ee
indicating that $ \eta_I $ from the Iwasawa decomposition needs to be evaluated as a function of $ k_1(\phi)$ and $a(\eta) $, from the Cartan decomposition. We note, for (\ref{eq:genindchK}), positivity of $e^{\eta_I}$ is consistent with $-\infty<\eta<\infty$, a fact turns out not to hold for the case of non-compact subgroup which we turn to next.

We take this opportunity to clarify some notations involving $\vec{t}$, defined for the rank-2 case in Eq. (\ref{eq:rho}). It was defined in the context of the Iwasawa decomposition, and we shall relabel it $\vec{t}(a_I)$ to avoid confusion with the Cartan decomposition, for example, Eq. (\ref{eq:InduceK}). From the equivalence of group decompositions discussed above, on the coset $G/K$, we have that $na_I = k_1a$. With $n$ triangular in a matrix representation, it is possible to write $\vec{t}(a_I)$ as a function of $k_1a$. That is
\begin{equation}
	\vec{t}(a_I) = \vec{t'}(k_1a)\ .
\end{equation}
Therefore, for the $SO(1,2)$ case discussed above, $\vec{t}(a_I) = \eta_I(k_1a) = \log(\cosh \eta + \cos \phi \sinh \eta)$. For the sake of brevity of notation, we will rewrite $\vec{t'}(k_1a)$ as $\vec{t}(k_1a)$.

\paragraph{Non-Compact Subgroup:} 

We next  deal with the more relevant case of non-compact $H$ as a subgroup of $SO(1,2)$, which serves a stepping stone for dealing with the case of $SO(d,2)$.
The $ H $ bi-invariant functions of interest  can be constructed by considering a slice of the null cone, $-\xi_1^2+\xi_2^2-\xi_3^2=0$, over which the non-compact subgroup $ H=SO(1,1) $ acts transitively. The plane can be chosen as $\Gamma_{2,+}= \left(1, \;\xi_2/\xi_1,\;  \xi_3/\xi_1\right)= (1, \cosh s, \sinh s) $, $\xi_1>0$, $\xi_2>0$, with $\tanh s= \xi_3/\xi_2$, which defines a hyperbolic section of the cone. (See Appendix \ref{app:F2} for $\Gamma_{2,\pm}$.) A basis for the space of functions parametrized by this slice, labelled by $\lambda\in \C$ and $p\in \mathbb{R}$, is given by
\be
f_{\lambda,p}(\xi)= \xi_1^{\lambda} \left(\frac{\xi_2+\xi_3}{ \xi_1}\right)^{ip} =\xi_1^{\lambda} e^{ip s}\ . \label{eq:homogeneous}
\ee
In the language of subgroups  introduced in Sec. \ref{sec:induced}, this can be interpreted as the induced representation written as $ f(\xi) = \pi(na)F(h) $. On the slice $ \Gamma_{2,+}$, we have $ \pi(na) = 1 $ and therefore, $ f_{\lambda,p}(\xi) = e^{ip s} $. A shift by $ a \in A $ takes this to the function $ f_{\lambda,p}(\xi') = (\xi_1')^\lambda \sum_{p'}\ t_{p,p'}\ e^{ip' s'} $.

For functions that are $ H $ bi-invariant, we need to consider a basis which is independent of $s$. It follows that we only need the component $p=p'=0$. This is accomplished   by projection over $H$, (Appendix \ref{app:F2}), leading to
\be
t_{0,0}(\eta)= \cosh \eta^{\lambda} \int_0^\infty ds (1+ \cosh s \tanh \eta)^\lambda\  \theta(\eta).  \label{eq:Q}
\ee
This integral is only well defined in the region dictated by the theta function, $\theta(\eta)$, and corresponds to  an integral representation for the Legendre function of the second kind, $Q_{-\lambda-1}(\cosh \eta)$, Eq. (\ref{eq:Qnoncompact}).

From the integrand, one has
\be
W_H=(\cosh \eta+ \cosh s \sinh \eta)^\lambda\theta(\eta),  \label{eq:InductiveC}
\ee
as the inductive character for $SO(1,2)$ over $SO(1,1)$. The restriction to $\eta\geq0$ corresponds to that appropriate for a semigroup. This is the unitary ``causal zonal spherical functions", with $\lambda = -1/2 + \widetilde{\lambda}$ and $ \widetilde{\lambda} $ imaginary.  (This was obtained earlier in a different context \cite{Viano:1980,Faraut:1986}. See Appendix \ref{app:sphr} and also \cite{Raben:2018rbn,Abarbanel:1971nj}.) The origin of this restriction can be traced back to the choice of the slice $\Gamma_{2,+}$, which for the Minkowski $CFT_1$ case is a causal restriction in the sense that all operators are timelike separated and do not commute. For the slice $\Gamma_{2,-}$, a separate treatment is required, with a complementary semigroup. A generalization of this is also the mathematical structure necessary to understand the causal constraint, (\ref{eq:causalconstraint}), when discussing $SO(d,2)$, $d\geq 2$.

\subsection{Generalized Inductive Character and Doubling Procedure}

\paragraph{Generalized Inductive Character:} The process of constructing induced unitary representations is essentially building up functions over the group $ G $ from known unitary representations of its subgroups. Of particular importance is the Cartan subalgebra which is made up of a direct sum of one dimensional commuting subalgebra such that they can be simultaneously diagonalized. For the principal series of $ SO(d,2) $, this is identified as $ SO(1,1)\times SO(1,1)\times SO(2)\times \cdots \times SO(2) $, for $d>2$. Each of the compact subgroups has a unitary representation of the form $ e^{i\ell\theta} $ with integer $ \ell $ and  each noncompact subgroup has a unitary representation of the form $ e^{\widetilde\lambda\eta} $ such that $ \widetilde\lambda $ is imaginary. Under the isomorphism $ SO(2)\cong U(1) $, these factors can be identified as the {character} (trace) associated with the representation. In the context of induced representations, as have already been mentioned above, for definiteness, we shall refer them as {generalized inductive characters}. 

For $SO(d,2)$, the UIR label can be expressed as a two-vector, ${\vec{\widetilde{\lambda}}} =(\widetilde \ell, -\widetilde \Delta)$~\footnote{This is the sign choice made in Eq. (\ref{eq:connection}).}, for MASG with positive roots ordered as $(L,D)$. The generalized  inductive character can be  expressed  as $W_{\vec{\lambda}}(a, h_1) =e^{(-\vec{\rho}+\vec{\widetilde{\lambda}})\cdot\vec{t}( h_1 a)}$, with $\vec{\lambda}= -\vec \rho +{\vec{\widetilde{\lambda}}} $.   (We have here made use of short-hand introduced earlier, Eq. (\ref{eq:rho}).) 
Once $W_{\vec{\lambda}}(a, h_1)$ is known, zonal spherical function can be   calculated  as an integral averaging over the non-compact subgroup $H$, i.e.,
\be
\varphi_{{\lambda}}(a )  =\varphi_{{\lambda}}(h_1 a h_2) =\int_{H} W_{\vec{\lambda}}(a, h_1) \text{d}h_1 = \int_{H} e^{(-\vec{\rho}+\vec{\widetilde{\lambda}})\cdot\vec{t}(h_1a)} \text{d}h_1. \label{eq:genindch}
\ee

\paragraph{Doubling Procedure:} For the split-rank 1 case,  the inductive character can be derived in several different ways~\footnote{A more conventional derivation can be found in \cite{Vilenkin:1991,Vilenkin1993,Vilenkin:1992}. An alternative derivation is to make use of ``Radon transform", or, more explicitly, Abel or Harish transform.}, including the explicit demonstration for $SO(1,2)$ in Appendix \ref{app:sphr}. Here, we present another  procedure, which we shall refer to as {``doubling"}. (See App. \ref{app:sphrDoubling} and \ref{app:sphrDoubling2}.)
This procedure most easily generalizes to rank 2 case we will be facing.

Let us start in the adjoint representation of the group and consider the space of symmetric matrices by ``doubling",  $ I(g)\equiv g\cdot g^{T} $, where $g^{T}$ stands for the transpose.  Of particular importance is the fact that $ k^{T}= k^{-1} $, which means that the subgroup $ K $ maps to identity in this space. Consider both Iwasawa and Cartan decompositions, 
$g = na_Ik$ and $g = k_1ak_2$,
where $ a_I $ and $ a $ have distinct parametrizations, with $\eta_I$ and $\eta$ respectively. Each group element maps to two equivalent product representations \begin{equation}
I(g)=n\, (a_I\cdot a_I^{T}) \,n^{T} \quad {\rm and} \quad  I(g)= k_1 \,(a\cdot a^{T}) \, k_1^{-1}\ .\label{eq:doubling}
\end{equation}
Note that $I(g)$ is independent of $k$ for the first equality and  $k_2$ for the second~\footnote{The second equality is analogous the standard procedure of SVD for diagonalizable matrices.}. Therefore, $ I(g) $ is associated with group elements explicitly on the coset $ G/K $. Furthermore, since, in a Cayley basis,  both $a_I$ and $ a  $ are diagonal,  Eq. (\ref{eq:Cayley3}), it follows that one can express $\eta_I$ in terms of $\eta$ and an angle $\phi$, which specifies $k_1$, leading to the same expression obtained earlier, $e^{\eta_I} = \cosh \eta+ \cos\phi \sinh \eta$, Eq. (\ref{eq:genindchK}).
(Alternatively, one can express $\eta_I$ in terms of $\eta$ and another parameter $x$ specifying $n$.)

In what follows, we will need to generalize this ``doubling procedure"  for obtaining the inductive character for the case of split-rank 2. This requires a more refined discussion on the root space structure for $SO(d,2)$. We will also need to transition to causal symmetric spaces where the subgroup $H$  is non-compact.
This requires an understanding of the Iwasawa decomposition $ NAH $ as well as Cartan decomposition $HAH$ for semigroups.

\subsection{Minkowski Causal $ \text{CFT}_2 $}\label{sec:MCFT2}

Recall from Sec. \ref{sec:4point} that the representation functions we are after are the $ H $ bi-invariant zonal spherical functions of the semigroup $ S $ such that:
\begin{equation}
	f(s) = f(ha_+h) = f(a_+)
\end{equation}
where $ a_+\in A = \{L,D\} $ and $ y-\eta>0 $. Our choice of basepoint is $ \xi_a = \{1,0;0,1\} $ which helps reduce the fourpoint function to a function over the semigroup of a single copy. Since $ A $ contains two commuting non-compact generators, the representation functions are labelled by two imaginary continuous parameters $ \widetilde{\lambda} = (\widetilde{\lambda}_1, \widetilde{\lambda}_2) $ and the Weyl vector is given by $ \vec{\rho} = (1,0) $, where the roots are ordered as $ (L,D) $ since $ L $ is the cone generating element.

In a direct generalization of $\Gamma_{2,+}$ discussed for $d=1$, we shall see that the characters themselves build in the semigroup structure by being well defined only over a restricted set of parameter values over the subgroup $a$. The resulting spherical functions therefore have a natural definition only over this restricted region of the semigroup.

In this rank-2 case, we have a homogeneity condition involving two parameters
\begin{equation}
	f(a_+\cdot \xi_a) = e^{(\widetilde{\lambda}_1 - 1)y_I}e^{\widetilde{\lambda}_2\eta_I}f(\xi_a).
\end{equation}
These homogeneous polynomials define an intertwining representation between a Hilbert space and the manifold. The inductive character $ W_H $ is
\be
	W_H= e^{(\widetilde{\lambda}_1-1)y_I}e^{\widetilde{\lambda}_2 \eta_I}, 
		\ee
where $ (y_I,\eta_I) $ are functions of the group element $ h(\alpha,\beta)a_+(y,\eta) $. That is, the zonal spherical functions can be obtained  by integrating the inductive characters $ W_H $ over a non-compact subgroup, $ H $.

Consider two alternate expansions for an element of semigroup, Iwasawa and Cartan-like, i.e., $g=n a_+(y_I,\eta_I) h$ and $g=h_1 a_+(y,\eta) h_2$. Notice that $ g\cdot g^{T} $ no longer takes us directly to the coset space $ G/H $ since $ h^{T} \neq h^{-1} $. For a coset construction, we therefore introduce a diagonal matrix $ P $ that has the property that $ h\cdot P\cdot h^{T} = P $. Since $ H $ is the isometry group of our basepoint, we can define $ P $ in terms of $ \xi_a $ such that $ P  = \text{diag}\{\xi_{a,-2}^2, \xi_{a,-1}^2,\cdots\}$. We now consider the doubling space $ I_P(g)=g\cdot P\cdot g^{T} $ to find that
\begin{equation}\label{eq:squareH}
	I_P(g)=na_I\cdot P\cdot a_I^{T}n^{T} \quad {\rm and } \quad I_P(g) = h_1 a\cdot P\cdot a^{T} h_1^{T}
\end{equation}
which again maps the coset of interest.  
Whereas the first representation is independent of $h$, the second is independent of $h_2$. For  $d=2$, subgroup $H$ is of dimension 2 and abelian, making the analysis relatively simple to handle. This analysis applies in general~\footnote{Our manifold $G/H$ is the non-compact analogue of a Grassmannian. Using projectors to define a coordinate system on them is standard \cite{bendokat2024grassmann}. The non-trivial aspect of this generalization is that the metric is not positive definite and we therefore need to keep track of it.}. In particular, our discussion on $ SO(1,2)/SO(1,1) $ can be recast in this language to give the same result as presented in the previous subsection. This is discussed in more detail in Appendix \ref{app:sphrDoubling}.

In the Cayley transformed representation~\footnote{We use the phrase Cayley transform somewhat loosely in our discussion. It is shorthand for a basis in which the Cartan subalgebra is diagonal. See App. \ref{app:sphrDoubling}.} where $ A $ is diagonal and of the form:
\begin{equation}
	d = c^{-1}ac = \text{diag}\{e^y,e^\eta,e^{-\eta},e^{-y}\},
\end{equation}
we can identify the characters $ e^{y_I} $ and $ e^{\eta_I} $ in terms of $ h_1a_+ $ by solving for them in Eq. (\ref{eq:squareH}).  Following a procedure analogous to one in~\cite{Vilenkin:1992},  since $ n $ is nilpotent with 1's on the diagonal, identifying $ e^{y_I} $ can be done simply by evaluating the first matrix element of the second representation. Solving for $ e^{\eta_I} $ can be done by considering the $ 2\times 2 $ minor matrix whose determinant is given by $ -e^{2y_I}e^{2\eta_I} $ (see App. \ref{app:sphrDoubling}, \ref{app:sphrDoubling2} for details).  Note that the root-ordering plays an crucial role in arriving at the desired result. 

It is interesting to note here the structure of the characters $ (e^{y_I},e^{\eta_I}) $. We find
\begin{align}\label{eq:chard2}
	e^{y_I}&\sim \Big(\cosh(y-\eta)\cosh(y+\eta)[1+\cosh(\alpha-\beta)\tanh(y+\eta)] \nonumber \\ 
 &\qquad \qquad \times [1+\cosh(\alpha+\beta)\tanh(y-\eta)]\Big)^{1/2},\\
	e^{\eta_I}&\sim \left(\frac{\cosh(y+\eta)}{\cosh(y-\eta)}\frac{[\tanh(y+\eta)\cosh(\alpha-\beta)+1]}{[\tanh(y-\eta)\cosh(\alpha+\beta)+1]}\right)^{1/2}.
\end{align}
where $\alpha$ and $\beta$ are two angles in specifying the abelian $h_1$. Recall that we are currently working in the causal region, $ 1<\sigma<w<\infty $. Evidently, the characters~\footnote{The character associated with dilatation is a ratio, whereas for the Lorentz boost it is a product. This is an important observation and will be the key insight when discussing the non-scattering regions, e.g., $M_s$, where $ 1<w<\sigma<\infty $ and continuation into the Euclidean region of $ |w|<1 $  in the follow-up paper \cite{Agarwal:2024}.} are well defined only in this region with $y>|\eta|$. This is exactly the generalization to $d=2$ of $\Gamma_{2,+}$ structure of Eq. (\ref{eq:InductiveC}) which is well defined only over the semigroup with $\eta>0$.

Plugging the values of $ e^{y_I} $ and $ e^{\eta_I} $ back into the integral form of zonal spherical functions, we get:
\begin{align}
	\varphi_{\vec{\lambda}}(\vec{t}) =&\ Q_{(-\widetilde{\lambda}_1-\widetilde{\lambda}_2-1)/2}(\cosh(y+\eta))Q_{(-\widetilde{\lambda}_1+\widetilde{\lambda}_2-1)/2}(\cosh(y-\eta)) \theta(y) \theta(y-|\eta|)  \label{eq:d2conformalblock}
\end{align} 
upto a normalization. As mentioned above, the integrals are well defined only for $ y>|\eta| $, which is our semigroup restriction (\ref{eq:causalconstraint}). These zonal spherical functions therefore have support only in this restricted region. Due to these causal restrictions, covering the full group necessarily requires us to define these functions piecewise for different regions. This is the central result of our work.

Taking into account Eq. (\ref{eq:connection}), with $d=2$, one arrives at
	\be
	\varphi_{\vec{\lambda}}(\vec{t})= Q_{(-\ell+\Delta)/2-1}(q)Q_{(-\ell-\Delta)/2}(\bar q). 
\ee
where we have shifted to variables $q=2/z-1$ and $\bar q= 2/\bar z-1$, with $z=\operatorname{sech}^2(y+\eta)/2)$  and $\bar z=\operatorname{sech}^2(y-\eta)/2)$,  (see Table \ref{tab:vars} for their relationship with other variables), with $1<\bar q<q<\infty$.
This agrees with the result for Minkowski conformal blocks first presented in Ref. \cite{Raben:2018rbn}, Eq. (III.13),  by directly solving the Casimir equation in terms  of variables $ (q,\bar{q}) $. Here we have provided an alternative derivation, which can be re-expressed in terms of $(u,v)$, with $0< z<\bar z<1$. They satisfy Minkowski boundary condition,
\be
G^{(M)}_{(\Delta, \ell)}(u,v) \sim \left(\sqrt{u}\right)^{1-\ell} \left(\frac{1-v}{\sqrt u}\right)^{1-\Delta} \label{eq:MCB}
\ee
This corresponds to an expansion about the point $T'$ in  Fig. \ref{fig:crsym22}.

We also note that, by evaluating zonal spherical functions defined in the  region $1<w<\sigma<\infty$, one is led to Euclidean conformal blocks. When further continued to $\bar z=z^*$, which corresponds to entering the region $-1<w<1$, (Fig. \ref{fig:crsym}), it satisfies boundary condition 
		\begin{equation}
			G^{(E)}_{(\Delta, \ell)}(u,v) \sim \left(\sqrt{u}\right)^{\Delta} 	\left(\frac{1-v}{\sqrt u}\right)^{\ell},\label{eq:ECB}
		\end{equation}
corresponding to an expansion about the point $T$  in  Fig. \ref{fig:crsym22}. Comparing   (\ref{eq:MCB}) and (\ref{eq:ECB}),  observe the $(\ell,\Delta)$ and $(1-\Delta, 1-\ell)$ swap, so necessary in various related treatments, e.g., \cite{caron2017analyticity,simmons2018spacetime}. This will be discussed further in \cite{Agarwal:2024}.

\subsection{Minkowski $ \text{CFT}_d $ for $ d>2 $}\label{sec:CFTd}

Harmonics for CFTs in $ d > 2 $ presents some new challenges. The case for $ d = 4 $ is particularly interesting for HEP applications. The Cartan subalgebra for this case is 3 dimensional and can be parametrized by $ \{L,D,R_{12}\} $. The harmonics are therefore labelled by three indices $(\widetilde \ell, \widetilde \Delta, \widetilde m)$: two continuous imaginary numbers and one integer.

The quantum number $\widetilde m$ allows one in principle to deal with functions with dependence on 2-dim vector $\vec b_\perp$. However, by imposing rotational symmetry in the $ x-y $ plane, we can restrict ourselves to functions depending on $b_\perp^2$ only. This corresponds to states with $ \widetilde{m} = 0 $, which is reflective of the fact that $ R_{12}\in H $. Furthermore, as discussed in Sec. \ref{sec:kinematics}, Lorentz invariance allows a further reduction in the dependence on $b_\perp^2$, leading to 
\begin{equation}
	\cosh\eta = \cosh\zeta + b_\perp^2.  \label{eq:bperp}
\end{equation}
This can be demonstrated in the embedding space in the following manner.

For the sake of notational brevity, let us consider $ d = 3 $ where we have a single transverse direction since one can always reduce to this case due to transverse rotational invariance, (that is invariance under $ R_{xy} $). The choice of basepoint analogous to Eq. \ref{eq:basepointab} is:
\begin{equation}
	\xi_a = \{1,0;0,0,1\};\quad \xi_b = \{1,0;0,0,-1\}.
\end{equation}
The fourpoint function is again characterized by Eq. (\ref{eq:points}). For finite impact parameter physics, let us consider the configuration where we have $ x_{1\perp} = x_{2\perp} = b_\perp $ and $ x_{3\perp} = x_{4\perp} =0 $. Therefore, we are led to consider the group elements
$
	g_{\text{left}} = a_{\text{left}}q^+(b_\perp)$ and $g_{\text{right}} = a_{\text{right}}
$,
where $q^+(b_\perp)$ generates translation. 
Notice that this frame is no longer antipodal but can still be characterized as a CM frame. We can therefore choose to parametrize $ a_{\text{left}} $ by $ (y/2,\zeta/2) $ and $ a_{\text{right}} $ by $ (-y/2,-\zeta/2) $, as done previously.  Reduction down to a single copy of the group is done by imposing global shift invariance of the fourpoint function under $ g_{\text{right}}^{-1} $. Therefore,
$
	g = g_{\text{right}}^{-1}g_{\text{left}} = q^+(b_\perp)a^2_{\text{left}}.
$
This group element again carries the decomposition $ g = hah' $. It is this new group element $ a (y,\eta)$ that is related to our $ (w,\sigma) $ variables. It follows that
$
	\cosh\eta = \cosh\zeta +   b^2_\perp,
$
as claimed.

This allows one to restrict to  conformal invariant scalar functions with dependence  only on  2 variables, $(w,\sigma)$, i.e., with scaling parameter $\zeta$ and transverse variable $\vec b_\perp$ combined into a single variable $\sigma=\cosh \eta$. In the complex $ w $ plane, the location of the cut at $ w=\sigma $ is therefore dictated by the impact parameter of the collision.

Group theoretically, this corresponds to having partial-wave amplitudes $ a(\widetilde  \ell ,\widetilde  \Delta, \widetilde m) $ independent of $\widetilde m$ so that the sum over them  can be carried out directly on the conformal harmonics, leading to an ``invariant spherical function", ${\cal G}_{(\bar\ell,\bar \Delta,0)} (w,\sigma)$, as appeared in Eq. (\ref{eq:double-mellin}), i.e.,  with a reduction of variable dependence from $(w,\cosh \zeta, \vec{b}_\perp)$ to $(w,\sigma)$.

In our procedure for constructing harmonics, we expect the Weyl vector to be given by $ \vec{\rho} = \{2,1\} $ and the representation labels
\begin{equation}
	\vec{\lambda} = \vec{\widetilde{\lambda}}-\vec{\rho} = \left(\widetilde\ell - 2, -\widetilde\Delta - 1\right)
\end{equation}
for the Minkowski case. Calculating the harmonics for $ d > 2 $ via an integral representation gets tedious, even if it is largely procedural. The structure of semigroups, ordering of roots and the Weyl transformations can also be viewed in the standard procedure of constructing conformal blocks using the Casimir equation. We shall explore this discussion elsewhere \cite{Agarwal:2024}. 

\subsection{Some Clarifying Comments on CFT$_1$ and Generalizations}\label{sec:more}

It should  also be stressed that  the principal series representation is kinematic; dynamics for CFT is encoded in the analytic structure of the partial-wave amplitude.  
Let us provide here first a simple illustration for $CFT_1$.  With $d=1$, the causal zonal spherical function is given by $Q_{-\ell}(w)$, (identifying $w=\cosh \eta$, $\sigma=1$ and $\lambda =-1/2 +\widetilde \ell=\ell-1$, or $\ell=\widetilde \ell +1/2)$). The partial-wave amplitude, upto a normalization factor, is given by
\be
A(\ell)= \int_1^\infty dw \, {\rm  Im}\,T(w) \, Q_{\ell-1}(w). \label{eq:partialwaved1}
\ee
With ${\cal \rm Im} T(w) =0(w^{1-\delta})$, this can be defined initially by keeping ${\rm Re}\, \ell>\ell_0$, ($\ell_0=2-\delta$),  and then analytically continued so that $A(\ell)$ is analytic to the right of the line  ${\rm Re}\, \ell=\ell_0=2$. Assuming further that ${\rm  Im}\,T(w)$ is power-behaved as $w^{1-\delta}$, this corresponds to a pole, $A(\ell) \simeq r/(\ell-\ell_0)$.   With $0\leq\delta\leq 1$, thus $1\leq \ell_0\leq 2$. this singularity lies in the RHP in $\widetilde \ell$. To apply Eq. (\ref{eq:double-mellin}) directly, as explained in Sec. \ref{sec:additions}, one can either define $A(\ell)$ by a subtraction, or, as is traditionally done in Regge theories,  one initially deforms the integration contour in $\widetilde \ell$ to the right of this leading singularity~\cite{Brower:2006ea,Cornalba:2006xm,Cornalba:2006xk,Brower:2007qh,Brower:2007xg,Cornalba:2007zb,Cornalba:2007fs,Hofman:2008ar,Strassler:2008bv,Cornalba:2008qf,Cornalba:2009ax,Costa:2012cb,Brower:2014wha,Belitsky:2013xxa,Belitsky:2013ofa,Banks:2009bj,Nally:2017nsp,Raben:2018rbn,carmi2020conformal}.

 Let us turn to an illustration for $d>1$.  It is generally accepted that, for CFT, given (\ref{eq:connection}), partial-wave amplitudes, for fixed $\ell$ is meromorphic in $\Delta$, with pole positions fixed by OPE. Conversely, at fixed $\Delta$, it is also reasonable to assume meromorphy in $\ell$, as expected in Regge theories. It has been  hypothesized in \cite{Brower:2006ea}, at least for large-$N$ based theories, these singularities can be simultaneously described by a family of spectral curves, $\widetilde \Delta_\alpha(\ell)$, i.e.,  one has a generalized OPE expansion where partial-wave amplitude formally admits an expansion
\be
A(\widetilde \ell,\widetilde \Delta)= \sum_\alpha \frac{r_\alpha(\ell)}{\widetilde \Delta^2 - \widetilde \Delta_\alpha(\ell)^2}
= \sum_\alpha \frac{r_\alpha(\ell)}{2\widetilde \Delta} \left(\frac{1}{\widetilde \Delta - \widetilde \Delta_\alpha(\ell)}+\frac{1}{\widetilde \Delta + \widetilde \Delta_\alpha(\ell)}\right)
\ee
This becomes immediately apparent in various examples where the spectral curves $ \Delta(\ell) $ are symmetric about $\widetilde \Delta=0$, or $ \Delta = d/2 $. As a simple illustration, consider the case of strong coupling limit for ${\cal N}=4$ YM for 4d, $\widetilde \Delta(\ell)^2= -2 + 2(\ell-2) /\sqrt \lambda$, with $\lambda$ the 't Hooft coupling.  Closing the contour in 
$\widetilde \Delta$ either to the left or to the right, directly leads to 
 a contribution with a Minkowski conformal block, obeying boundary condition, Eq. (\ref{eq:MCB}). (See \cite{Brower:2006ea,Brower:2007qh,Brower:2007xg,Cornalba:2006xm,Cornalba:2007fs,Cornalba:2008qf,Cornalba:2009ax,Costa:2012cb,Raben:2018rbn,Caron-Huot:2020nem,Basso:2011rs,gromov2014quantum,Gromov:2015wca} for further discussion.)

We also see some of the discussion in Ref. \cite{Raben:2018rbn} in new light, e.g.,  inversion for the principal series.  In particularly, the Bethe-Salpeter type equation for SYK-like models can be interpreted in the language of semigroups. Early works, motivated by works of Toller {\it et al.} on Regge physics \cite{Toller:1973} via 3 dimensional Lorentz group such as \cite{Toller3,Viano:1980,Faraut:1986,cronstrom:1972,Hermann}, are particularly interesting from current perspective. The integral kernel is of Volterra type and preserves the causality constraints of the scattering region. For $ d = 1 $, it was shown in Ref. \cite{Raben:2018rbn} that the partial wave amplitude, (\ref{eq:partialwaved1}), is given by $ A(\ell) = A_1(\ell)/(1-k(\ell)) $, with $k(\ell)$ given by a causal kernel, leading to $A(\ell)$ free of singularity to the right of $\ell_0$. In general, this diagonalization corresponds  to a  Laplace transform.   Conversely, given the partial wave amplitude, the equivalent discontinuity can be reconstructed explicitly via an inverse Laplace transform.  This requires  invoking the the relationship between Legendre functions of first and second kind,
\be
P_{\widetilde\lambda(z)-1/2}(q) = \frac{\cot \widetilde\lambda \pi}{\pi} \left(  Q_{\widetilde\lambda-1/2}(q)  - Q_{-\widetilde\lambda-1/2}(q)\right).  \label{eq:PQ}
\ee
(See~\cite{Mack:2019akh} as well as Appendix E of Ref. \cite{Raben:2018rbn}.) This inversion procedure can be generalized for $d>1$, framed in the language of spherical Laplace and Abel transforms \cite{Hilgert:1996}.   This is an interesting application of the semigroup structure developed here. A related issue is the relation of the so-called Froissart-Gribov partial-wave to that obtained directly via $SO(2,1)$ group theoretically, the so-called Toller analysis \cite{Toller3,Hermann}. This as well as the question of inversion for our causal harmonics  will  explored further in~\cite{Agarwal:2024}.

\section{Discussion}\label{sec:conclusion}

The goal of  this study is to clarify the suggestion made in Refs.  \cite{Raben:2018rbn,kravchuk2018light,Mack:2019akh,mack2007simple} that Minkowski CFT 4-point amplitudes can be treated directly in terms of the principal series representation, Eq. (\ref{eq:double-mellin}), and  to amplify the work of Ref. \cite{Raben:2018rbn} on Minkowski conformal blocks. We have tried to limit the scope of this study with emphasis on adopting a direct embedding space approach to $SO(d,2)$ for 4-point Minkowski CFT amplitudes. In this final section, we highlight some of the key issues and their interplay with the results presented here.

The single most important issue for this study involves causality constraints associated with Minkowski space-time.   We introduce a set of variables, $(w,\sigma)$ in Sec. \ref{sec:kinematics},  directly related to cross ratios $(\sqrt u,\sqrt v)$.  By extending to the whole $\sqrt u$-$\sqrt v$ plane, from Minkowski perspective, the $w$-$\sigma$ plane is most useful in specifying the causal scattering regions. By associating these variables with Minkowski antipodal frames, they can be identified with Minkowski $t$-channel OPE, which in turn allows a connection to Regge limit for CFT.

\begin{figure}[ht]
	\centering
	\begin{subfigure}{0.25\textwidth}
		\includegraphics[width=\textwidth]{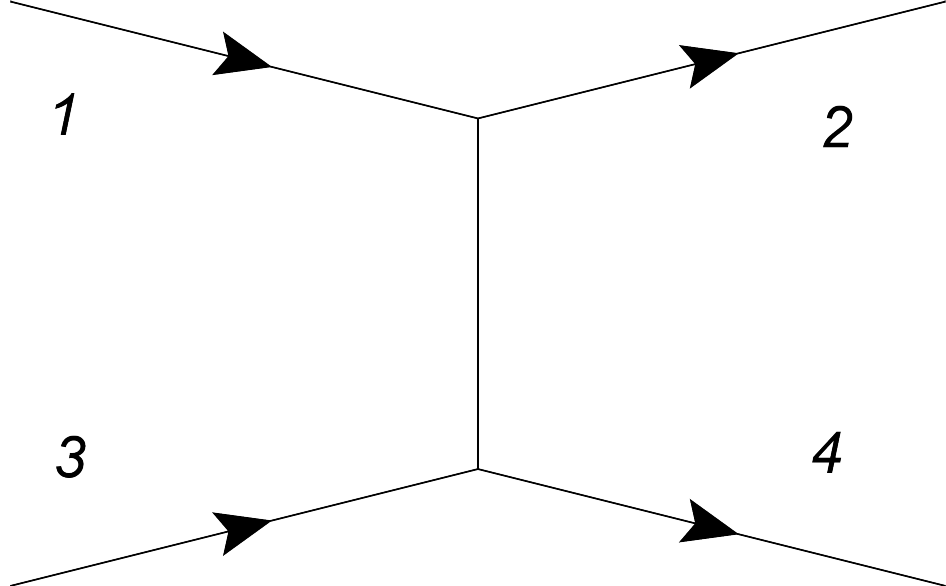}
		\caption{}
	\end{subfigure}
	\hfil
	\begin{subfigure}{0.25\textwidth}
		\includegraphics[width=\textwidth]{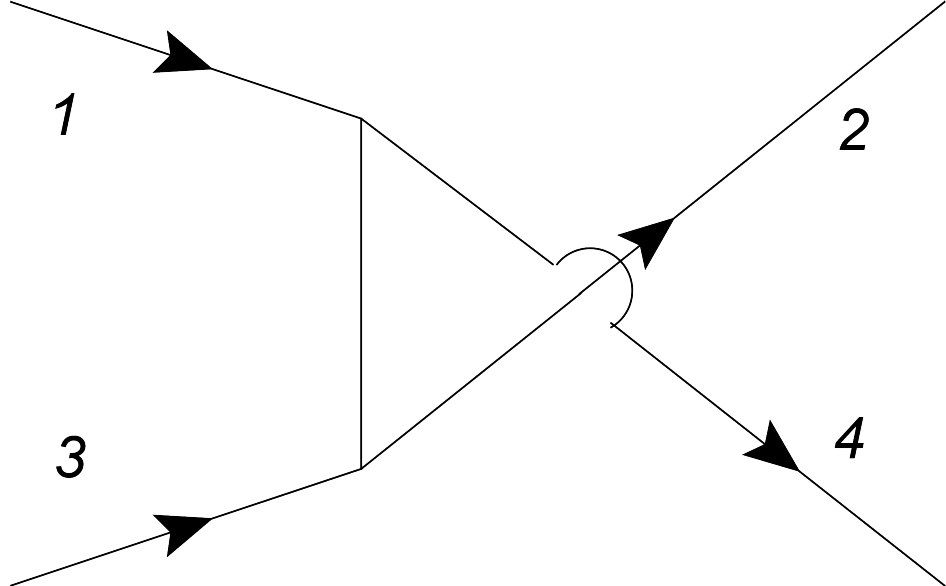}
		\caption{}
	\end{subfigure}
	\hfil
	\begin{subfigure}{0.25\textwidth}
		\includegraphics[width=\textwidth]{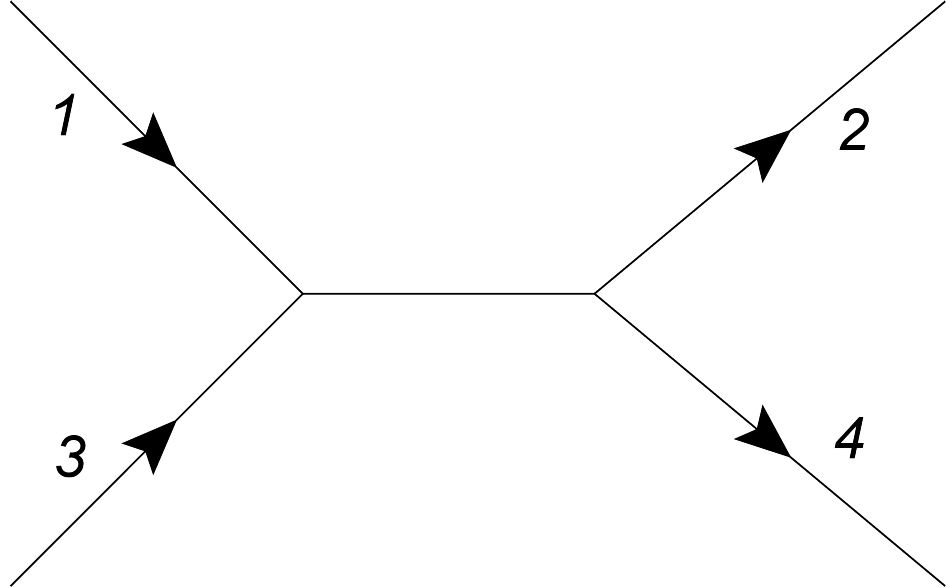}
		\caption{}
	\end{subfigure}
	\caption{ Three alternative descriptions for Minkowski $s$-channel scattering, via OPE connections. (Figure taken from \cite{Raben:2018rbn}.)  They can be associated with  three distinct regions, $M_s^{(t)}$, $M_s^{(u)}$ and $M_s^{(s)}$, in the extended $\sqrt u$-$\sqrt v$ plane, Fig. \ref{fig:crsym22}(a), and also in Fig. \ref{fig:crsym22}(b).  }\label{fig:crsym3}
\end{figure}

\begin{figure}[ht]
	\centering
	\begin{subfigure}{0.4\textwidth}
		\includegraphics[width=\textwidth]{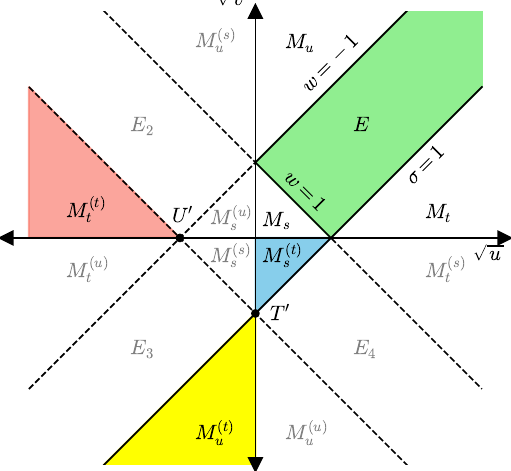}
		\caption{}
	\end{subfigure}
	\hfil
	\begin{subfigure}{0.4\textwidth}
		\includegraphics[width=\textwidth]{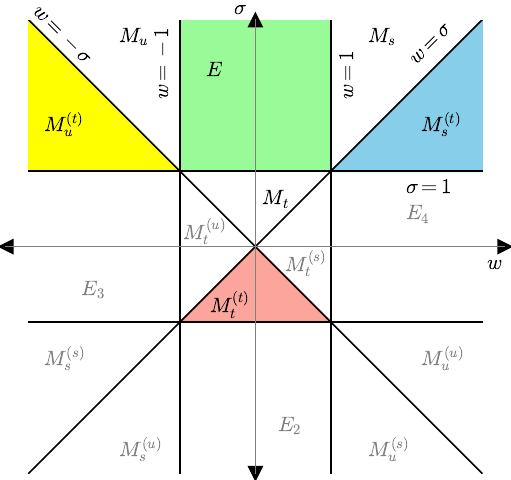}
		\caption{}
	\end{subfigure}
	\caption{ By extending  to  the whole $\sqrt{u}$-$\sqrt{v} $ plane, it  leads to 16 distinct regions in Fig. \ref{fig:crsym2}(a).  In terms of new variables $(w,\sigma)$, the corresponding 16 regions are shown in Fig. \ref{fig:crsym2}(b). Special causal regions $M_s^{(t)}$, $M_u^{(t)}$ $M_t^{(t)}$ are shaded.}\label{fig:crsym22}
\end{figure}

Let us next turn  to the question of  alternative procedure in specifying causal scattering regions. Let us focus on $s$-channel scattering, (\ref{eq:4phi}). As pointed earlier, to reach {\it causal scattering regions}, it is necessary to cross singularities at $\sqrt u=0$ and/or $\sqrt v=0$, (or at $+\infty$).   There can be three equivalent modes of descriptions, associated with three possible Minkowski OPE limits \cite{Raben:2018rbn}. This is schematically illustrated by Fig. \ref{fig:crsym3}. (This corresponds to three equivalent expansions for Minkowski amplitude, $T(x_i)$.)  In terms of the $\sqrt u$-$\sqrt v$ plane, 
they correspond to three distinct regions, $M_s^{(t)}$, $M_s^{(u)}$ and $M_s^{(s)}$, with superscripts associated with the corresponding Minkowski $t$-channel, $u$-channel and $s$-channel OPE's respectively.   (See Fig.  \ref{fig:crsym22}(a). In terms of the $u$-$v$ plane, Fig. \ref{fig:crsym}, they are  connected respectively to regions  $M_t$, $M_u$ and $M_s$ holonomically.)   As stress in Sec. \ref{sec:antipodal}, our choice of antipodal frame singles out the Minkowski $t$-channel description and it leads to our working with the region $M_s^{(t)}$, i.e., that associated with the $t$-channel OPE. This in turn leads to (\ref{eq:scausality}) as our causality condition.

The alternative and equivalent description would be to work with the region $M_s^{(u)}$, in the second quadrant in  Fig. \ref{fig:crsym22}(a), corresponding changing (\ref{eq:scausality}) to 
\be
\langle 0|[\phi(x_2),\phi(x_1)] [\phi(x_4),\phi(x_3)]|0\rangle\neq 0,
\ee
i.e.,  having $x_{12}$ and $x_{34}$ timelike.  Operationally, this is simply $2\leftrightarrow 4$ exchange, Fig  \ref{fig:crsym3}(b).  Both  can be associated with a CFT Regge limit~\cite{Brower:2006ea,Cornalba:2006xm,Cornalba:2006xk,Brower:2007qh,Brower:2007xg,Cornalba:2007zb,Cornalba:2007fs,Cornalba:2008qf,Cornalba:2009ax,Costa:2012cb,Brower:2014wha,Raben:2018rbn}, e.g., with Minkowski $t$-channel and $u$-channel OPE points $T'=(0,-1)$ and $U'=(-1,0)$ labelled in  Fig. \ref{fig:crsym22}(a). Both  choices allow a group theoretic interpretation.  The third option leads to $M_s^{(s)}$, which can best be understood in  a direct-channel picture.  Group theoretically, with $w$ and $\sigma$ bounded in the range $[-1,1]$, it suggests that, in place of the principal series, the discrete series for $SO(d,2)$ is at play. This is a tantilazing possibility worth pursuing further.  

The consideration above  can equally apply to  the $t$- and $u$-channel scatterings, e.g., leading to regions $M_t^{(s)}$, $M_t^{(u)}$, etc. This is consistent with the map from  the first quadrant of the $\sqrt u$-$\sqrt v$ into the whole plane~\footnote{For related discussions from a different perspective on how they are all related holonomically, see \cite{caron2021dispersive,qiao2022classification}.}, resulting in 16 distinct regions, labelled appropriately in Fig. \ref{fig:crsym22}.   We emphasize that we have adopted (\ref{eq:scausality}) and (\ref{eq:ucausality}) in this paper as the causality conditions,  associated with Minkowski $t$-channel OPE,  by working with regions $M_s^{(t)}$ and $M_u^{(t)}$. We postpone to \cite{Agarwal:2024} a more detailed  discussion on the connection to CFT Regge limits, together with their group theoretic interpretations.

Our results in this study develop a framework for discussing CFTs by putting the embedding space and the structure of causal symmetric spaces at its core. Our procedure of coset construction restricts us to a single Poincar\'e patch, avoiding the issues of infinite time windings on the AdS cylinder~\footnote{We have implicitly assumed the existence of these windings so that causal ordering is well defined. In the notation of \cite{Todorov:967138}, $N(\tau)$ is the same for our 4 points.}. Restricting ourselves to representations that trivially depend on the compact part of the Cartan subalgebra (i.e. $ \widetilde{m} = 0 $) made the reduction of variable $ (\zeta,\vec{b}_\perp)\rightarrow\eta $ possible. For applications to DIS or near forward scattering, an explicit dependent on $ \vec{b}_\perp $ is warranted~\footnote{In the short distance limit, it has long been known that BFKL physics \cite{Kotikov:2013xu,kovchegov} involves  aspects of Euclidean $CFT_2$. For a comprehensive review, see \cite{Braun:2003rp}. This has also been done in various related studies such as \cite{Cornalba:2006xk,Cornalba:2006xm,Cornalba:2007fs,Cornalba:2007zb,Cornalba:2008qf,Cornalba:2009ax,Brower:2006ea,Brower:2007qh,Antunes:2021,Shenker:2014cwa,mueller:1995,li1995high,liu2022rapidity} among others. Transverse dependence in $ d = 4 $ comes in via representation functions on $ H_3 = SO(3,1)/SO(3) $. From an $ \text{AdS}_5 $ perspective, the coordinates on $ H_3 $ are $ (\vec b_\perp,\beta) $, where $ \beta $ is the bulk direction and $ \vec{b}_\perp = (x,y) $.}. Furthermore, studying the geometry of AdS in this perspective can also be very fruitful, e.g., the relationship between Legendre functions of first and second kind, (\ref{eq:PQ}), can be understood by relating the geometry of $ H_2 = SO(1,2)/SO(2) $ to the geometry of $ \text{AdS}_2 = SO(1,2)/SO(1,1) $, (Appendix \ref{app:sphr}).

Lastly, let us next turn to some issues on how to relate our work to that where one starts from Euclidean CFT.  We have in mind that of  \cite{caron2017analyticity,carmi2020conformal,mazavc2021basis,caron2021dispersive} and other related works, e.g.,  \cite{Caron-Huot:2020nem}. These  are beyond the scope here, but nevertheless worth noting.    First, it should be pointed out that what we considered as  ``discontinuity", (\ref{eq:reduced}),  in a direct Minkowski treatment, has been identified as a ``double-discontinuity" in \cite{caron2017analyticity}. (See Appendix \ref{app:dDisc}.) Therefore, starting from a Minkowski treatment,  one cannot obtain the Euclidean fourpoint function directly by a one-dimensional dispersion relation, e.g., by that in complex $w$-plane with  $1<\sigma$  fixed. That is, in terms of Figure \ref{fig:crsymb}, a dispersion integral at fixed $\sigma$, over $1<\sigma<|w|<\infty$,  does not directly leads to Euclidean CFT amplitude in region $ E $, nor its natural continuation into regions $M_s$ and $M_u$. It at best can be related to appropriate continuation and/or discontinuity of  the Euclidean fourpoint function.  
Indeed, in Ref. \cite{carmi2020conformal}, an interesting and a relatively involved  analysis was carried out in relating the causal discontinuity, (\ref{eq:reduced}),  to Euclidean fourpoint function in the region $ E $, where $|w|<1<\sigma<\infty$.   It would be  interesting if the procedure of \cite{carmi2020conformal} can be adopted by making use of simultaneous analyticity in $(w,\sigma)$.  Equally interesting is the group theoretic interpretation of the work  of \cite{Caron-Huot:2020nem} where one arrives at a Minkowski inversion formula, equivalent to ours. Closer examination on the relation between these different approaches can be fruitful and some will be addressed further elsewhere.


\paragraph{Note Added:} After the completion of this work, we were made aware of the series of papers by Buri\'c, Isachenkov, and Schomerus \cite{Schomerus2017,Schomerus2018,Isachenkov2018,Buric2020,Buric:2022ucg}. Some of the key constructions presented here are discussed in their work in the context of Euclidean CFTs. In particular, their $ K $-spherical functions are the Euclidean analogues to our Minkowski $ H $-bi-invariant functions, where their subgroup $ K = SO(d)\times SO(1,1) $ is a Wick rotation away from our $ H = SO(d-1,1)\times SO(1,1) $. The reduction of the Casimir equation to a Calogero-Sutherland Hamiltonian evolution can be understood in terms of our $ (y,\eta) $ variables and was briefly discussed in \cite{agarwal2022application}.

The key distinction is the role played by causality in the Lorentzian limit. In the Euclidean setting, the reduced root space is 1-dimensional and therefore, the ``shift",  required in the contour integration of Euclidean analogue of Eq. (\ref{eq:double-mellin}), i.e., (\ref{eq:connection}) for $\Delta$,  is unambiguously determined. In the Minkowski setting, the root space is two dimensional, making this shift ambiguous. It has been known that the result for conformal harmonics in the causal region we are studying is not a direct analytic continuation of the Euclidean result, and requires a ``swap'' in the eigenvalue. We show that causality and semigroups dictate the appropriate ordering of roots discussed in \cite{Isachenkov2018,kravchuk2018light}, which then determines the required shift. 

Another key distinction is the construction of causal  ``conformal blocks".  It is interesting to consider  a generalization of the work of \cite{Faraut:1986}. There, the authors considered an $\text{AdS}_2$ manifold and construct zonal spherical functions in order to diagonalize the absorptive parts of the scattering amplitude. For the $\text{CFT}_1$ case, this was presented in \cite{Raben:2018rbn}. Analogous treatment for CFTs in general $d$ is a challenge because of two problems that present simultaneously: the split-rank 2 nature of the conformal group (studied extensively in mathematical literature \cite{knapp1986,knapp2002lie,Vilenkin:1991,Vilenkin1993,Vilenkin:1992}) and the implementation of causality via semigroups (also studied extensively for the split-rank 1 cases \cite{Viano:1980,Faraut:1986,Hilgert:1996,helgason2022groups}). The work of Schomerus et al. \cite{Schomerus2017,Schomerus2018,Isachenkov2018,Buric2020,Buric:2022ucg}, emphasizing conformal Casimir equation, offers some key insights into the first of these two issues, even as they are mostly restricted to Euclidean CFTs.  In such an approach,  boundary conditions have to be supplied separately.  In our integral approach which is done using inductive characters, causal boundary conditions are imposed automatically via semigroup.  The leading behavior of the amplitude is then a group theoretic result, and not a constraint imposed to solve the problem. These play a central role in understanding Eq. (\ref{eq:double-mellin}), e.g.,  the problem of inversion to determine partial wave amplitudes. This and further discussion in related directions is currently being prepared \cite{Agarwal:2024}.

\paragraph{Acknowledgements:}
The work of RCB is supported by the U.S. Department of Energy (DOE) under Award No. DE-SC0015845. PA acknowledges support from the NUS Research Scholarship for this work. We would like to thank S. Caron-Huot for correspondence and for commenting on a preliminary draft of this paper. We would also like to thank S. Rychkov, I. Buri\'c, M. Isachenkov, and V. Schomerus for their comments. RCB and CIT would also like acknowledge earlier collaboration with J. Polchinski and M. Strassler which paved the way for the current CFT investigations.

\newpage

\appendix
\addtocontents{toc}{\protect\setcounter{tocdepth}{1}}

\section{Summary of Notations}\label{app:notations}
In this appendix, we highlight our notational usage, e.g., clarifying the notion of functions over $SO(d,2)$, $F(g)$, to that over the physical space-time, $f(x_\mu)$  and/or the embedding space, $F(X_\alpha)$. 

\begin{itemize}
\item{Physical spacetme:} $x_\mu$: $\mu=0,1,\cdots, d-1$.  Lorentzian metric with signature: $( -,+,+,\cdots)$. Occasional alternative notations:  $x_\mu=(x_t,x_\perp,x_z)$, and lightcone coordinates: $x_\mu=(x^+,x^-,x_\perp)$ with $x^\pm=x_t\pm x_z$.

\item{Embedding space:}  $X_\alpha$: $\alpha=-2,-1, 0, \cdots, d-1$. Lorentzian metric with signature: $(-,+; -,+,+,\cdots)$.  The standard practice  is to define the physical spacetime coordinates, $x_\mu$, as an appropriate projection (a ``slice") of the null-cone, where $X^2=0$, thus reducing from $(d+2)$ to $d$ degrees of freedom. 

\item{Function over the physical space-time:}  $f(x)$, generically using lower letters.

\item{Function over the embedding space:}  $F(X)$, generically using capital letters, with $X$ on the null-cone. 

\item{Lifting from  physical spacetime to embedding space:}  A  function over the Minkowski space-time, $f(x)$, can often be ``lifted" as a function over the embedding space, $F(X)$, with $X$ on the null-cone.  When the context is clear, we often will use $F(X)$ and $f(x)$ interchangeably, e.g., the left-hand side of (\ref{eq:double-mellin}).

\item{Functions over $G=SO(d,2)$:} For each function over the null-cone, $F(X)$, we can introduce a function over $G$ as follows.  There are several equivalent ways of doing this, similar to the choice of ``active" versus ``passive" view in transformation for ordinary vectors. For our purpose, we shall adopt $X= g\cdot \xi_0$, with $\xi_0$ a fixed {\it base-point} on the null-cone. Since $G$ acts transitively, $F(X)$ can be treated as  a function  over $G$, i.e., $F_g(\xi)=F(g^{-1}\cdot\xi)$.  We will often adopt the same notation $F(g)=F_g$ as the group function, with $ \xi_0$ implicit.

\item{Group Representation by ``Shifting":}  Given $\xi_0 = g^{-1}\cdot X$, consider another mapping $g_1$. Define
\be
T(g_1) F_g = F_{g'}\leftrightarrow F_{g'}( \xi)
\ee
where $g'=g_1g$ and $F_{g'}( \xi)$ can also be expressed  as $F_g( \xi')$, with $\xi'= g_1^{-1}\cdot \xi$.
This leads to a representation by ``left-shift" where $T(g_1)T(g) = T(g_1g)$.

\item{Left-coset:}  Given a subgroup $H$, the ``left coset" is defined by an equivalence relation where for some $g_1,g_2\in G$, we have $g_1\cong g_2H$. This is often denoted by the notation $G/H$.
A corresponding right-coset $H\backslash G$ can also be defined similarly.

\item{Functions of several variables:}
The above constructions generalize to the case of functions of several variables, i.e., $F(X_1,X_2,\cdots )\rightarrow F(g_1, g_2, \cdots)$. In order  to avoid confusion with components $X_\alpha$ for individual embedding coordinate, we will  often change notation for embedding variables from $X_i$ to $\xi_i$, $i=1,2,\cdots.$ We also allow the option of having a different base-point for each $g_i$, as done  in Sec. \ref{sec:4point}.

\item{Lie Groups and Associated Algebra:}  Will generically  adopt capital letters, e.g., $G$, $K$, $Q^\pm$, $\cdots$ for groups, subgroups, etc.. Their corresponding algebra by capital script letters,  ${\cal G}$, ${\cal K}$, ${\cal Q}^\pm$,$\cdots$, e.g., generators for $G=SO(d,2)$ are ${\cal L}_{A,B}$, $A,B=-2,-1,0,\cdots ,d-1$.  This is not followed religiously for various reasons: special considerations, historical, etc., (see below.)

\item{Basics for $SO(d,2)$:} 
There are $(d+2)(d+1)/2$ independent generators for $SO(d,2)$,  the same as for $SO(d+1,1)$, with ${\cal L}_{AB}=-{\cal L}_{BA}$.  Commutation relations for these generators   are standard, which can be expressed compactly as, 
\be
[L_{AB},L_{CD}] =  \eta_{AC} L_{BD} +\eta_{BD} L_{AC} -\eta_{AD} L_{BC} -\eta_{BC} L_{AD}, 
\ee
with signature $\eta$. Considered as a diagonal matrix, we have ${\rm diag}\, \eta=(-,+; -, +,+,+,\cdots)$.  We follow the convention where  the subgroup generated by ${\cal L}_{AB}$ is expressed as  $g(\xi)=e^ {\xi {\cal L}_{AB}}$, with  $\xi$ real, i.e., the convention typically adopted  in mathematical texts.  These commutation relations follow from the fact that $SO(d,2)$ leaves the quadratic form, 
$
X^2=X^T \eta X =-X_{-2}^2 +X_{-1}^2-X_0^2+ X_1^2 + X_2^2+\cdots+X_{d-1}^2$
 invariant. In adjoint representation, matrix elements for each  generator are real, ${\cal L}_{AB}=-{\cal L}_{BA}$, with \be
\left ({\cal L}_{AB}\right)_{CD} =\eta_{AD}\delta_{CB} - \eta_{BD}\delta_{CA}.  \label{eq:ME4generators}
\ee

\item{Compact and Non-Compact Subgroups:}  Generators in adjoint representation, (\ref{eq:ME4generators}), are real. The group contains compact subgroups, with generators as  {\it  anti-symmetric} matrices, $\left ({\cal L}_{AB}\right)_{CD}=-\left ({\cal L}_{AB}\right)_{DC}$.  For non-compact generators,   their matrix elements  are real and {\it  symmetric}, i.e., $\left ({\cal L}_{AB}\right)_{CD}=\left ({\cal L}_{AB}\right)_{DC}$. 

\item{Rotations and Boosts:} In the usual physics usage, compact generators can be characterized as ``rotations" and non-compact generators as``boosts".  We will often adopt the convention where compact generators (rotations) are indicated by ${\cal K}=\left( {\cal R}_{\alpha,\beta}\right)$,  and non-compact generators (boosts) are indicated by ${\cal P}=\left ({\cal B}_{\alpha,\beta} \right)$. Although for the most part we will not need these explicit forms,  
it is nevertheless important to discuss some structural aspects. Illustrative examples for $d=1,2$, etc. will be provided below in Appendix \ref{app:Iwasawa}.

\item{Special Designations for $SO(d,2)$:}  The generator for dilatation, ${\cal L}_{-1,-2}=-{\cal L}_{-2,-1}$ and the generator for Lorentz boost along the ``longitudinal" direction, ${\cal L}_{d-1,0}=  -{\cal L}_{0, d-1}$, are designated as  $D$ and $L$  respectively. Some other special designations are also provided in Appendix \ref{app:root-2}.

\end{itemize}

\section{Kinematics of  Minkowski 4-Point CFT Amplitudes}\label{app:MinkowskiCFT}
This appendix serves several  purposes. In \ref{app:B1}, we summarize relations among various sets of conventional variables, with emphasis on continuation beyond the Euclidean region, E. (Table-\ref{tab:vars}(a).)   In \ref{app:B2}, we discuss the relation between our  group theoretic motivated variables, $(w,\sigma)$ -- expressed in terms of group parameters $(y,\eta)$) -- with the standard sets of variables. (Table-\ref{tab:vars}(b).) In \ref{app:B3}, we provide a more qualitative description for the separation of scattering vs non-scattering regions in a Minkowski light-cone diagram. In \ref{app:dDisc}, a brief description of  the ``double discontinuity",  ${\rm dDisc} G(\rho,\bar\rho)$, of Ref. \cite{caron2017analyticity} and how it relates to ${\rm Im} T(w,\sigma)$ adopted in this work is provided.  

\subsection{Cross Ratios $(u,v)$ and Alternative Representations}\label{app:B1}

In place of $(u,v)$, or, equivalently, the pair $(\sqrt u,\sqrt v)$,  several  equivalent sets of independent variables can be introduced, e.g.,  $(z,\bar z)$, $(\rho,\bar \rho)$ and $(q,\bar q)$. In terms of the pair $(z,\bar z)$, their relationships are summarized in  Table-\ref{tab:vars}a. 

\begin{table}[ht]
	\renewcommand{\arraystretch}{1.5} 
	\begin{subtable}{0.5\textwidth}
		\begin{tabular}{ |c|c|c| } 
			\hline
			$ (f_1,f_2) $ & $ f_1(z,\bar z)$ & $ f_2(z,\bar z) $\\	\hline \hline
			$(u,v)$ & $z \,{\bar z}$ & $ {(1-z)}{(1-\bar z)} $\\ \hline
			$(\sqrt u,\sqrt v)$ & $\sqrt z \sqrt {\bar z}$ & $ \sqrt {(1-z)}\sqrt{(1-\bar z)} $\\ \hline
			$(\rho,\bar \rho)$ & \Large{ $\frac{z}{(1+\sqrt {1-z})^2} $}& \Large{$\frac{\bar z}{(1+\sqrt {1-\bar z})^2}$}\\[2pt] \hline
			$(\sqrt \rho,\sqrt{\bar \rho})$ & \Large{ $\frac{\sqrt z}{(1+\sqrt {1-z})} $}& \Large{$\frac{\sqrt{\bar z}}{(1+\sqrt {1-\bar z})}$}\\[2pt] \hline
			$(q,\bar{q})$ & $\frac{2-z}{z} $ & $ \frac{2-\bar z}{\bar z} $\\ \hline
		\end{tabular}
		\caption{}
	\end{subtable}
	\hfil
	\begin{subtable}{0.5\textwidth}
		\begin{tabular}{ |c|c|c| } 
			\hline
			$ (g_1,g_2) $ & $ g_1(y,\eta)$ & $ g_2(y,\eta) $\\
			\hline \hline
			$(\sqrt \rho,\sqrt{\bar{\rho}})$ & $ e^{-(\eta\pm y)/2} $ & $ e^{-(\eta\mp y)/2} $\\ \hline
			$(\sqrt u,\sqrt v)$ & \Large{$\frac{2}{(\cosh\eta+\cosh y)} $} &  \Large{$\frac{(\cosh \eta-\cosh y)}{(\cosh\eta+\cosh y)} $}\\[2pt] \hline
			$(\sqrt {\rho\bar \rho},\sqrt {\rho/\bar \rho})$ & $e^{-\eta}$ & $e^{-y}$\\ \hline
			$(q,\bar{q})$ & $ \cosh(\eta\pm y) $ & $ \cosh(\eta\mp y) $\\ \hline
			$(z,\bar{z})$ & $ \operatorname{sech}^2((\eta\pm y)/2) $ & $ \operatorname{sech}^2((\eta\mp y)/2) $\\ \hline
		\end{tabular}
		\caption{}
	\end{subtable}
	\caption{Relationship among variables commonly used in CFT studies. Option for signs in (b) are fixed by convention. With $\rho<\bar\rho$, we adopt the upper sign, e.g., (\ref{eq:rhotr}).}
	\label{tab:vars}
\end{table}
\paragraph{Euclidean Region $E$ and OPE Limits:}
By expressing $(z,\bar z)$ in terms of $(u,v)$ while staying within the Euclidean region $E$,  one finds that  $\bar z=z^*$ and $0<|z| <\infty$. It also follows that  $\bar \rho=\rho^*$ and $\bar q= q^*$. In particular,   in the Euclidean region, one finds 
\be
\rho=\bar \rho^*=r e^{i\theta},  \quad 0<r<1, \quad -\pi<\theta<\pi. \label{eq:rhoE}
\ee
Under the convention adopted in our analysis, the  $t$-channel  OPE limit, i.e., approaching point $T$ in Fig. \ref{fig:crsym}, associated with the partition $(1,2)(3,4)$,  corresponds to $x_{12}^2, x_{34}^2\rightarrow 0$, leading to $u\rightarrow 0$, $ v\rightarrow 1$, or equivalently,  the limit $(z,\bar z)\rightarrow (0,0)$.  For the Euclidean $s$-channel OPE, i.e.,  approaching point $S$, associated with the partition $(1,3)(2,4)$,  
one has  $x_{13}^2, x_{24}^2\rightarrow 0$,  corresponding to $u\rightarrow +\infty$, $ v\rightarrow +\infty$, or  $(z,\bar z)\rightarrow (\infty,\infty)$, and for Euclidean $u$-channel OPE, approaching point $U$, associated with the partition $(14)(2,3)$, with $x_{14}^2, x_{23}^2\rightarrow 0$,  $u\rightarrow 1$, $ v\rightarrow 0$, or  $(z,\bar z)\rightarrow (1,1)$.

The relations between $(z,\bar z)$ and $(\rho, \bar \rho)$ can also be expressed as
\be
\rho=\frac{1-\sqrt {1-z}}{1+\sqrt {1-z}}, \quad {\rm and}\quad  \bar \rho=\frac{1-\sqrt {1-\bar z}}{1+\sqrt {1-\bar z}}
\ee
One can  avoid singularities at $z=0$ and $z=1$ by drawing branch cuts along $(-\infty,0)$ and $(1,\infty)$. The ``first-sheet" for the complex $z$-plane  corresponds to the unit disk in $\rho$~\footnote{Although this is  well-understood, it nevertheless is worth stating here that Euclidean OPE points, $T$, $S$ and $U$,  should be approached by staying within the first-sheet.}, Eq. (\ref{eq:rhoE}).

\paragraph{Lorentzian Extension and Beyond:} In moving outside of  the region $ E $, e.g., into regions $M_s$, $M_u$ and $M_t$, variables $(z,\bar z)$ are to be treated as independent variables. Access beyond the first quadrant in the $\sqrt u$-$\sqrt v$ plane requires crossing $\sqrt u=0$ and/or $\sqrt v=0$.  This is also equivalent to encircling singular points in $(z,\bar z)$ at $(0,0)$ and/or $(1,1)$ appropriately. That is, with $(u,v)$ as functions of $(z,\bar z)$, it leads to a multiple-sheeted structure.   (See Refs. \cite{Cornalba:2006xm,Cornalba:2007fs,Cornalba:2008qf,Cornalba:2009ax,Costa:2012cb,caron2021dispersive,qiao2022classification}). 

This structure can be resolved  by considering 
$\sqrt \rho$ and $\sqrt {\bar \rho}$.
By inversion,  one finds
$
z(\rho) = z(\rho^{-1})$ and  $\bar z(\bar \rho) = \bar z({\bar \rho}^{-1})
$. The mapping  is 1-to-2, dividing  the $\rho$-plane into two regions, $|\rho|<1$ and $|\rho|>1$, reflecting again a two-sheeted structure.
As functions of $\sqrt \rho$ and $\sqrt {\bar \rho}$, we have
\bea
\sqrt z = {2  }{(\sqrt \rho^{-1}+\sqrt \rho)^{-1}}, &\quad {\rm and} \quad   &\sqrt {\bar z} ={2  }{(\sqrt {\bar \rho}^{-1}+\sqrt {\bar \rho})^{-1}},
\nn
\frac{1-\sqrt v}{\sqrt u }= \frac{1}{2} \Big(\sqrt  \frac{\rho}{\bar \rho}+ \sqrt  \frac{\bar \rho}{ \rho}\Big),   &\quad {\rm and} \quad   &
\frac{1+\sqrt v}{\sqrt u} =   \frac{1}{2}  \Big(\sqrt  {{\rho}{\bar \rho}} + \frac{1}{\sqrt  {{\rho}{\bar \rho}} }\Big).
\eea
This also allows an extension, though less obvious,   into the whole $\sqrt u$-$\sqrt v$ plane.

The pair of variables $q$ and $\bar q$ also leads to kinematic simplification (Ref. \cite{Raben:2018rbn}). Here we list just one useful relations
$q=(\rho+\rho^{-1})/2 =   (\sqrt \rho+{\sqrt \rho}^{-1})^2/2-1$ and  $\bar q=(\bar \rho+\bar\rho^{-1})/2=   (\sqrt {\bar \rho}+{\sqrt{\bar  \rho}}^{-1})^2/2-1.$

\subsection{Group Theoretic Motivated Variables and Antipodal Frames}\label{app:B2}

We next turn to a discussion of {\bf antipodal frames}, thus providing a more  intuitive understanding for invariants $(w,\sigma)$, Eq. (\ref{eq:newvariables}),
\be
\sqrt u = \frac {2 }{\sigma+w}, \quad \sqrt v = \frac {\sigma-w}{\sigma+w}  \quad\quad \Leftrightarrow \quad\quad  w\equiv \frac{1-\sqrt v}{\sqrt u}, \quad \sigma \equiv  \frac{1+\sqrt v}{\sqrt u},
\ee
equivalently, in terms of $(\rho,\bar \rho)$,
\be
w \equiv \cosh y = \frac{1}{2}\Big(\sqrt  \frac{\rho}{\bar \rho}+ \sqrt  \frac{\bar \rho}{ \rho}\Big),   \quad {\rm and} \quad \sigma\equiv \cosh \eta=  \frac{1}{2}\Big(\sqrt  {{\rho}{\bar \rho}} + \frac{1}{\sqrt  {{\rho}{\bar \rho}} }\Big). \label{eq:groupparameters}
\ee
\paragraph{Euclidean  Antipodal Frame:}
For the Euclidean region $E$,  the symmetry group is $ SO(d+1,1) $.  
Consider  the frame where $ x_1 = -x_2 $ and $ x_4 = -x_3 $. 
Each pair corresponds to a diameter within  a d-sphere, $S^{d}$, with varying radius.  Let the relative angle between these two diameters be $\theta$, e.g., $\cos \theta = \hat x_1 \cdot \hat x_4$. One can also treat the radius for each sphere as a new variable, and denote  their ratio by $e^{\tau}$.
These configurations  can be visualized as a {\bf cylinder} of length $\tau$, with two  unit spheres as ends.  This will be referred to as the {\bf Euclidean antipodal frame}~\cite{Hogervorst:2013sma}. Group theoretically, $\tau$ can be identified with a non-compact boost in $SO(d+1,1)$, leading to {\bf radial-quantization}, and $\theta$ with a compact rotation angle. (Eq. (\ref{eq:rhoE}) provides an alternative definition.) 
In this frame,  $ \sqrt u = 2/(\cosh \tau+\cos \theta)$ and $\sqrt v =(\cosh \tau-\cos \theta)/(\cosh \tau+ \cos \theta),$
i.e., $\sigma=\cosh \tau$ and $w=\cos \theta=\hat x_1\cdot \hat x_4$ (defined shortly via 
a Wick rotation), as done in ~\cite{brower2021radial}. 
Thus, in  region $E$,  
$
-1\leq  w \leq 1$ and $1<\sigma\, <\infty\, 
$, Eq. (\ref{eq:regionE}).

\paragraph{Minkowski Antipodal Frame:}

For the kinematics of scattering in the Minkowski limit, (\ref{eq:4phi}),  we can  again adopt a special frame where
the time-components of $x_i$  are ordered oppositely,   $- x^{(t)}_1=x^{(t)}_2>0$ and $-x^{(t)}_3=   x^{(t)}_4>0\, .$ 
To specify their spatial components, we identify a longitudinal direction, e.g., $ z $-axis, and denote perpendicular components by $x_{\perp,i}$.   This naturally lends to a {\bf lightcone} description. (See Fig. \ref{fig:DLC-limit}, for a schematic representation.)

To be definite, we parametrize the time and longitudinal coordinates explicitly, with  $x_1=-x_2$ and $x_3=-x_4$, leading to  light-cone coordinates, $x^\pm= x_t\pm x_z$,
\be
x_1^{\pm}= - x_2^{\pm} = \mp r_1e^{\pm y/2}\quad {\rm  and} \quad  x_3^{\pm}= - x_4^{\pm} = \pm r_3 e^{\mp y/2},  \label{eq:Santipodal}
\ee
or, $ (x^{(t)}_1, x^{(z)}_1)=-(x^{(t)}_2, x^{(z)}_2)=(- r_1\sinh y/2, -r_1 \cosh y/2)$,  and $ (x^{(t)}_3, x^{(z)}_3)=-(x^{(t)}_4, x^{(z)}_4)=(- r_3\sinh y/2, r_3 \cosh y/2)$. 
For transverse components, we choose, for simplicity,
\be
x_{\perp, 1}=x_{\perp, 2}=b_\perp,\quad{\rm and} \quad x_{\perp, 1}=x_{\perp, 2}=0. \label{eq:Santipodalperp}
\ee
We shall refer to this as {\bf Minkowski Antipodal Frame}. (In \cite{Raben:2018rbn},  this  is also referred to as the {\bf double-lightcone} frame.)
In this frame, evaluating $u,v$ using (\ref{eq:Santipodal}) and (\ref{eq:Santipodalperp}), cross ratios  can again be expressed as in Eq. (\ref{eq:newvariables}) and Eq. (\ref{eq:sigma}), i.e.,
\be
1<w\equiv  \cosh y <\infty, \quad{\rm and} \quad 1<\sigma \equiv  \cosh \eta = \cosh \zeta + b_\perp^2<\infty. \nn
\ee
Here, $y$  parametrizes a Lorentz  boost, $\zeta$ parametrizes a scale transformation, $\cosh \zeta=({r_1/r_3}+{r_3/r_1})/2$, and $b_\perp$ is the transverse separation between $(12)$ and $(34)$. (We have normalized $2r_1r_3=1$ with $b_\perp$ dimensionless.  See   Ref. \cite{Raben:2018rbn} for more details.)
A similar analysis for the $u$-channel configurations, with $w\rightarrow -w$.

\paragraph{Relationship  between  Standard CFT Variables and Group Variables $(w,\sigma)$:}
With $w=\cosh y$ and $\sigma=\cosh \eta$, 
we can express other invariants commonly used in CFT studies as functions of group parameters ($y,\eta$). This is summarized in Table-\ref{tab:vars}b.

As a consistency check, as stated above, by starting with Minkowski antipodal frame, one can directly verify that
$ w=\cosh y= (1-\sqrt v)/\sqrt{u}    \quad{\rm and} \quad 
\sigma=\cosh \eta =(1+\sqrt v)/\sqrt{u} 
$
i.e., Eq. (\ref{eq:newvariables}),
with $\cosh \eta = \cosh \zeta + b_\perp^2 $.
Note that this corresponds to a reduction to two invariants, i.e., either  $(w,\sigma)$ or one of the set from $(\sqrt u,\sqrt v)$, $(z,\bar z)$, $(\rho,\bar{\rho})$ or $(q,\bar q)$. For $d=2,4$, the choice $(\rho,\bar{\rho})$ or $(q,\bar q)$ can also be understood group theoretically, in terms of the concept of Weyl chambers.

\paragraph{Wick Rotation:} Consider transition between regions $E$ and $M_s$. Starting in $M_s$, we adopt the convention, for group parameters $y$ and $\eta$,   $\rho=e^{-(\eta+y)}$ and $\bar \rho=e^{-(\eta-y)}$, with $0<y<\eta$. (This also fixes the $\pm$ sign convention  in Table-\ref{tab:vars}b.) For later convenience,  let 
\be
t= \sqrt \rho/\sqrt {\bar \rho}=e^{-y}, \quad r= \sqrt \rho \sqrt {\bar \rho}=e^{-\eta}. \label{eq:rhotr}
\ee 
Performing next a Wick rotation where $y\rightarrow i \theta$ and also re-express $\eta\rightarrow \tau$, one is led back to region $E$, with
$
\rho=\bar\rho^*= e^{\textstyle -\tau+i\theta} 
$, (\ref{eq:rhoE}), 
$
\sqrt u=2/(\sigma+w)=2/(\cosh\tau+\cos \theta)$ and  $  \sqrt v= (\sigma-w)/(\sigma+w)=(\cosh \tau-\cos \theta)/(\cosh\tau+\cos \theta) $, as expected. 

\subsection{Minkowski  CFT and Light-cone Diagram} \label{app:B3}

The Minkowski kinematics have often been  represented schematically by the use of {\bf light-cone diagrams}~\cite{Brower:2006ea,Brower:2007qh,Brower:2007xg,Cornalba:2006xm,Cornalba:2007fs,Cornalba:2008qf,Cornalba:2009ax,Costa:2012cb,Brower:2014wha,Raben:2018rbn,simmons2018spacetime,kravchuk2018light}. Relative to any coordinate origin,  the limit indicated in Fig. \ref{fig:DLC-limit} covers the $s$-channel near-forward scattering, with $x_{14}$ and $x_{23}$ timelike.  However, it can be shown that the diagram also involves {\it non-causal (Minkowski) scattering regions}, i.e., $x_{14}$ and $x_{23}$ spacelike. (That is, region $M_s$ in Fig. \ref{fig:crsym}. See also Fig. 3 in Ref. \cite{simmons2018spacetime}.)  

\begin{figure}[ht]
	\centering
	\begin{subfigure}{0.35\textwidth}
		\includegraphics[width=\textwidth]{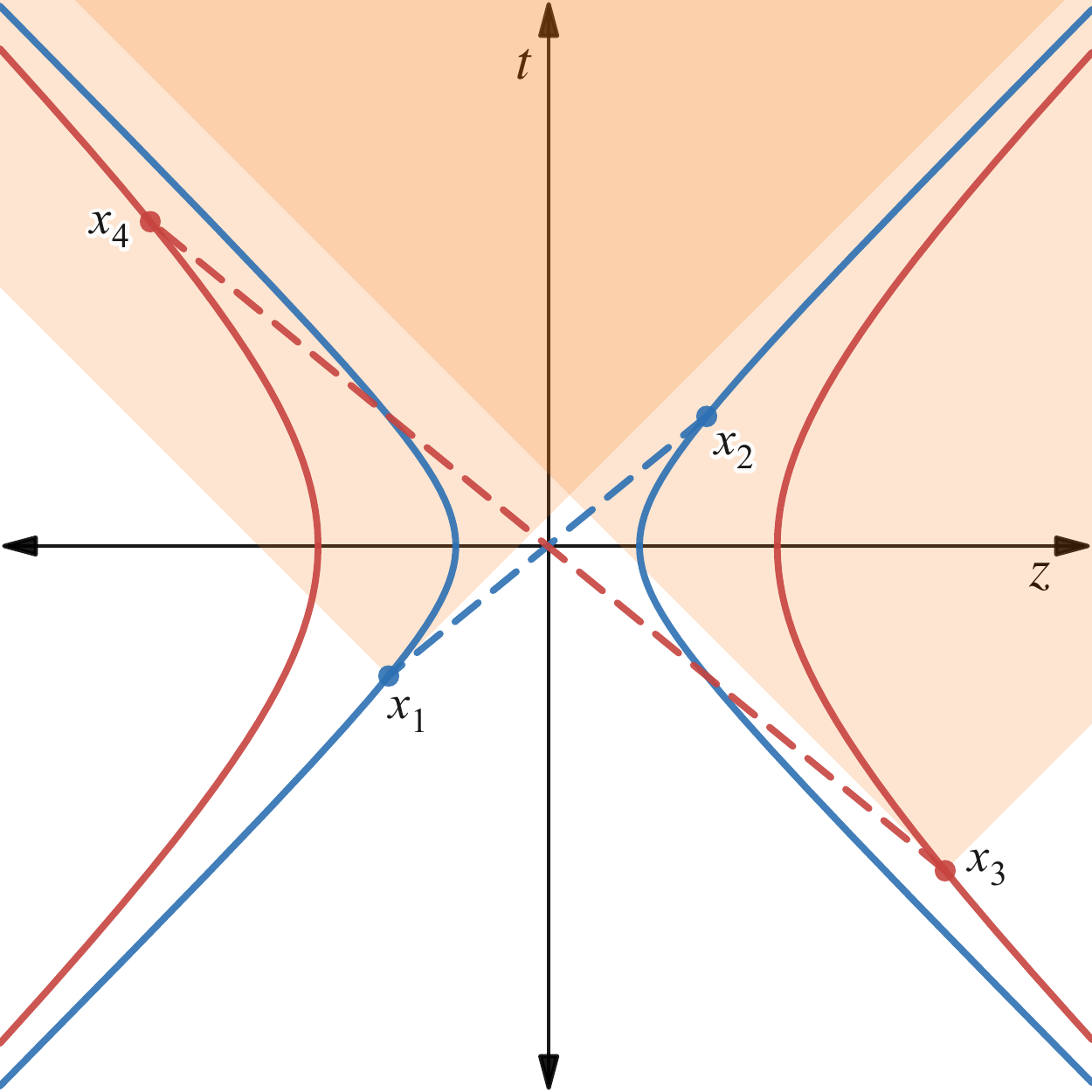}
		\caption{}
	\end{subfigure}
	\hfil
	\begin{subfigure}{0.35\textwidth}
		\includegraphics[width=\textwidth]{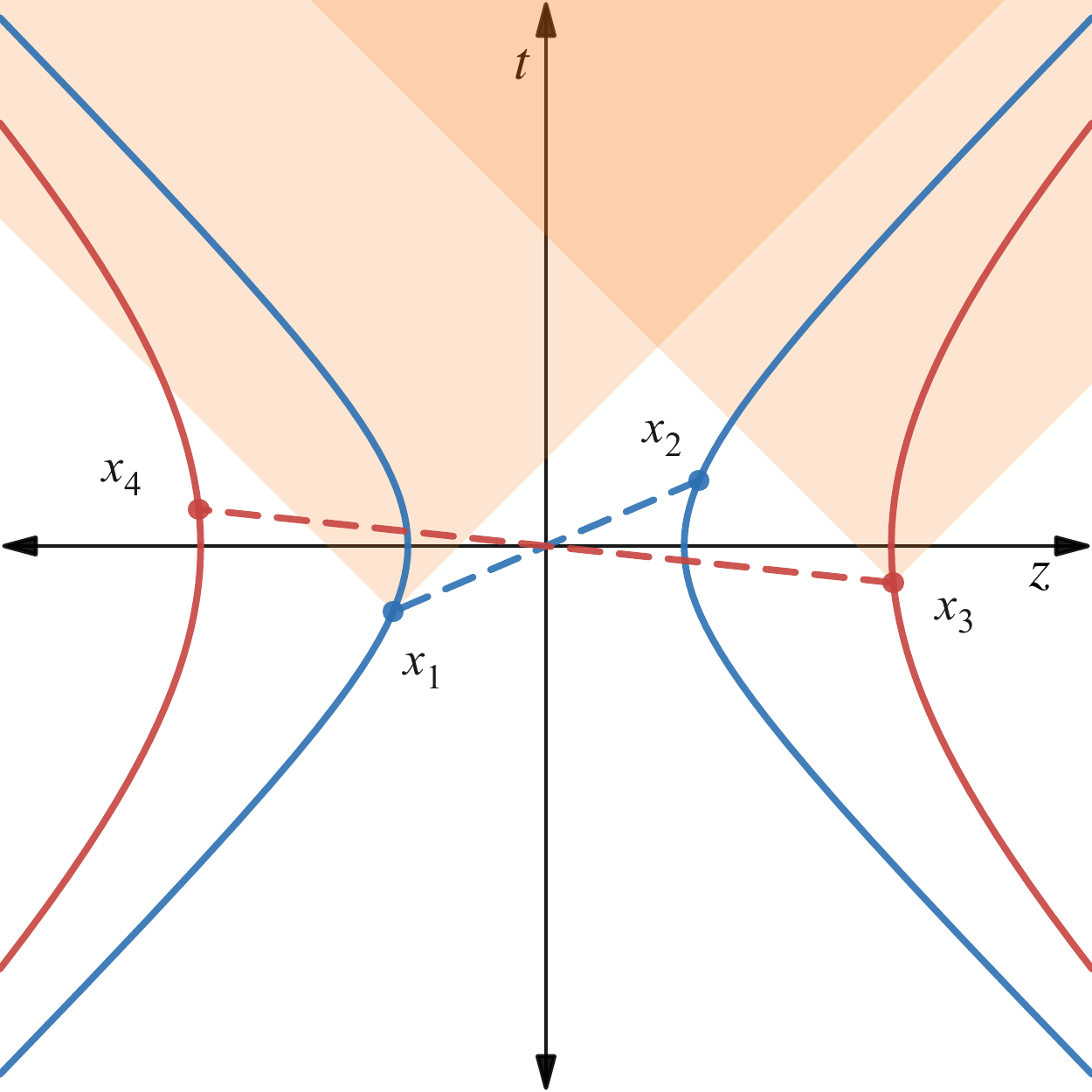}
		\caption{}
	\end{subfigure}
	\caption{Antipodal frame in the  Minkowski causal-scattering and non-scattering settings respectively.}
	\label{fig:antipodal}
\end{figure}

A precise identification of these regions can be made either through the $\sqrt u$-$\sqrt v$ plane or through our new set of variables, $(w,\sigma)$,  Eq. (\ref{eq:newvariables}).  In an antipodal frame, Eq. (\ref{eq:Santipodal}), one easily verifies the causal condition, $x_{14}^2= x_{23}^2 =  (\sigma -w)$ and $0<\eta<y<\infty$, Eqs. (\ref{eq:condition}) and (\ref{eq:semi}). Here, we provide a qualitative evidence for this fact. 

Fig. \ref{fig:antipodal}a-b showcase a  scattering and non-scattering setting respectively for $d=2$. In Fig. \ref{fig:antipodal}b, lightcones at $ x_1$ and $x_3 $ are shaded in orange to demonstrate that in the non-scattering case, all points are spacelike separated, with $x_2$ and $x_4$ outside of the respective lightcone.  For the scattering case, Fig. \ref{fig:antipodal}a, $( x_1,x_4 )$ and $( x_3,x_2 )$ are timelike separated. 

It is possible to  include transverse components schematically by treating the case of $d=3$, e.g., Fig. 4 of Ref. \cite{simmons2018spacetime}, with Fig. \ref{fig:antipodal} as the 2-dim projection onto the $t$-$z$ plane. Going from a scattering configuration to non-scattering requires increasing $ \sigma $  or lowering $ w $ thus broadens the difference in curvature of the two hyperbolae or brings the points in closer respectively. One moves from one region to the other by crossing the line $w=\sigma$. For $u$-channel, one interchanges $x_3$ and $x_4$.

\subsection{Double Discontinuity,  ${\rm dDisc} \,G(\rho,\bar\rho)$, and ${\rm Im}\, T(w,\sigma)$}\label{app:dDisc}
In Ref. \cite{caron2017analyticity},  Minkowski discontinuity, ${\rm Im}\,T$, Eq. (\ref{eq:double-mellin}), has been related to analytically continued Euclidean fourpoint function $G$ on the ``second sheet" through the notion of ``double-discontinuity", denoted by {dDisc} $G$. Although we work exclusion with Minkowski CFTs in this study, it is nevertheless useful to  spell out this connection thus allowing to clarify further  our group theoretically motivated variables $(w,\sigma)$ in this context.  It is convenient to first make use of $(\rho, \bar \rho)$ before turning to $(w,\sigma)$ through the use  $(y,\eta)$.

Given $G(\rho, \bar \rho)$ in  the Euclidean region, $\bar \rho=\rho^*$, we first extend into Region $M_s$ where we adopt the convention $\rho=e^{-(\eta+y)}$ and $\bar \rho=e^{-(\eta-y)}$, with $0<y<\eta<\infty$, i.e., $0<\rho<\bar \rho<1$.  Continuation into $s$-channel scattering region corresponds to switching to $0<\eta<y<\infty$, leading to $0<\rho<1<\bar \rho<\infty$. This corresponds to crossing the line segment $L_s$, i.e., $1<\sigma=w$. Whereas $G$ is real in the region $M_s$, it turns complex into the scattering region, one must distinguish values below or above the cut starting at $\bar\rho=1$.

Let's try to be more precise. Let $ G(\rho, \bar \rho, \pm i\epsilon)$ be two alternative continuations, from $0<\bar \rho<1$ to $1<\bar \rho<\infty$, with  $0<\rho<1$. In analogy with usual $S$-matrix,
$S(\pm i \epsilon)  = {\bf I}  \pm  i {T}(\pm i \epsilon)$,
one defines Minkowski scattering amplitude ${T}(\pm i \epsilon)$ through
\be
G(\rho, \bar \rho, \pm i\epsilon) = {G}_{E} \pm i {T}(\rho, \bar \rho, \pm i \epsilon)
\ee
The first factor $G_E$ is evaluated at $(\rho, 1/\bar\rho)$ where it is real.
${T}(\rho, \bar \rho, \pm i \epsilon)$ over the cut can be expressed in terms of its real and imaginary parts,
\be
{T}(\rho, \bar \rho, \pm i \epsilon) = {\cal R} \pm i {\cal I}.
\ee
It follows that 
\be
{\cal I }= {G}_{E}  - \frac{1}{2} [ G(\rho, \bar \rho, + i\epsilon)+G(\rho, \bar \rho, - i\epsilon)]
\ee
and
\be
{\cal R }=  \frac{1}{2i} [ G(\rho, \bar \rho, + i\epsilon)-G(\rho, \bar \rho, - i\epsilon)]
\ee
In \cite{caron2017analyticity}, $  {\cal I }(\rho,\bar \rho)$ is referred to as a double discontinuity, designated as ${\rm dDisc}\, G$. (In \cite{caron2017analyticity}, one makes one additional notational change by replacing $\bar \rho$ by $\bar \rho^{-1}$ for $G(\rho, \bar \rho, \pm i\epsilon)$. We prefer not to do so to emphasize that causal discontinuity is evaluated in the region $0<\rho<1<\bar \rho<\infty$.)  One can next express $(\rho,\bar \rho)$ in terms of $(w,\sigma)$, through $(y,\eta)$, e.g., (\ref{eq:groupparameters}) and (\ref{eq:rhotr}). When ${\rm dDisc }\,G(\rho,\bar \rho)$ is expressed in terms of $(w,\sigma)$, it leads  to ${\rm  Im }\,T(w,\sigma)$ in this study.

\subsection{Illustrations for $ \text{dDisc} $ from a Spacetime Perspective}

\paragraph{Ising Model in $ 1+1 $ dimensions:} Consider the fourpoint correlator $ \braket{\sigma\sigma\sigma\sigma} $. This is given upto a normalization by:
\begin{equation}
	{\cal G}_E = \left|\frac{1}{(1-\rho^2)^{1/4}}\right|^2+\left|\frac{\sqrt{\rho}}{(1-\rho^2)^{1/4}}\right|^2 = \left(\frac{1+\sqrt{u}+\sqrt{v}}{2(\sqrt{v})^{1/2}}\right)^{1/2}
\end{equation}
for Euclidean CFTs where $ \bar\rho = \rho^* $. In going to the Minkowski space, we make $ \rho,\bar\rho $ real and independent such that both are bounded at 1. In taking the double discontinuity in the $ s $-channel scattering region step by step, the $ (u,v) $ variables prove to be more useful.

Consider first the points where $ x_{14} $ become timelike separated. This takes $ v\rightarrow -v $ and we get:
\begin{equation}
	{\cal G} = \left(\frac{1+\sqrt{u}\pm i|\sqrt{v}|}{2(\pm i|\sqrt{v}|)^{1/2}}\right)^{1/2},
\end{equation}
where $ |\sqrt{v}| $ is positive in the region where a single commutator is non-vanishing. Therefore,
\begin{align}
	i\ \text{Disc}_{14}{\cal G} &\equiv  {\cal G}(+i|\sqrt{v}|) - {\cal G}(-i|\sqrt{v}|)\nonumber\\
	&=\left(\frac{1+\sqrt{u}+ i|\sqrt{v}|}{2(|\sqrt{v}|)^{1/2}}\right)^{1/2}e^{-i\pi/8} - \left(\frac{1+\sqrt{u} -i|\sqrt{v}|}{2(|\sqrt{v}|)^{1/2}}\right)^{1/2}e^{i\pi/8}.
\end{align}

For the double discontinuity, we take $ x_{23} $ timelike as well which takes us from $ -v\rightarrow v $. This leads to another cut in the analytic continuation of $ \text{Disc}_{14}{\cal G} $ such that $ \pm i|\sqrt{v}|\rightarrow \pm|\sqrt{v}| \Rightarrow \pm|\sqrt{v}|\rightarrow\mp i|\sqrt{v}| $. Notice the sign change to $ \mp $ is indicative of being in the scattering region. Therefore,
\begin{align}
	2i\ \text{dDisc}\,{\cal G} &\equiv \text{Disc}_{14}{\cal G}(+\sqrt{v}) - \text{Disc}_{14}{\cal G}(-\sqrt{v})\nonumber\\
	&= 2i\ \left(\frac{1+\sqrt{u}+|\sqrt{v}|}{2(|\sqrt{v}|)^{1/2}}\right)^{1/2} - 2i\ \cos(\pi/4)\left(\frac{1+\sqrt{u} -|\sqrt{v}|}{2(|\sqrt{v}|)^{1/2}}\right)^{1/2}
\end{align}
Here, $ |\sqrt{v}| $ is in the scattering region. Notice that the first term is visually similar to $ {\cal G}_E $, but is evaluated with $ |\sqrt{v}| = (\cosh y - \cosh\eta)/(\cosh y + \cosh\eta) $. That is, it is evaluated under a ``swap'' given by $ (w,\sigma)\rightarrow(\sigma,w) $, which is analogous to the map $ \bar\rho\rightarrow 1/\bar\rho $.

Therefore,
\begin{align}
	\text{dDisc}\,{\cal G} &= \frac{1}{\sqrt{2}}\left(\frac{1+\sqrt{u}+|\sqrt{v}|}{(|\sqrt{v}|)^{1/2}}\right)^{1/2} - \frac{1}{2}\left(\frac{1+\sqrt{u} -|\sqrt{v}|}{(|\sqrt{v}|)^{1/2}}\right)^{1/2}\nonumber\\
	&= \frac{\sqrt{2}\cosh y/2-\cosh\eta/2}{(\cosh^2y-\cosh^2\eta)^{1/4}} = \frac{\sqrt{2}(1+w)^{1/2}-(1+\sigma)^{1/2}}{(4w^2-4\sigma^2)^{1/4}}
\end{align}
which is in agreement with \cite{caron2017analyticity}.

\paragraph{Generalized Mean Field Theory:} We can perform a similar analysis for the generalized Mean Field Theory where the fourpoint function in the Euclidean setting is given by:
\begin{equation}
	{\cal G}_E = 1 + (u)^{\Delta} + \left(\frac{u}{v}\right)^{\Delta}.
\end{equation}
The double discontinuity is given by:
\begin{equation}
	\text{dDisc}\, {\cal G} = \frac{2^{1+2\Delta}\sin^2(\Delta\pi)}{(w-\sigma)^{2\Delta}}.
\end{equation}

\section{Some Useful Aspects of \texorpdfstring{$SO(d,2)$}{SO(d,2)}}\label{app:Iwasawa}

We quickly introduce special features for $SO(d,2)$  in \ref{app:C1},  leading to Iwasawa decomposition. In \ref{app:root-2}, we illustrate several important properties and, in particular, we discuss its root structure. In \ref{app:C3}, we discuss various useful group decompositions, e.g., generalized Cartan-type, $G=HAH$. This is done by making use of the more familiar example of $SL(2,R)$.

\subsection{Iwasawa Decomposition for $SO(d,2)$} \label{app:C1}

The group $SO(d,2)$ contains a maximal compact subgroup (MCSG),  $K=SO(d)\times SO(2)$, with compact generators, ${\cal K}$. It also contains a rank-2 non-compact  Maximal Abelian Subgroup (MASG), $A$, with generators ${\cal A}$, i.e.,  dim$({\cal A})=2$.  It is useful to express its Lie Algebra in a Cartan decomposition, ${\cal G}={\cal K} \oplus {\cal P}$.   With dim$({\cal K})=d(d-1)/2+1$,  it follows that dim$({\cal P})=  2d$.   It can be shown that the space ${\cal G}$ can be spanned by $ {\cal K}$, ${\cal A}$ and ${\cal N_+}$, with dim$({\cal N}^+)=2d-2$. This subset, ${\cal N}^+$,  can be chosen to be a set of ``positive roots",   leading to a set of nilpotents. 
(It is worth noting that,  for $SO(d+1,1)$, one has  dim$({\cal K})=d(d+1)/2$, dim$({\cal A}) = 1$, and dim$({\cal N}_+)=d$.) This leads to a unique group decomposition
\be
G= PK = N^+A K , 
\ee
known as {\bf Iwasawa decomposition}, where $A$ and $N^+$ are generated by ${\cal A}$ and ${\cal N}^+$ respectively.  (See Appendix \ref{app:C3} for a simple  illustration for $d=1$.)

This decomposition can be considered as a generalization of the Gram-Schmidt procedure for organizing the Lie algebra. Schematically, in this picture, for each group element, one can choose $A$ to be diagonal, with positive entries, $N^+$ is upper triangular, with 1 on the diagonal and 0 in the lower half, and $K$ consists of orthogonal matrices. The details in arranging for $N^+$, the set of positive roots, is the same as that done for compact groups which involves specifying its root-space structure.  This will serve as our starting point for constructing induced representations for non-compact Lie groups.

\subsection{Illustrative Features and Root Space}\label{app:root-2} 
We provide  here explicit illustrations for  $d=2$. In anticipating  generalizing to $d=4$, we shall denote the embedding coordinates by $(X_{-2},X_{-1}; X_0, X_3)$, with the understanding that $X_3$ is associated with the longitudinal coordinate, i.e., the z-axis. (Instead of subscripts 0 and 3, we will write these  occasionally as $t$ and $z$ when the context is clear. When moving to $d=4$, we will occasionally replace subscripts $1$ and $2$ by $x$ and $y$.) 

The dimension for the algebra is now $6$, with dim$({\cal K})=2$ and dim$({\cal P})=  4$. Each generator can be represented by a real $4\times 4$ matrix.   Let us designate compact rotational generators in the two-dimensional space-like ($X_{-1}$-$X_3$) and time-like ($X_{-2}$-$X_0$)   planes as $K_1={\cal L}_{3,-1}=-{\cal L}_{-1,3}$ and $K_2={\cal L}_{-2,0}=-{\cal L}_{0,-2}$ respectively.   From (\ref{eq:ME4generators}), each is an anti-symmetric $4\times 4$ matrix, with mostly vanishing elements except for two, e.g., $K_1=\left(\begin{array}{ll}
0 &1\\
-1&0\\
\end{array}\right)$,  acting on the space-like plane $X_{-1}$-$X_3$. Similarly, one also has $K_2=\left(\begin{array}{ll}
0 &1\\
-1&0\\
\end{array}\right)$,  acting on the $X_{-2}$-$X_0$ plane.  

Of four noncompact generators, we have already indicated that we denote the generator for scaling as $D={\cal L}_{-1,-2}=-{\cal L}_{-2,-1}$ and the generator for Lorentz boost along ``longitudinal" direction as $L={\cal L}_{3,0}=-{\cal L}_{0,3}$.   Other two ``mixed" boosts are  $B_1={\cal L}_{3,-2}=-{\cal L}_{-2,3}$ and $B_2={\cal L}_{-1,0}=-{\cal L}_{0,-1}$.   All are  symmetric $4\times 4$ matrices, with mostly vanishing elements except two. They all take on the form $\left(\begin{array}{ll}
0 &1\\
1&0\\
\end{array}\right)$ when acting on their respective two-dimensional Lorentzian plane.

From these, we note the following special features:
\begin{itemize}
\item{\bf Commuting pairs and Cartan Subalgebras:}
There are three commuting pairs, 
 \be
[K_1,K_2]=[L_{3,-1},L_{-2,0}]=0, \quad [B_1,B_2]=[L_{3,-2},L_{-1.0}]=0,\quad [D,L]=[L_{-1,-2},L_{3,0}]=0,
\ee
which can be identified with the three sets of allowed Cartan subalgebras, with $(D,L)$ and  $(B_1,B_2)$ as the possible generators  for the (non-compact) MASG.  For CFT considerations, we shall adopt ${\cal A}=(D,L)$ as that for the MASG.  (The option of adopting the pair $(B_1,B_2)$ will re-emerge when we consider causal symmetric space below.)

\item{\bf Basis for ${\cal G}$:}  With dim$({\cal G})=  6$, the natural choice for its basis would be the three pairs $(K_1,K_2)$, $(D,L)$ and $(B_1,B_2)$. However, there are several more meaningful options, e.g., by forming the following combinations,
\be
T_z=-K_1+B_1, \quad T_t=-K_2+B_2, \quad C_z=  K_1 + B_1,\quad C_t=K_2+B_2.
\ee
  In the context of CFT, $T_t$ and $T_z$ serve as generators of translations and $C_t$ and $C_z$ as that for special conformal transformations. When represented as $4\times 4$  matrices, 
 \be
T_t=\begin{pmatrix}
	0 & 0 & -1 & 0\\
	0 & 0 & 1 & 0\\
	1 & 1 & 0 & 0\\
	0 & 0 & 0 & 0
\end{pmatrix}, \quad T_z =\begin{pmatrix}
0 & 0 & 0 & 1\\
0 & 0 & 0 & -1\\
0 & 0 & 0 & 0\\
1 & 1 & 0 & 0
\end{pmatrix},   \label{eq:Ttz}
  \ee
with   $C_t$ and $C_z$ given by the transpose of $T_t$ and $T_z$.
It is easy to check that these are nilpotent, i.e., $T_t^3=T_z^3=C_t^3=C_z^3=0$.
 
Another  useful set  obtained by forming lightcone combinations,   is  $T_{\pm}=T_t\pm T_z$ and $C_\pm=C_t\pm C_z$. These lead to a simple set of commutation relations with $D$ and $L$,
\be
 [L,T_\pm]=\pm T_\pm, \quad [L,C_\pm]=\mp C_\pm, \quad [D,T_\pm]=-T_\pm , \quad  [D,C_\pm]=C_\pm
\ee

\item{\bf Root Space and Positive Roots:}   As a linear vector space, the basis for ${\cal G}$ can also be chosen to consist of independent eigenvectors of linear transformation on ${\cal G}$. 
Denote elements of ${\cal A}$ by ${\cal A}_i$, e.g., with ${\cal A}_1=L$ and ${\cal A}_2=D$.  The action of each ${\cal A}_i$ with other generators $X$ via commutators now serves as a linear map for the algebra. Since $L$ and $D$ commute, ${\cal G}$ can be spanned by a set of simultaneous eigenvectors, $ \{ {\cal L}_\alpha\} $, each  associated with  a root vector, $ \vec \alpha =(\alpha_1,\alpha_2)$, via
\begin{equation}
	[{\cal A}_i,{\cal L}_\alpha] = \alpha_i {\cal L}_\alpha.  \label{eq:roots}
\end{equation}
It is well known that the root vectors are one dimensional subspaces. That is, each root vector corresponds to a unique generator in the algebra. 
It can also be shown that, for $SO(d,2)$, the root space spectrum of each element of $ {\cal A} $ takes on values $ \{-1,0,1\} $.

It is useful to first focus on a decomposition according to eigenvalues relative to each ${\cal A}_i$.
For instance, for $L$,  
\begin{equation}
	{\cal G} = {\cal G}(-1,L) \oplus {\cal G}(0,L) \oplus{\cal G}(1,L),   \label{eq:GL}
\end{equation}
where ${\cal G}(\lambda,L)$ is the set with eigenvalue $\lambda=0,\pm 1$.   Clearly, ${\cal G}(0,L) =(L, D)$. From commutation relations given above, one has ${\cal G}(1,L) =(T_{+}, C_{-})$ and ${\cal G}(-1,L) =(T_{-}, C_{+})$.   Similarly, 
$	{\cal G} = {\cal G}(-1,D) \oplus {\cal G}(0,D) \oplus{\cal G}(1,D),  $ \label{eq:GD}
with ${\cal G}(0,D) =(L, D)$, ${\cal G}(1,D) =(C_+, C_-)$ and ${\cal G}(-1,D) =(T_+, T_-)$. 

This also leads to a natural decomposition of the group algebra dictated by its spectrum on each generator in the Cartan subalgebra, ${\cal A}$, i.e., associating generators as simultaneous eigenvectors of $(L,D)$, with eigenvalue, (root-vector), $(\alpha_1,\alpha_2)$. The allowed non-zero roots are $(1,1)$, $(1,-1)$, $(-1,1)$ and $(-1,-1)$, associated with generators, $C_-$ and $T_+$, $C_+$ and $T_-$ respectively. Together with $L$ and $D$, associated with the zero root, $(0,0)$, they can serve as a basis for the group algebra. 

Given a specific ordering, $(\alpha_1,\alpha_2)$, it selects  out the set of positive roots, defined by: 
\begin{align}
	\alpha_1>0,\quad & \alpha_2\in\{-1,0,1\},\quad \nonumber\\
	\text{or}\quad\alpha_1=0,\quad & \alpha_2>0\ .
\end{align}
Associated with  positive roots $(1,1)$ and $(1,-1)$ are $C_-$ and $T_+$ respectively.  Other root vectors are $C_+$ and $T_-$, associated with roots $(-1,1)$ and $(-1,-1)$.

\item{\bf Cayley Basis:} We can perform a change of basis such that the MASG $ A $ is diagonal. 
A familiar example of this is the map from $ SO(2,1) $ to $ SL(2,R) $ which we shall discuss in the next subsection. (See Eq. (\ref{eq:SL2Ra}.) For $ SO(2,2) $, however, there is a choice in the ordering of MASG. Let us adopt the order $(L,D)$, where, in this basis, the  Lorentz boost genrator $L$  only has non-vanishing diagonal in the first and fourth entries, and scaling $D$ has non-vanishing second and third entries.  The nilpotents, corresponding to positive roots, $N^+_t(t_+)= e^{t_+(T_t + T_z)}$ and  $N^+_c(c_-)=  e^{c_-(C_t - C_z)}$,  can be shown to take upper triangular forms. That is, with this choice for the Cayley basis, $a_L(y) =e^{y L} $, $a_D(\xi)=e^{\xi D}$, $N^+_t(t_+)$ and $N^+_c(c_-)$ are 
\begin{equation}
	\begin{pmatrix}
		e^y & 0 & 0 & 0\\
	0 & 1 & 0 & 0\\
		0 & 0 & 1 & 0\\
		0 & 0 & 0 & e^{-y}
	\end{pmatrix},\, 
	\begin{pmatrix}
	1 & 0 & 0 & 0\\
	0 & e^{\xi}& 0 & 0\\
	0 & 0 & e^{-\xi} & 0\\
	0 & 0 & 0 & 1
	\end{pmatrix},\, 
	\begin{pmatrix}
		1 & -2t_+ & 0 & 0\\
		0  & 1 & 0 & 0\\
		0 & 0 & 1 & 2t_+\\
		0 & 0 & 0 & 1
	\end{pmatrix},\, 
	\begin{pmatrix}
		1 & 0 & -2c_- & 0\\
		0 & 1 & 0 & 2c_-\\
		0 & 0 & 1 & 0\\
		0 & 0 & 0 & 1
	\end{pmatrix}. \label{eq:Cayley4}
\end{equation}
Let us stress that the choice of Cayley basis corresponds to a particular ordering for positive roots, e.g., $(L,D)$, which affects which of the nilpotents are in this upper triangular form. This is equivalent to saying that the ordering of roots dictates the set $ N^+ $. The Iwasawa decomposition is best understood in this Cayley basis where the Cartan subalgebra becomes diagonal.

\end{itemize}

\subsection{Group Decompositions}\label{app:C3}
We provide a simple illustration for the difference between Iwasawa and Cartan decompositions  by considering $SO(2,1)$.   Instead of working with $3\times 3$ matrices,  we will instead considering the equivalent $SL(2,R)$ by working with real $2\times 2$ matrices. There are three generators, one compact  and two non-compact. Consider
\be
X_0=(1/2)\left(\begin{array}{ll}
0&1\\
-1& 0\\
\end{array}\right), \quad X_1=(1/2)\left(\begin{array}{ll}
1&0\\
0&-1\\
\end{array}\right), \quad X_2=(1/2)\left(\begin{array}{ll}
0&1\\
1&0\\
\end{array}\right). \label{eq:SL2Ra}
\ee
  \begin{itemize}
 \item{\bf Iwasawa Decomposition:}
Let us adopt $A=X_1$ as the generator for the MASG and  $X_0$ as the generator for the compact subgroup $K$.  Consider 
$
 {\cal N}^{\pm}=X_0 \pm X_2
$, i.e.,
$
{\cal N}^+=\left(\begin{array}{ll}
0&1\\
0&0\\
\end{array}\right) \label{eq:SL2Rb}
$, which is upper triangular.
One easily veries that $ {\cal N}^\pm $ are  root vectors,
$
 [X_1,{\cal N}^{\pm}]=\pm {\cal N}^\pm
$,
 with ${\cal N}^+$ being the positive root. For each element of the group, Iwasawa decomposition corresponds tox $ g(\theta,\eta, \xi)= k(\theta) a(\eta) n(\xi) $
 where 
 \be 
 k(\theta)=\left(\begin{array}{ll}
\,\,\cos \frac{\theta}{2} &\sin  \frac{\theta}{2}  \\
-\sin  \frac{\theta}{2} & \cos \frac{\theta}{2} \\
\end{array}\right), \; 
a(\eta)= \left(\begin{array}{ll}
e^{\frac{\eta}{2}}&0\\
0& e^{-\frac{\eta}{2}}\\
\end{array}\right) , \; n(\xi) =e^{\xi {\cal N}^+}= \left(\begin{array}{ll}
1&e^{{\xi}}\\
0& 1\\
\end{array}\right) \; .
\ee

\item{\bf Cartan Decomposition:}
In addition to Iwasawa decomposition, another useful decomposition is of the Cartan type.  Consider generalizations of Eulerian decomposition for $SO(3)$, $g(\phi,\theta, \psi)=k_z(\phi) k_y(\theta) k_z(\psi)$.
For $SO(2,1)$, there are two possibilities.  Again fixing ${X_1}$ for the MASG, one involves $K$ and the other involves $H$, generated by $X_2$, i.e.,
\be
g(\phi,\eta, \psi)=k(\phi) a(\eta) k(\psi), \quad {\rm and} \quad 
g(\xi,\eta, \xi')=h(\xi) a(\eta) h(\xi').
\ee
where 
\be 
h(\xi)=e^{\xi X_2}=\left(\begin{array}{ll}
\cosh  \frac{\xi}{2} &\sinh \frac{\xi}{2}\\
\sinh \frac{\xi}{2}& \cosh\frac{\xi}{2}\\
\end{array}\right)\in  H \; .
\ee
These are Cartan type of decompositions. The first is of the type $G=KAK$, i.e., it involves a compact subgroup $K$. The second involves a non-compact subgroup, $H$, and will be referred to as the $G=HAH$ type.

\end{itemize}

\section{Causal Symmetric Spaces}\label{app:embed}

In this appendix, we provide a background discussion to {\it non-Riemannian symmetric spaces}, or, more precisely, \textit{causal symmetric spaces} \cite{Hilgert:1996}. This step is necessary in formulating {\bf causality} for CFT group theoretically. In \ref{app:symmetricspace},  we discuss spacetime as a Symmetric Space. In  \ref{app:semigroup}, we provide a first look at the Causal Semigroup.    In \ref{app:SG4point}, we finally discuss causal semigroup for four-point CFT Amplitude.

The defining property of Riemannian spaces is that the separation between any two points on the manifold is positive definite.  Although instructive, these spaces are not sufficient for discussing Lorentzian physics directly as the salient feature of Minkowski spacetime is that it admits a causal structure. That is, on the spacetime manifold, separation between two events can be positive, null or negative. Therefore, these spaces admit a lightcone description of causality. As stated earlier, these  are {\it non-Riemannian causal symmetric spaces}.  To understand spherical harmonics appropriate for studying Minkowski CFTs, we need to understand these causal symmetric spaces from a group theoretic perspective. 

\subsection{Spacetime as a Symmetric Space}\label{app:symmetricspace}

The group $ SO(d,2) $ can be realized via the exponentiation of $(d+2)(d+1)/2 $ generators, (see Appendix-\ref{app:Iwasawa}). In Sec. \ref{sec:induced}, a {\it Cartan decomposition} for  the Lie algebra, ${\cal G}={\cal K}\oplus {\cal P}$ was introduced, where ${\cal K}$ stands for generators for the maximal compact subgroup. We have also identified a particular set of non-compact generators, ${\cal A}=\{D,L\}$, ${\cal A}\subset {\cal P}$,  as the preferred set of  maximal Abelian subgroup in $G$. 

The standard Cartan decomposition, ${\cal G}={\cal K}\oplus {\cal P}$, corresponds to having an involution, $s$, with $s^2=1$, where $s({\cal K})={\cal K}$ and $s({\cal P})=-{\cal P}$. In this section, we will introduce another decomposition, ${\cal G}={\cal H}_0\oplus {\cal Q}_0$, which is more useful in dealing with causality issues~\cite{Mack:2019akh,mack2007simple}. This corresponds to introducing another involution, $\Theta$, where $\Theta({\cal H}_0)={\cal H}_0$ and $\Theta({\cal Q}_0)=-{\cal Q}_0$, with ${\cal G}={\cal H}_0\oplus {\cal Q}_0$.  This new involution is compatible with that for the standard involution $s$ in the sense that its MASG, ${\cal A}_0$, can be chosen to also lie in ${\cal P}$, i.e., ${\cal A}_0 \subset {\cal P} \cap{\cal Q}_0$ \cite{Hilgert:1996}.

It is important to consider the role of coordinate origin, $x_\mu=0$, which lies on the null-cone  at $ \xi_0=(1,1;0,0,\cdots) $. By construction, we have $Q^{-} \xi_0=\xi_0$. 
Special attention should also  be paid to the subgroup $ {H_0} $ associated with the algebra $ {\cal H}_0 = {\cal L}\oplus D$, which serves as the isometry group of the point  $ \xi_0=(1,1;0,0,\cdots) $,  upto an overall scaling. That is, an element $ h\in {H_0} $ acting on $\xi_0 $ gives $ h\cdot\xi_0 = (a,a;0,0,\cdots)= a\,\xi_0$.  For $d=4$, the set $ {\cal H}_0 = {\cal L}\oplus D$ consists of 7 elements,
\begin{equation}
	{\cal H}_0 = {\cal L}\oplus D=\{B_{01},B_{02},B_{03}=L;R_{12},R_{31},R_{23}\}\oplus \{D\}.
	\label{eq:HforSapcetime}
\end{equation}
That is, ${\cal L}$ consists of 6 generators, 3 for rotations and 3 for boosts. In particular, $D$ is central in ${\cal H}_0$, i.e., it commutes with all other generators. State it more simply, for 
$d=4$, $ {\cal H}_0$ consists of Lorentz generators and dilations.  For $d=2$, however, the set $ {\cal H}_0$ is much smaller, consisting of only two elements $ {\cal H}_0 = \{L,  D\} $. We also note that the subalgebra ${\cal Q}_0$, consists of $ {\cal Q}_0 ={\cal Q}^-\oplus {\cal Q}^+ $. Both associated subgroups $Q^\pm$ are abelian.

The pair $ ({\cal G}; {\cal H}_0) $ can be used to define a symmetric space~\footnote{The structure theory of symmetric spaces can be rather involved \cite{Hilgert:1996}. Here, we try to present the minimum required for our discussion. In particular, it needs to be shown that the following commutation relations hold: $ [{\cal H},{\cal H}]\subset{\cal H}; [{\cal H},{\cal Q}]\subset{\cal Q}; [{\cal Q},{\cal Q}]\subset {\cal H}$. This is true for all variations of $ ({\cal G},{\cal H}) $ presented in this work.}. In particular, the quotient space
\begin{equation}
	\mathcal{M} = G/H_0 Q^-
\end{equation}
can be associated with our spacetime manifold. The construction here is the Lorentzian counterpart for the Euclidean spacetime identification $ \mathcal{M}_E = G/MAN^- $ made in \cite{Dobrev:1977} where the subgroup $ MAN^- $ is maximal parabolic in $ SO(d+1,1) $. Similarly, $ {H_0}{Q^-} $ is a maximal parabolic in $ SO(d,2) $.

Generators of translations, traditionally denoted as ${\cal T}_\mu$, are realized in the embedding space as discussed in App. \ref{app:Iwasawa}.
To understand this identification intuitively,  this means that the group parameters for $Q^+$ are the coordinates on the spacetime manifold, and they therefore act transitively on it. A generic element of $ Q^+,Q^- $ is given by:
\begin{equation}
	Q^+(a^\mu) = \begin{pmatrix}
		1+a^2/2 & a^2/2 & a_\mu\\
		-a^2/2 & 1-a^2/2 & -a_\mu\\
		a^\mu & a^\mu & \text{diag}(\mathbf{1}_d)
	\end{pmatrix}; \quad Q^-(b^\mu) = \begin{pmatrix}
		1+b^2/2 & -b^2/2 & (b^\mu)^t\\
		b^2/2 & 1-b^2/2 & (b^\mu)^t\\
		(b_\mu)^t & -(b_\mu)^t & \text{diag}(\mathbf{1}_d)
	\end{pmatrix}.
\end{equation}
Transitive action of $ Q^+ $ on the spacetime manifold can be shown by projecting down onto $ \xi_0 $, thus recovering our parabolic slice:
\begin{equation}
	{Q^+}(x^\mu)\cdot\xi_0/2=  {Q^+}(x^\mu)\cdot \left(\frac{1}{2},\frac{1}{2}; 0, \cdots\right) =  \bigg(\frac{1+x^2}{2},\ \frac{1-x^2}{2};\ x^\mu\bigg) .
\end{equation}

\subsection{A First Look at the Causal Semigroup}\label{app:semigroup}
Our spacetime manifold $ \mathcal{M} $ thus obtained admits a causal structure in the sense that it carries a natural definition of lightcones. That is, for any given point $ a\in\mathcal{M} $, we have a set of points $ b $ in the future of $ a $ such that $ (a-b)^2 \leq 0 $ and $ a_t\leq b_t $. The set of group transformations that take $ a\rightarrow b $ form a semigroup in $ SO(d,2) $. A semigroup structure is characterized by the absence of an inverse. Intuitively, this can be understood by the fact that the future point $ b $ has a past lightcone which contains $ a $, along with all other points $ c $ such that $ (c-b)^2\leq0 $ and $ b_t\geq c_t $. (That is, the point $ b $ itself does not have any information about its past.)

For our choice of basepoint, this condition is trivially satisfied by time translations, Lorentz group and dilatations under the constraints that $ x_t>0 $, where $ x_t $ is the time translation parameter. That is, the semigroup relative to the origin $\xi_0 $, 
\begin{equation}\label{eq:semigroup}
	S_0 = \{g\in G\ |\ \xi_0 \leq g\cdot \xi_0 \}.
\end{equation}
can be expressed as
\begin{equation} 
	S_0 = {H}_0\exp x_t{\cal T}_0.
\end{equation}
where the generators for ${H}_0$ are $ {\cal H}_0={\cal L}\oplus D$, given earlier, (\ref{eq:HforSapcetime}).  Any point $ x_a $ in the forward lightcone of the basepoint can be written as $ \xi_a = S_0\cdot \xi_0 $.

Recall that the generator of time translations in the embedding formulation can be realized as $ {\cal T}_0 =  {\cal B}_{-1,0}  +{\cal R}_{-2, 0} $.   These two generators, ${\cal B}_{-1,0}$ and ${\cal R}_{-2, 0} $, have a special place in the structure theory of causal symmetric spaces.  The non-compact generator $ {\cal B}_{-1, 0} $ is often called the \textit{cone generating element} in $ {\cal Q }_0 $.  The compact rotation $ {\cal R}_{-2, 0} $ defines a gradation
on $ \cal G $ such that $ {\cal G}(0,R_{-2, 0}) = {\cal K}$ are generators for the maximal compact subgroup. (That is, we have a decomposition analogous to Eq. (\ref{eq:GL}) such that $ {\cal G} = {\cal G}(-1,R_{-2, 0})\oplus {\cal G}(0,R_{-2, 0}) \oplus{\cal G}(1,R_{-2, 0}) $.) Together with $D$, they  form an $ SO(2,1) $ subalgebra. In general, given the symmetry group $ SO(d,2) $, $ R_{-2, 0} $ is the central element of $ \cal K $ in $ \cal K $. With a choice of $ {\cal H} $, there is a central element $ Y $ of $ \cal H $ in $ \cal H $. Taking the commutator $ [R_{-2, 0}, Y] $ and completing the $ SO(2,1) $ algebra gives the cone generating element.

The cone generating element,  $ {\cal B}_{-1, 0} $,  also belongs to the set ${\cal A}_0$, which allows an analogous Iwasawa-like decomposition
\be
S_0\subset N_0^+A_{0} H_0.  \label{eq:HAN}
\ee
The set of positive roots $ N_0^+ $ is ordered such that the cone generating element is the leading root \cite{Hilgert:1996}. That is, fixing $ \cal H $ not only determines $ \cal A $ completely, it also tells us the ordering with respect to which positive roots are defined in the induced representation picture. Therefore, fixing the cone-generating element determines the Weyl vector $ \vec{\rho} $ and hence plays a central role in determining the leading behavior of representation functions defined over the semigroup (see Sec. \ref{sec:4point}). We also note that this is where the ambiguity pointed out earlier (in section 3.2)
	is resolved. The subalgebra $ {\cal A} $ used for induced representations is maximal abelian in $ {\cal Q}\cap {\cal P} $.

For the purposes of harmonic analysis, a particularly important decomposition of the causal semigroup is a Cartan-like decomposition of the form:
\begin{equation}
	S_0 = {H}_0 {A}_{0+}{H}_0  \label{eq:HAH}
\end{equation}
where $ {A}_{0+}$ is a subset of $ {\cal A}_0 = \{{\cal B}_{-1, 0 },{\cal B}_{-2, 3}\} $,  the maximal abelian subalgebra in $ {\cal Q}_0\cap {\cal P}_0 $ introduced earlier. As was outlined in Sec. \ref{sec:4point}, for CFTs, we will be interested in studying the $ H $-bi-invariant zonal spherical functions. A Cartan-like decomposition makes the identification of these functions more natural.
The restriction to $ {A}_{0+} $ is such that boost parameters satisfy the inequality  
\begin{equation}
	\text{param}({\cal B}_{-1, 0 })-\text{param}({\cal B}_{-2, 3})>0 \label{eq:restriction}
\end{equation}
to ensure the final point lies in the forward lightcone. This can be shown by considering  points $ \xi = {A}_0\cdot \xi_0 $ and imposing the constraint that $ (x-x_0)^2<0 $.  It is worth stressing that the choice of maximal abelian subalgebra is associated with the choice of baseloint $\xi_0$.

Let the subgroup $ A_0 $ be parametrized by $ (y_0,\eta_0) $. The basepoint of interest is $ \xi_0 = \{1,1;0,0\} $, which maps to the origin of the spacetime manifold. Under action of $ A_0 $, this point moves on the spacetime manifold to:
\begin{equation}
	x = (t,z) = \left(\frac{\sinh y_0}{\cosh\eta_0+\cosh y_0},\frac{\sinh\eta_0}{\cosh\eta_0+\cosh y_0}\right).
\end{equation}
Requiring this point to lie in the forward lightcone of $ x_0 = (0,0) $ leads to:
\begin{align}
	x^2 - x_0^2 &< 0\\
	-\sinh^2y_0 + \sinh^2\eta_0 &< 0
\end{align}
which leads to our causal condition:
\begin{equation}
	y_0 - \eta_0 > 0;\quad y_0>0\ .
\end{equation}
This analysis applies for general $d$ in the antipodal frame where $\vec{b}_\perp = 0$. For non-zero impact parameter, we discussed similar constraints in Sec. \ref{sec:CFTd}.

\subsection{Causal Semigroup for Four-point CFT Amplitude}\label{app:SG4point}

As we have shown in Sec. \ref{sec:4point} that, for causal consideration CFT amplitude, it is necessary to consider semigroup relative to a basepoint other than  $\xi_0$.  A generic point $ x_a $ on the manifold can be arrived at from the basepoint by some transformation $ \tilde{g} \in G$. Therefore, we find that  the semigroup $S_a $  with respect to the basepoint $ \xi_a= \tilde{g}\cdot\xi_0$ given by
\be
S_a  =\tilde{g}\cdot S_0\cdot \tilde{g}^{-1}
\ee
That is, the semigroup at any point $ \xi_a $ on the manifold can be obtained via a shift by $ \tilde{g} $ where $ \tilde{g}\cdot \xi_0 = \xi_a $.  With similarly shifted generators, one has a different set of maximal abelian subgroup, $ {\cal A}_a= \tilde{g}{\cal A}_0 \tilde{g}^{-1}$,  cone generating element, etc.

As was discussed in Sec. \ref{sec:emantipodal}, the relevant basepoint can be chosen to be  the point $ \xi_a = \{1,0;0,0,0,1\}$, which can be translated from $\xi_0$ by  
\begin{equation}\label{eq:rotation}
	\tilde g =\exp(\frac{\pi}{2} {\cal R}_{-1, 3}). 
\end{equation}
This translates from $ S_0 $ to $ S_a $ by conjugation, i.e., $ S_a = \tilde{g}\tilde{S_0}\tilde{g}^{-1} $. Therefore, in terms of algebra we have that, after conjugating ${\cal H}= \tilde{g} {\cal H}_0\tilde{g}^{-1}$, for $d=2$, 
\be
 {\cal H} = \{B_{-2,z},B_{-1,t}\}.\label{eq:H2}
 \ee
For $d=4$ ,
\be
{\cal H} = \{B_{-2,z},B_{-1,t},R_{-1,x},R_{-1, y},B_{tx},B_{ty},R_{xy}\}\label{eq:H4}
\ee
where $B_{-2,z}$ is central.  More explicitly, the causal condition for the semigroup relative to $\xi_a$ becomes
\begin{equation}
	S_a = \{g\in G\ |\ \xi_a \leq g\cdot \xi_a \}.
\end{equation}
and it can be expressed as
\begin{equation} 
	S_a= {H}\exp x_tT_a,
\end{equation}
where $ T_a = \tilde{g} ( R_{-2, t} + B_{-1, t } ) \tilde{g}^{-1}=R_{-2, t} + L $, with $L$ as
the cone generating element. 
The generators for the MASG, $A$, are $ {\cal A} = \tilde{g} {\cal A}_0\tilde{g}^{-1}= \{L,D\} $.
More interestingly, the semigroup can now be expressed as
\begin{equation}
	S_a = HA_+H
\end{equation}
The restriction on $ A_+ $, changing from (\ref{eq:restriction}), is $ y-\eta > 0 $. Note that this is precisely the condition defining the physical causal scattering discussed in Sec. \ref{sec:GAF}.

\section{Inductive Characters \& Zonal Spherical Functions}\label{app:sphr}

In this appendix,  we begin by considering zonal spherical functions for $SO(1,2)$ relative to its subgroups, following the procedure as done for $SO(3)$ via Euler-type group expansions~\footnote{For integral representations of special functions and their further generalizations, see \cite{Vilenkin:1991,Vilenkin1993,durand1978product}}, (also often described as Cartan decomposition). We next introduce a ``doubling procedure" in App. \ref{app:sphrDoubling}, which facilitates our treatment for the rank 2 case  in App. \ref{app:sphrDoubling2}.

There are two possible zone spherical functions, one for compact subgroup $K=SO(2)$ and the second for non-compact subgroup $H=SO(1,1)$. We are only interested in the latter; however, the treatment for both are instructive. The associated zonal spherical functions are
\bea
\varphi_K(\eta,\lambda)&=& P_{\lambda}(\cosh \eta)=
\int_0^{2\pi} d\phi  (\cosh \eta + \cos \phi \sinh \eta)^{\lambda}, \label{eq:Pcompact}  \\ 
\varphi_H(\eta,\lambda)&= &Q_{-\lambda-1}(\cosh \eta)= 
\int_0^\infty ds (\cosh \eta+ \cosh s \sinh \eta)^\lambda\ . \label{eq:Qnoncompact}
\eea
The integrands are  the inductive characters, $W_K(\cosh \eta,\phi;\lambda)$ and  $W_H(\cosh \eta, s; \lambda)$,  for $SO(1,2)$ over $SO(2)$ and $SO(1,1)$ respectively, with $\lambda =  -1/2 + \widetilde \ell$. Let us begin by listing various basic  facts, with $d=1$.

\paragraph{Homogenous Functions and Null-Cone:}
Consider $SO(1,2)$ acting transitively on the null-cone, $-\xi_1^2+\xi_2^2-\xi_3^2=0$. In anticipation of a CFT reduction, we introduce a space of homogeneous functions on the null cone,   
${\cal F}_\lambda= \left \{ F(\xi)\Big | F(\sigma \xi) = \sigma^\lambda F(\xi)\right \} $. This homogeneity condition can be realized with different ``slices" of the null-cone, dictated by  certain subgroup of $SO(1,2)$, acting transitively on the slice.

For instance, with $\sigma = \xi_2^{-1}>0$, the slice $\gamma_1$ is
$\Gamma_1=\{ \xi\Big | \xi(\phi)=(\cos \phi, 1, \sin \phi ),\quad \phi\in [0,2\pi]\}$
and the compact subgroup $K=R_{13}= SO(2)$, i.e., rotations in $1$-$3$ plane, acts transitively on  $\Gamma_1$. With this choice, the physical coordinate 
$x=\xi_3/(\xi_1+\xi_2)$ can take on both signs.

Another choice, $\sigma = |\xi_1|^{-1}$,  leads to a difference slice,
$\Gamma_2=  \Gamma_{2,+}\cup \Gamma_{2,-}, \quad  \Gamma_{2,\pm} =\{ \xi \Big | \xi_\pm(s)=(\pm 1,\cosh s, \sinh s),\quad s\in R\}
$, with a non-compact subgroup $H=B_{23}=SO(1,1)$, e.g., boosts in $2$-$3$ plane, acting transitively on  $\Gamma_2$.  With this choice, the physical coordinate 
$x=\xi_3/(\xi_1+\xi_2)$ is positive for $s>0$ and negative for $s<0$.

The space of homogeneous functions for each slice can be characterized by a reduced function, $f_i$, defined on  $\Gamma_i$, $i=1,2$,  respectively, 
\be
{\cal F}_{\lambda,1}=\left \{ F(\xi)\Big | F(\xi) = \xi_2^\lambda f_1(\xi_1/\xi_2,\xi_3/\xi_2)\right \}, \,
{\cal F}_{\lambda,2}= \left \{ F(\xi)\Big | F(\xi) = |\xi_1|^\lambda f_2(\xi_2/\xi_1,\xi_3/\xi_1)\right \}.
\ee

\paragraph{Choice of Basis for ${\cal F}_{(\lambda,i)}$:}
Each ${\cal F}_{(\lambda,i)}$ is still infinite-dimensional; a natural choice for each consists of  eigenfunctions of the relevant subgroup, i.e., $K$ for $\Gamma_1$ and $H$ for $\Gamma_2$.  For ${\cal F}_{(\lambda,1)}$,   we choose the discrete Fourier basis,   labelled by $|\lambda, m\rangle$, $m$ integers, $-\infty\leq m \leq \infty$,  corresponding to  $K=R_{13}$  diagonal. For ${\cal F}_{\gamma_2}$, the basis for ${\cal F}_{(\lambda,2)}$ can be labelled by a continuous Fourier basis, $|\lambda, p\rangle$,  $-\infty\leq p \leq \infty$, with $H=B_{23}$ non-compact  and diagonal. 	

Consider  ${\cal F}_{(\lambda,1)}$ first. Let us adopt an Eulerian parametrization where $g= R_{13}(\phi) B_{12}(\eta) R_{13}(\psi)$,  with $B_{12}$ serving as boosts and $R_{13}$ and $R_{13}$ serving as rotations. That is, we consider a Cartan decomposition for $SO(1,2)$ where $g\in K A K$, with $B_{12}(\eta)$ serving as the MASG.  With $K=R_{13}$  diagonal,   the action of a group element can be represented as $\langle \lambda,m|T_\lambda (g)|\lambda,n \rangle= e^{-i m \phi} e^{i n \psi } t_{\lambda; m,n}(\eta)$, where
\be
t_{\lambda;m,n}(\eta)= \langle \lambda,m|A(\eta)|\lambda,n\rangle = \langle \lambda,m|B_{12}(\eta)|\lambda,n\rangle.   \label{eq:tmn}
\ee
We need to find matrix elements $t_{\lambda;m,n}(\eta)$, leading to a group representation.

Consider  ${\cal F}_{(\lambda,2)}$ next. With $H$ non-compact, we consider an alternative group decomposition.   Staying with $A= B_{12}(\eta)$, consider $g= B_{23}(\zeta) B_{12}(\eta) B_{23}(\zeta')$, i.e., it is of  the type $HAH$. With $B_{23}(\zeta)$ and $B_{23}(\zeta')$ diagonal  on  the continuous basis,  $|\lambda, p\rangle$,  $-\infty\leq p \leq \infty$,  we have explicit representation
$
\langle \lambda, p |T_\lambda (g)|\lambda, p'\rangle = e^{-i p \zeta} e^{i p' \zeta' } t_{\lambda;p,q}(\eta)
$
again with  
\be
t_{\lambda;p,p'}(\eta) = \langle \lambda,p|A(\eta)|\lambda,p'\rangle =\langle \lambda, p|B_{12}(\eta)|\lambda, p'\rangle. \label{eq:tpq}
\ee	
The challenge is again in evaluating $t_{\lambda;p,p'}(\eta)$.

Before proceeding, we note the unusual feature for the case of ${\cal F}_2$. In lifting  from $\Gamma_{2,\pm}$, the null cone is divided into two separate components, labelled by the sign for $\xi_1$.  Acting transitively on each also divides  boosts $B_{12}$ into two components, each can be identified separately as a semigroup. See discussion after Eq. (\ref{eq:InductiveC}).

\paragraph{Unitary Irreducible Representation:} 
By specifying a group  action on each functional space via ``shifting", e.g.,
\be
T_\lambda (g) F_{(\lambda,i)}(\xi) = F_{g; (\lambda,i)}(\xi) = F_{(\lambda,i)}( g^{-1} \xi ),   \label{eq:shiift1}
\ee
where $F_{(\lambda,\gamma_i)}(\xi)  \in  {\cal F}_{\gamma_i}$, it can be shown that this leads to a representation,  $T_\lambda$, for $SO(1,2)$.  	One can also show that, with $\lambda$ complex, each leads to a distinct representation in general, with  $\lambda$ and $-1-\lambda$ being equivalent.  By restricting the scaling label $\lambda = -1/2 + \widetilde \lambda$ where $\widetilde \lambda$ is purely imaginary, an inner product can be introduced, leading to a unitary irreducible principal series representation. 	Representations for different slices can be shown to be equivalent.  

\subsection{Spherical Functions for $SO(1,2)/SO(2)$}\label{app:F1}
A function  in $ {\cal F}_{(\lambda,1)}$ can be treated as a vector $|F\rangle$. As a function over the null-cone, it can be expanded  in a Fourier series,
$
F(\xi)=\langle \xi|F\rangle= \xi_2^\lambda \sum_n a_n e^{in \phi}
$,
where $e^{i\phi} = ( \xi_1+i\xi_3)/\xi_2$, i.e., $a_n$ are Fourier coefficients for the function $\widetilde F(\xi)=\xi_2^{-\lambda} F(\xi)$. Consider next the action of a group element $g$ on $F(\xi)$ specified by (\ref{eq:shiift1}). It is sufficient for us to consider $g=B_{12}(\eta)$, i.e., a boost in the $1$-$2$ plane. It leads to a new function $F_g(\xi)$, which can again be expressed as 
$
F_g(\xi) = {\xi_2}^\lambda \sum_n a'_n e^{in \phi}
$,
with  $a'_n$ a  new set of Fourier coefficients.   On the other hand, by (\ref{eq:shiift1}), one has
$ F_g(x) 
=  {\xi'_2}^\lambda \sum_n a_n e^{in \phi'},$ with $\xi'=g^{-1}\xi=(\xi'_1,\xi'_2,\xi'_3)=\xi_2'(\sin \phi',1, \cos \phi')$, where 
$
\xi_2'= (\cosh \eta + \cos \phi_1 \sinh \eta) \xi_2\, $ and $\quad \phi'= \tan^{-1} [(\sin \phi)/(\sinh \eta+\cos \phi \cosh \eta)].$   

Note that $\xi_2'/\xi_2>0$, i.e., given $\xi_2>0$ initially, $\xi_2'$ remains positive for $-\infty<\eta<\infty$. 
The new set of Fourier coefficients $a'_n$ are related to the initial set, $a_n$, linearly,
$a'_m=\sum_{n} t_{\lambda,(m,n)} a_n$, with $t_{\lambda;m,n}$ given by the matrix elements (\ref{eq:tmn}). 
A direct calculation leads to
\be
t_{\lambda;m,n}(\eta)=\int d\phi  (\cosh \eta + \cos \phi \sinh \eta)^\lambda e^{-im \phi + i n \phi'}, 
\ee
with $\phi'$ a function of $\eta$ and $\phi$ given above and    $\lambda=-1/2 +\widetilde \lambda$.  Let us make the following observations:
\begin{itemize}
	\item{\bf Spherical functions:} Representation via shifts allows the identification of a subspace of ${\cal F}_{\lambda,1}$ which is $K$-right-invariant, i.e., $F(g k) =F(g)$.  It follows that all $a_n$ vanish except for $a_0$. This subspace can be identified with the coset, $SO(1,2)/SO(2)$. It follows that the integrand of (\ref{eq:tmn}), with  $n=0$  serves as the generating function for spherical functions for $SO(1,2)$ over a compact subgroup $SO(2)$.
	\item{\bf $K$-bi-invariance and Zonal spherical functions:} One can further identify  the $K$-bi-invariant subspace, i.e., $F(k g k') =F(g)$. it follows that only $t_{\lambda;0,0}$ survives,  leading  to zonal spherical function,  Eq. (\ref{eq:Pcompact}).
	From the integrand, the inductive character, $W_K(\eta,\phi)$ for $SO(1,2)/SO(2)$,  can be expressed as in (\ref{eq:genindch}).
\end{itemize}

\subsection{Spherical Functions for $SO(1,2)/SO(1,1)$}\label{app:F2}
The analysis follows analogously as for the compact case, with some added subtleties. The main difference is the fact that a function  in $ {\cal F}_{(\lambda,2)}$, treated as a vector $|F\rangle$,  can be expanded  in a Fourier integral, 
$
F(\xi)=\langle \xi|F\rangle=|\xi_1|^\lambda \int dp \,  a(p)  e^{i p \zeta}
$,
where $\xi=(\xi_1,\xi_2,\xi_3)= |\xi_1|(\pm 1,\cosh \zeta, \sinh \zeta)$, $e^{\zeta} = (\xi_2+\xi_3)/(|\xi_1|)$ and  $\pm $ stands for ${\rm sign} \, \xi_1$. It is again sufficient to consider action of $B_{12}(\eta)$ on $F(\xi)$, leading to a new function $F_g(\xi)$, where
$
F_g(\xi) =F(\xi')= |\xi_1|^\lambda \int d p\, a'(p) e^{ip \zeta } = |\xi'_1|^\lambda \int d p' \, a(p') e^{ip' \zeta' }
$,
with $\xi'=g^{-1}\xi= (\xi_1',\xi_2',\xi_3')= |\xi_1'| (\pm 1,\cosh \zeta', \sinh \zeta')$, 
$
\xi_1'= (\pm \cosh \eta + \cosh \zeta \sinh \eta)\, |\xi_1|$ and \, $\zeta'= \tanh^{-1} (\sinh \zeta)/(\pm \sinh \eta+\cosh \zeta\cosh \eta)$.  
The Fourier coefficients $a'(p)$ are relate to $a(p)$ linearly. As the case for $SO(1,2)/SO(2)$, 
$a'(p)=\int dp' \, t^{\lambda}(p,p';\eta)\, a(p')$, where $t^{\lambda}(p,p';\eta)$ is given by the matrix elements (\ref{eq:tpq}),
\be
t_{\lambda;p,p'} (\eta)=\int_{-\infty}^\infty d \zeta  \left(\cosh \eta \pm \cosh \zeta \sinh \eta\right) ^\lambda e^{\textstyle -i p \zeta + i p' \zeta'},  \label{eq:tpp'}
\ee
with $\zeta'$ a function of $\eta$ and $\zeta$ and $\lambda=-1/2 +\widetilde \lambda$.  The requirement that signs for $\xi_1'$ and $\xi_1$ remain unchanged noted earlier leads to non-trivial constraints, (to be discussed below).

It is now possible to make  the following observations:
\begin{itemize}
	\item{\bf Spherical functions:} Representation via shifts leads to a subspace of ${\cal F}_{\lambda,2}$ which is $H$-right-invariant, i.e., $F(g h) =F(g)$. It follows that $a(p)=0$ except for $p=0$, i.e., it is given by a delta-function. It follows that (\ref{eq:tpq}), with $p'=0$ and $\lambda=\ell=-1/2 + \widetilde \ell$, serves as the generating function for spherical functions over the coset, $SO(1,2)/SO(1,1)$. 
	\item{\bf $H$-bi-invariance and zonal spherical functions:} For the $H$-bi-invariant subspace, i.e., \, $F(h g h') =F(g)$,  only $t_{\lambda;0,0} (\eta)$ survives, and the integration can be restricted to $0<\zeta<\infty$ since the integrand is now even. For functions defined on $\Gamma_{2,+}$, it leads to   Eq. (\ref{eq:Q}), i.e., Legendre function of the second kind, $Q_{-1/2+\widetilde \ell}(\cosh \eta)$. From the integrand, one obtains the inductive character $W_H(\eta,\zeta)$ for $SO(1,2)/SO(1,1)$, given by (\ref{eq:InductiveC}), as claimed. 
\end{itemize}
We also note, due to the appearance of $\pm$ sign, care must be exercised in interpreting (\ref{eq:tpp'}), leading to restriction on region where it is defined. More specifically, Eq. (\ref{eq:tpp'}) is well defined for the positive sign only when $\eta>0$, thus defined over $\Gamma_{2,+}$,  and for the negative sign when $\eta<0$, thus on $\Gamma_{2,-}$. Physically, this restriction emerges because of lightcone semigroups.  

The coset $SO(1,2)/SO(1,1)$ can be thought of in two distinct ways:
\begin{itemize}
	\item CFTs on the null cone: This is the interpretation that we have worked out here for $\text{CFT}_1$. Going from $\eta>0$ to $\eta<0$ corresponds to a change in time ordering of the four points. A restriction to $\eta>0$ in Eq. (\ref{eq:InductiveC}) maintains the causal orientation for the scattering configurations. (Same holds  true by restricting to $\eta<0$.)

	\item $\text{AdS}_2$: This corresponds to moving off the null cone, and was worked out in \cite{Faraut:1986} with similar results when compared  to the ones presented here. There, the restriction $\eta>0$ is explicitly to maintain the lightcone structure on the manifold. We have a similar interpretation of these semigroup restrictions when discussing $\text{CFT}_2$.
\end{itemize}


\subsection{Generalized Doubling Procedure for  Inductive Character}\label{app:sphrDoubling}
As explained in Sec. \ref{sec:4point}, our ultimate goal is to calculate group harmonics for $ H $-bi-invariant functions with split-rank 2  which can be  obtained by the appropriate inductive character via Eq. (\ref{eq:genindch}) for $d=2$. In what follows, we introduce an alternative procedure of arriving at the inductive character,  which can be generalized more easily to the case of split-rank 2. 

As shown earlier, zonal spherical functions for $SO(1,2)/SO(1,1)$ are given by matrix element $t^\lambda(p,p')$ with $p=p'=0$, which can be thought as the generalization for the case of $SO(3)/SO(2)$ where $P_\ell(\cos \theta)= \langle \ell,0|R_z(\phi) R_y(\theta) R_z(\psi)|\ell, 0\rangle$.  The absence of $\psi$ for dependence $SO(3)/SO(2)$ can be understood as due to action of $R_z(\psi)$ on a basepoint on the z-axis. As such, the inductive character, $W_K(\theta,\phi)$ serves as the generator for spherical function, $Y_{\ell,m}$. The zonal spherical function can be obtained by averaging the inductive character $W_K(\theta,\phi)$ over the subgroup, $R_z(\phi) $. 

Similarly, In moving to $SO(1,2)/SO(1,1)$, we replace $R_z(\phi) R_y(\theta) R_z(\psi)$ by $B_{23}(\zeta) B_{12}(\eta) B_{23}(\zeta')$. The absence of $\zeta'$ dependence should be understood due to the action of  $B_{23}$ on the basepoint $\xi_a = (0,1,1)$. Notice that we have moved to the null cone and $B_{23}$ can be treated as an isotropy only upto a scale and the dependence on $ \zeta' $ does not drop out immediately. To overcome this technical issue, let us define a space of matrices $X = g\cdot P \cdot g^{T}$ where $P$ is diagonal, with $\text{diag} (P)=(0,1,-1)$. The important feature of $ P $ is that $ h\cdot P \cdot h^{T} = P $. That is, the dependence on $ \zeta' $ drops out explicitly. Further, because of this property, the space of matrices $ X $ carries an image of each element of the coset $ g\in G/H $.

Since we are now working with non-Riemannian manifolds, in the framework of induced representations, we decompose a semigroup element $ g = na_+h $. The induced representation of this group element is given by
\begin{equation}
	{ F}(g) = e^{\lambda\eta_I}F(h) = e^{\lambda\eta_I}e^{ip\zeta}
\end{equation}
where we write $ \eta_I $ to remind ourselves that this is the parametrization of $ a $ in the Iwasawa-like decomposition. Further, $ \lambda = \widetilde{\lambda} - \rho = \widetilde{\lambda} - 1/2 $. Setting $ p = 0 $ takes us to the coset $ G/H $.

The same semigroup element $ g $ also carries the Cartan-like decomposition $ g = h_1a_+h_2 $. The function $ {F}(g) = e^{\lambda\eta_I} $ is on the one-sided coset $ G/H $. For $ H $-bi-invariant functions, we can expand this function in a basis such that $ h_1 $ is diagonal. We can then project out the invariant vector by
\begin{equation}
	\varphi_{{\lambda}}(a) = c'_{\lambda, 0} \int_{h_1} e^{\lambda\eta_I}\ \text{d}h_1\ ,
\end{equation}
much in analogy with the $ SO(3) $ discussion. The challenge then is to find the dependence of $ \eta_I $ on $ h_1(s) $.

Recall that under $ SO(2)\cong U(1) $, the inductive character $ e^{i\ell\theta} $ is simply the trace of the one dimensional representation. This identification can be made in general by performing a Cayley transform so that $A$ is diagonal. For the compact case, this is often referred to as the maximal torus. We can generalize this appropriately for the non-compact case and write more explicitly. After the Cayley transformation, $ A $ can be written as:
\begin{equation}
	A = \begin{pmatrix}
		e^{\eta} & 0 & 0\\
		0 & 1 & 0\\
		0 & 0 & e^{-\eta}   \label{eq:Cayley3}
	\end{pmatrix}.
\end{equation}
This is in the form of a $ 3\times3 $ matrix since we are working with the adjoint representation of $ SO(1,2) $.

In the space of matrices $ X $ for the two decompositions of $ g $ discussed, we have that 
\begin{equation}\label{eq:double}
	na^I_+\cdot P \cdot a_+^{I\ T}n^{T} = h_1a_+\cdot P\cdot a_+^{T}h_1^{T}.
\end{equation}
We can solve this equation for $ e^{\eta_I} $. Notice that in the left hand side, the first entry in the matrix is $ e^{2\eta_I} $. We can therefore find the square of the inductive character by finding the first matrix element on the right hand side. We therefore have:
\begin{equation}
	\varphi_H(\eta,\lambda)= Q_{\lambda}(\cosh \eta)= 
	\int ds (\cosh \eta+ \cosh s \sinh \eta)^\lambda\ .
\end{equation}
in agreement with that obtained earlier.
\subsection{Rank 2 Case: $ SO(2,2) $}\label{app:sphrDoubling2}

For the rank-2 case, the same construction applies. The adjoint representation now is in terms of $ 4\times 4 $ matrices. We need to solve Eq. (\ref{eq:double}) for two characters $ e^{y_I},e^{\eta_I} $. In the Cayley basis of Eq. (\ref{eq:Cayley4}), we have
\begin{equation}
	a\cdot P\cdot a^{T} = \begin{pmatrix}
		e^{2y} & 0 & 0 & -1\\
		0 & -e^{2\eta} & -1 & 0\\
		0 & -1 & -e^{-2\eta} & 0\\
		-1 & 0 & 0 & e^{-2y}
	\end{pmatrix}\ .
\end{equation}
It is sufficient to consider the $ 2\times2 $ minor of $ na\cdot P\cdot a^{T}n^{T} $ which is given by
\begin{equation}
	\text{Minor}_1 = \begin{pmatrix}
		e^{2y_I}
	\end{pmatrix};\quad\quad\text{Minor}_2 = \begin{pmatrix}
		e^{2y_I} & -2e^{2y_I}t_+\\
		-2e^{2y_I}t_+ & 4e^{2y_I}t_+^2 -e^{2\eta_I}
	\end{pmatrix}\ .
\end{equation}
The determinant is given by $ |\text{Minor}_2| = -e^{2y_I}e^{2\eta_I} $. We can therefore solve for both $ e^{y_I} $ and $ e^{\eta_I} $ in terms of $ h_1(\alpha,\beta) $. This leads to  Eq. (\ref{eq:chard2}).

\bibliographystyle{unsrt}
\bibliography{conformald.bib}

\end{document}